\documentclass[aps,prd,twocolumn,superscriptaddress,showpacs]{revtex4}

\usepackage{graphicx} \usepackage{color} \usepackage{natbib}

\usepackage{amsmath,amssymb,amsfonts}
\usepackage{epsf}
\usepackage{epsfig}
\usepackage{longtable}
\begin{document}


\title{Equation-of-state dependence of the gravitational-wave signal from the ring-down phase of neutron-star mergers}


\author{A.~Bauswein}
\affiliation{Max-Planck-Institut f\"ur
  Astrophysik, Karl-Schwarzschild-Str.~1, 85748 Garching, Germany}
\author{H.-T.~Janka}
\affiliation{Max-Planck-Institut f\"ur
  Astrophysik, Karl-Schwarzschild-Str.~1, 85748 Garching, Germany}
\author{K. Hebeler}
\affiliation{Department of Physics, The Ohio State University,
Columbus, OH 43210, USA}
\author{A.~Schwenk}
\affiliation{ExtreMe Matter Institute EMMI, GSI Helmholtzzentrum f\"ur
Schwerionenforschung GmbH, 64291 Darmstadt, Germany}
\affiliation{Institut f\"ur Kernphysik, Technische Universit\"at
Darmstadt, 64289 Darmstadt, Germany}


\date{\today}

\begin{abstract}
Neutron-star (NS) merger simulations are conducted for 38 representative microphysical descriptions of high-density matter in order to explore the equation-of-state dependence of the postmerger ring-down phase. The formation of a deformed, oscillating, differentially rotating very massive NS is the typical outcome of the coalescence of two stars with 1.35~$M_{\odot}$ for most candidate EoSs. The oscillations of this object imprint a pronounced peak in the gravitational-wave (GW) spectra, which is used to characterize the emission for a given model. The peak frequency of this postmerger GW signal correlates very well with the radii of nonrotating NSs, and thus allows to constrain the high-density EoS by a GW detection. In the case of 1.35-1.35~$M_{\odot}$ mergers the peak frequency scales particularly well with the radius of a NS with 1.6~$M_{\odot}$, where the maximum deviation from this correlation is only 60~meters for fully microphysical EoSs which are compatible with NS observations. Combined with the uncertainty in the determination of the peak frequency it appears likely that a GW detection can measure the radius of a 1.6~$M_{\odot}$ NS with an accuracy of about 100 to 200~meters. We also uncover relations of the peak frequency with the radii of nonrotating NSs with 1.35~$M_{\odot}$ or 1.8~$M_{\odot}$, with the radius or the central energy density of the maximum-mass Tolman-Oppenheimer-Volkoff configuration, and with the pressure or sound speed at a fiducial rest-mass density of about twice nuclear saturation density. Furthermore, it is found that a determination of the dominant postmerger GW frequency can provide an upper limit for the maximum mass of nonrotating NSs. The effect of variations of the binary setup are investigated and corresponding functional dependences between the peak frequency and radii of nonrotating NSs are derived. With higher total binary masses, correlations are  tighter for radii of nonrotating NSs with higher masses. The prospects for a detection of the postmerger GW signal and a determination of the dominant GW frequency are estimated to be in the range of 0.015 to 1.2 events per year with the upcoming Advanced LIGO detector.

\end{abstract}


\maketitle

\section{Introduction}
Neutron stars (NSs) are the smallest, densest, and most compact
stellar objects known to exist in the universe. The conditions in the
cores of NSs at several times nuclear saturation density ($
\rho_0=2.7\cdot10^{14}~\mathrm{g}/\mathrm{cm}^3$) have remained a mystery
since the discovery of these objects. In particular, the composition
and the equation of state (EoS) of high-density matter are only
incompletely known~\cite{2007PhR...442..109L,2007ASSL..326.....H,1996csnp.book.....G}. This results from the impossibility to directly
access NS matter by laboratory experiments, although some constraints
are provided by the properties of neutron-rich nuclei and by heavy-ion
collisions~\cite{2007PhR...442..109L,2011IJMPE..20.2077H}, as well as
from uncertainties in nuclear forces and in solving the strongly
interacting many-body problem (see e.g.~\cite{1996csnp.book.....G,2007PhR...442..109L,2007ASSL..326.....H}).

From the astrophysical point of view the EoS is crucial because it uniquely determines the NS structure and therefore the mass-radius relation of NSs, via the stellar structure equations (see e.g.~\cite{2007coaw.book.....C,2007PhR...442..109L,2007ASSL..326.....H,1996csnp.book.....G}). This one-to-one correspondence between the EoS and the mass-radius relation of NSs allows in principle to determine or at least to constrain the EoS from the simultaneous measurement of the mass and the radius of a NS. While NS masses have been measured very precisely in binary systems~\cite{1999ApJ...512..288T,2007coaw.book.....C,2007PhR...442..109L,2011A&A...527A..83Z}, the determination of radii has not yet been achieved with high precision (see e.g.~\cite{2007ASSL..326.....H,2007PhR...442..109L,2010ApJ...722...33S,2012arXiv1201.1680G} and references therein). However, mass measurements alone already constrain the possible EoSs because all mass-radius relations feature a maximum-mass configuration, where more massive objects are unstable and collapse to a black hole (BH). The detection of a pulsar with $(1.97 \pm 0.04)~M_{\odot}$ via the determination of the Shapiro delay in a NS-white dwarf binary rules out a variety theoretical descriptions of the EoS~\cite{2010Natur.467.1081D}.

To date about ten NS binaries are known, which contain at least one star visible as a pulsar. The stellar masses in these systems have partially been measured very accurately and cluster at about 1.35~$M_{\odot}$~\cite{1999ApJ...512..288T,2007coaw.book.....C,2007PhR...442..109L,2011A&A...527A..83Z}. Because of observational selection effects one expects many more NS binaries to exist, which is also predicted by theoretical models of the NS population~\cite{2008ApJ...680L.129B}. Observationally is has been confirmed that the NSs in a binary slowly approach each other because of the emission of gravitational waves (GWs), which are generated by the orbital motion and which carry away angular momentum and energy. As a consequence of these losses, it is the fate of (isolated) NS binaries to inspiral and finally to merge on timescales which are found to be of the order of several hundred Megayears for the known systems~\cite{2004ApJ...601L.179K,2007coaw.book.....C}.

NS mergers have been identified as events whose dynamics depend crucially on the high-density EoS (see e.g.~\cite{2009CQGra..26k4004F,2010CQGra..27k4002D,2010arXiv1012.0912R} for a review). Consequently, in numerical models potentially observable signals of NS coalescences have been found to show an EoS dependence. For instance, short gamma-ray bursts are speculated to originate from the BH-torus system forming milliseconds after the merging of the binary components~\cite{2007PhR...442..166N}. In particular the survival time of the merger remnant before BH formation and the torus mass providing the energy input to the burst has been found to depend on the binary masses and the EoS~\cite{2006MNRAS.368.1489O,2006PhRvD..73f4027S}, where also the accurate treatment of thermal effects plays a role~\cite{2010PhRvD..82h4043B}. In the most extreme cases, the direct collapse of the merger remnant to a BH or a very long phase of stability before it becomes gravitationally unstable, might disfavor the gamma-ray burst production, a possibility that depends sensitively on the mass of the hypermassive remnant relative to the NS mass limit of the EoS. It has also been shown that the amount of matter becoming gravitationally unbound from the merger site, for a given binary configuration is influenced by the adopted model of supranuclear matter~\cite{2007A&A...467..395O}. These ejecta (typically a few $10^{-3}~M_{\odot}$) are observationally relevant because nucleosynthetic reactions in the outflow can form the so-called r-process elements, heavy neutron-rich nuclei whose astrophysical production site has not been identified yet (see~\cite{2011ApJ...738L..32G} for calculations which take into account in particular the detailed properties of the EoSs). Observationally not only the nucleosynthesis yields are important but also the possibility to detect an electromagnetic counterpart of a NS merger~\cite{1998ApJ...507L..59L,2005astro.ph.10256K,2010MNRAS.406.2650M,2011ApJ...736L..21R,2011ApJ...738L..32G}. The signal, which is powered by the radioactive decay of the freshly synthesized nuclei, may provide information about the host galaxy of an event.

However, it appears unlikely that the aforementioned observations can significantly constrain the properties of the high-density matter because the EoS effects are too indirect or still not fully explored, and additional information e.g. about the involved binary masses is required. More promising in this respect is the detection of GWs from inspiraling and merging NS binaries by Advanced LIGO~\cite{2010CQGra..27h4006H} or Advanced Virgo~\cite{Acernese:2006bj} going into operation within the next years. Indeed, NS mergers are among the prime targets for these detectors with expected event rates of 0.4 to 400~\cite{2010CQGra..27q3001A}. It is clear that the determination of the involved NS masses will be possible by the matched filtering technique~\cite{1994PhRvD..49.2658C}. But, it is crucial for constraining the EoS to measure simultaneously the mass and the radius of a NS. Information on the radius can be obtained from the late inspiral signal when finite-size effects become important and cause deviations from a point-particle behavior and finally terminate the inspiral phase (see~\cite{2002PhRvL..89w1102F,2009PhRvD..79l4033R,2010PhRvL.105z1101B} and references therein). In~\cite{2009PhRvD..79l4033R} it was worked out that NS radii can be determined with an uncertainty of about 1~km. Only symmetric binaries with two NSs of 1.35~$M_{\odot}$ have been considered there, employing simplified EoSs. The effects of the initial mass ratio, the total binary mass, and especially of microphysical EoSs have been left unexplored. Apart from these open questions, alternative methods for determining NS radii are desirable in any case.

Besides the impact on the inspiral signal, different EoSs have been recognized to have a strong influence on the postmerger GW signal in the cases when the merging stars do not directly form a BH~\cite{1994PhRvD..50.6247Z,2005PhRvL..94t1101S,2005PhRvD..71h4021S,2006PhRvD..73f4027S,2007PhRvL..99l1102O,PhysRevD.81.024012,2011PhRvD..83l4008H,2011PhRvL.107u1101S}. Based on calculations for a limited set of EoSs it has been shown that the EoS determines the total binary mass for which such a prompt collapse does not occur, which may provide a handle on the maximum mass of NSs~\cite{2005PhRvL..94t1101S,2011PhRvD..83l4008H}. This procedure, however, requires additional knowledge of the radius of a 1.4~$M_{\odot}$ NS, which is not yet available. Furthermore, in the simulations of~\cite{2011PhRvD..83l4008H} shock heating was assumed to be very inefficient.

In order to identify the EoS influence on the postmerger GW emission the full functional dependence of the GW signal on the EoS has to be determined. With a systematic survey based on a large set of representative candidate EoSs we have demonstrated a strong correlation between the dominant GW frequency of the postmerger phase and the radii of nonrotating NSs~\cite{2012PhRvL.108a1101B}. This relation allows to deduce the supranuclear EoS when the dominant postmerger GW frequency is observationally determined, and therefore represents a complementary approach to methods relying on the inspiral signal.

The purpose of this paper is 1) to extend our recent study of the EoS effects on the postmerger GW signal by considering an enlarged set of 38 microphysical EoSs, 2) to investigate the influence of the total binary mass and the initial mass ratio, 3) to explore the scatter inherent to the relations between the postmerger GW frequency and NS radii and thus to specify the accuracy to which NS radii can be determined, 4) to evaluate additional dependences between the late-time GW signal and EoS features, and finally 5) to present details on the employed EoSs, the detectability of the GW signature, and the involved uncertainties.

The paper is organized as follows. First we introduce the code used for our study focussing in particular on the implementation of the EoS. Sect.~\ref{sec:eos} outlines details on the considered EoSs and the corresponding stellar properties of NSs. Then we describe the simulations and the analysis of the GW emission including a discussion of the detectability (Sect.~\ref{sec:simulations}). In Sect.~\ref{sec:stellar} the relations between features of the GW signal and stellar parameters are described. Here, we also discuss the spread in the presented relations and interpret the findings. Sect.~\ref{sec:thermo} examines direct correlations between the dominant GW frequency of the postmerger phase and microphysical properties of the EoS. Variations of the binary mass and the mass ratio are investigated in Sect.~\ref{sec:binpara}. Finally, Sect.~\ref{sec:con} presents a summary and our conclusions.

Note that in this paper we refer to the gravitational mass of an object in isolation if not specified otherwise.

\section{Methods and Initial Data} \label{sec:methods}
The simulations of our survey are conducted with a three-dimensional relativistic Smoothed Particle Hydrodynamics (SPH) code, where the gravitational fields are obtained by solving the Einstein equations within the conformal flatness approximation~\cite{1980grg..conf...23I,1996PhRvD..54.1317W}. A detailed description of the code can be found in~\cite{2002PhRvD..65j3005O,2007A&A...467..395O,PhysRevD.81.024012}. Within the SPH approach the stellar fluid is modelled by a set of particles with constant rest mass, which are considered not to be pointlike but to be spread out over a small spatial domain. Hydrodynamical quantities are assigned to the particles and are evolved comoving with the fluid, i.e. according to the Lagrangian formulation of hydrodynamics. Specifically, we evolve a set of so-called conserved quantities, the conserved rest-mass density $\rho^*$, the conserved specific momenta $\hat{u}_i$ and the conserved energy density $\tau$, by a system of coupled ordinary differential equations. These conserved quantities are functions of the metric potentials and the ``primitive'' quantities, the rest-mass density $\rho$, the coordinate velocities $v_i$, the specific internal energy $\epsilon$, and the pressure $P$. In every timestep the primitive variables are recovered from the evolved quantities ($\rho^*$,$\hat{u}_i$,$\tau$) by an inversion procedure, which requires to invoke the EoS $P=P(\rho,Y_{\mathrm{e}},\epsilon)$ with the electron fraction $Y_{\mathrm{e}}$ to close the system of hydrodynamical equations. This step is necessary because the pressure explicitly occurs in the evolution equations. The electron fraction is evolved according to $\frac{dY_{\mathrm{e}}}{dt}=0$, i.e. advected with the fluid, because the timescale of lepton number transport through neutrinos is long compared to the dynamical timescale (at least for the bulk of the stellar matter).

Concerning technical aspects the EoSs employed in our survey can be grouped into four categories.
\begin{itemize}
\item[(i)] Some EoSs provide the full dependence $P=P(\rho,Y_{\mathrm{e}},T)$, where the temperature $T$ is obtained by inverting $\epsilon=\epsilon(\rho,Y_{\mathrm{e}},T)$. These EoSs are available in the form of tables, where we use a trilinear interpolation scheme.

\item[(ii)] Many published EoSs describe high-density matter at zero temperature and in equilibrium with respect to weak interactions (so-called neutrinoless beta-equilibrium with the neutrino chemical potential $\mu_{\nu}=0$). Thus, for these ``barotropic'' EoSs the pressure is given as a unique function of the rest-mass density or, alternatively, the energy density $e=\rho(1+\epsilon)$. In these cases a thermal component of the EoS can be taken into account approximatively by means of an ideal gas-like ansatz. This is necessary because temperature effects become important when the NSs collide. To this end we employ a scheme detailed in~\cite{2010PhRvD..82h4043B}, where a thermal contribution of the specific internal energy is computed via $\epsilon_{\mathrm{th}}=\epsilon-\epsilon_{\mathrm{cold}}(\rho)$. The rest-mass density $\rho$ and the specific internal energy $\epsilon$ are given by the evolution of the hydrodynamical scheme. The cold contribution $\epsilon_{\mathrm{cold}}$ is determined by $\rho$ only by calling the ``cold'' barotropic EoS. The thermal contribution to the pressure reads $P_{\mathrm{th}}=(\Gamma_{\mathrm{th}}-1)\rho\epsilon_{\mathrm{th}}$ with the ideal-gas index $\Gamma_{\mathrm{th}}$. It is added to the cold component to yield the total pressure $P=P_{\mathrm{cold}}(\rho)+P_{\mathrm{th}}$. Here, the cold part for a given rest-mass density is again obtained only from the barotropic EoS. The validity and the involved uncertainties of this approach have been discussed in~\cite{2010PhRvD..82h4043B}. There it has been shown that in the context of NS mergers a value of $\Gamma_{\mathrm{th}}=2$ represents a suitable choice to mimic thermal effects of nuclear matter. We therefore use $\Gamma_{\mathrm{th}}=2$ if not mentioned otherwise. In particular, the collapse timescale of the merger remnant is reproduced very well by this choice of $\Gamma_{\mathrm{th}}$.

\item[(iii)] Additional barotropic EoSs which are not available to us as tables
can be included by their parametrized versions provided
in~\cite{2009PhRvD..79l4032R} in form of piecewise polytropes. Also
for this class of models we make use of the thermal ideal-gas extension outlined
above. One should bear in mind that the polytrope fits imply approximations that will have to be considered later in our discussion. First, the parametrized versions do not perfectly match the underlying microphysical EoSs (see~\cite{2009PhRvD..79l4032R}) and hence simulation results may differ because of the use of the piecewise polytrope instead of the original EoS. Second, the properties (e.g. radii) of cold NSs obtained from the fitted versions are slightly different, thus shifting our results when presented in dependence of these NS properties.

\item[(iv)] The fourth class of EoSs is based on the calculations of~\cite{2010PhRvL.105p1102H}. Detailed computations of the EoS below nuclear saturation density using chiral effective field theory interactions are extended by piecewise polytropes for higher densities requiring a maximum mass of NSs above 1.97~$M_{\odot}$. At densities below $5\cdot 10^{10}$g/cm$^3$ we extend the original tables with an additional crust. The resulting barotropic relations are again supplemented with the ideal-gas component for thermal effects. 
\end{itemize}

Note that the classes (ii) to (iv) involve an additional approximation beside the treatment of thermal effects because matter is assumed to adjust instantaneously to beta-equilibrium as given by the barotropic EoSs (see the discussion in~\cite{2010PhRvD..82h4043B}). In these cases the electron fraction is not used as an input for the EoS call because the pressure is entirely determined by the rest-mass density and the specific internal energy.

The details of the different EoSs employed here, their properties and the underlying microphysical models are discussed in Sect.~\ref{sec:eos}. Note that for indicating the different classes of EoSs in figures we use the same color scheme throughout this paper (class (i) in red, class (ii) in black, class (iii) in green, class (iv) in blue).

General relativistic gravity cannot be treated within the SPH scheme. Therefore, the five coupled nonlinear elliptic equations resulting from the Einstein equations by imposing a confromally flat spatial metric are solved on an overlaid grid covering well the binary orbit and the postmerger remnant at later times. By the conformal flatness approximation the effects of GWs are neglected, for which reason an additional GW backreaction scheme is implemented in the code to account for the loss of energy and angular momentum by the emission of gravitational radiation. This procedure requires to solve additional elliptic equations to compute corrections to the conformally flat metric. Details are reported in~\cite{2007A&A...467..395O}.

Magnetic fields or neutrino transport physics are not included in our model, because it has been shown that these effects have only little influence on the dynamics of a merger as long as the initial field strength inside the stars is below $~10^{14}$~G and only a few dynamical timescales are tracked after the merging~\cite{2008PhRvD..78b4012L,2009MNRAS.399L.164G,2011PhRvD..83d4014G,1996A&A...311..532R,1997A&A...319..122R}.

Our simulations start from a quasi-equilibrium orbit about two to three revolutions before the actual merging of the binary components. Initially we impose a nonrotating (irrotational) velocity profile of the NSs, because the inspiral time is not sufficient to enforce synchronous rotation \cite{1992ApJ...400..175B,1992ApJ...398..234K}. This state can be considered to be a good approximation also for NSs spinning with millisecond periods because even in such cases the rotation is still slow compared to the orbital period. Furthermore, the initial stars are assumed to have zero temperature and to be in neutrinoless beta-equilibrium. The two NSs are set up typically with about 340,000 SPH particles, but we perform additional higher resolved calculations to ensure that our results do not depend on the resolution (see Subsect.~\ref{ssec:gwana}).

\section{Equations of state and stellar properties} \label{sec:eos}
The EoSs used in our survey are listed in Tab.~\ref{tab:models}, where it is also indicated to which technical category a particular EoS belongs. Seven out of the 38 models include thermal effects consistently (class i). The remaining EoSs have been computed for high-density matter at zero temperature and in equilibrium with respect to weak interactions (barotropic EoSs).

The EoSs of classes (i), (ii) and (iii) have been derived within
different frameworks, including nonrelativistic and relativistic
nuclear energy-density functionals, Brueckner-Hartree-Fock
calculations, and phenomenological models, such as the liquid-drop model (see e.g.~\cite{1996csnp.book.....G,1999Weber,2007ASSL..326.....H} for an overview). All except for two microphysical models
describe interacting neutrons, protons and electrons (possibly also
muons), where some also include hyperons (Glendnh3, H3, H4). MIT60 and
MIT40 consider strange quark matter and are discussed below. In
addition, the nonzero temperature EoSs include positrons and photons.
Also the models for the (nuclear) interactions have been varied, which
implies that the EoSs explore a range of nuclear experimental data.
Some EoSs take into account three-body forces.

\begin{figure}
\includegraphics[width=7.cm]{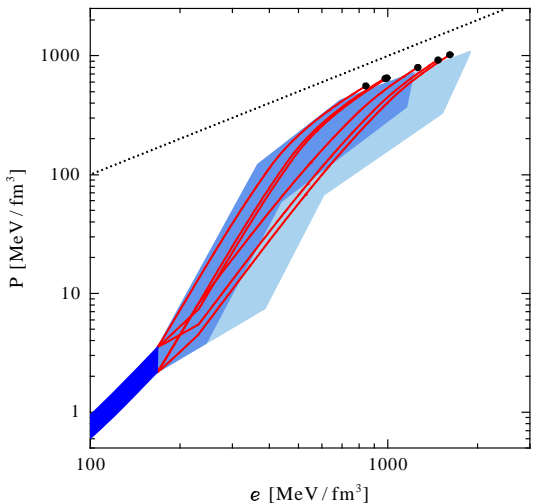}
\caption{\label{fig:heb}Pressure versus energy density for the 6 EoSs of class (iv) (red
lines). The dark blue band at low densities shows the results of microscopic calculations based on chiral effective field theory interactions~\cite{2010PhRvC..82a4314H,2010PhRvL.105p1102H}. The light blue extension band covers possible polytropic extensions compatible with causality and NSs of $M \geqslant 1.97 M_\odot$. The darker blue extension is for $M \geqslant 2.4 M_\odot$~\cite{2010PhRvL.105p1102H,Hebelerprep}. The black dots mark the central density of the
maximum-mass configuration and the dotted line indicates the causal limit $P = e$.}
\end{figure}

Figure~\ref{fig:heb} shows the 6 EoSs of class (iv). The dark blue band at low
energy densities $e$ shows the uncertainty band of the pressure
obtained from microscopic calculations based on chiral two- and three-nucleon interactions~\cite{2010PhRvC..82a4314H,2010PhRvL.105p1102H}. This band is extended in a general way to higher densities using piecewise polytropes as a function of the rest-mass density~\cite{2010PhRvL.105p1102H,Hebelerprep}. The extensions are
constrained by the causality of the polytropes and the heaviest observed NS~\cite{2010Natur.467.1081D}. The light blue extension band covers all possible EoSs that are
compatible with the constraints based on chiral effective field theory
interactions at low densities and are able to support NSs of mass $M
\geqslant 1.97 M_\odot$~\cite{2010PhRvL.105p1102H,Hebelerprep}. In addition, we overlay a dark blue extension band, which is obtained by requiring a maximum NS mass
$M \geqslant 2.4 M_\odot$. The 6 EoSs of class (iv) within these bands are a representative sample of EoSs characterized by different stiffness fulfilling these constraints. For our choice of polytopes the EoSs become acausal (the speed of sound exceeds the velocity of light) beyond a density $\rho_{\mathrm{limit}}$ where the speed of sound is $v_s (\rho_{\mathrm{limit}})= c$. We replace this unphysical regime by $P=e - {\rm const.}$, which is by construction causal, and choose the constant in a way that the pressure is continuous. The revised parts of the EoSs can be recognized by the short pieces of the lines sticking out of the blue bands at high densities. (The upper edge of the blue bands (parallel to the dotted $P=e$ line) indicates $\rho_{\mathrm{limit}}$.) Note that the EoSs are only replaced at very high densities slightly below the central density $\rho_{\mathrm{max}}$ of the maximum-mass configuration of static NSs shown by the black dots. Because of this, the density regime above $\rho_{\mathrm{limit}}$ does not play a role during the evolution of NS mergers for the cases considered here, and the replacements of the acausal part of the EoSs affect only the exact values of stellar properties of very massive static NSs.

\begin{figure}
\includegraphics[width=8.9cm]{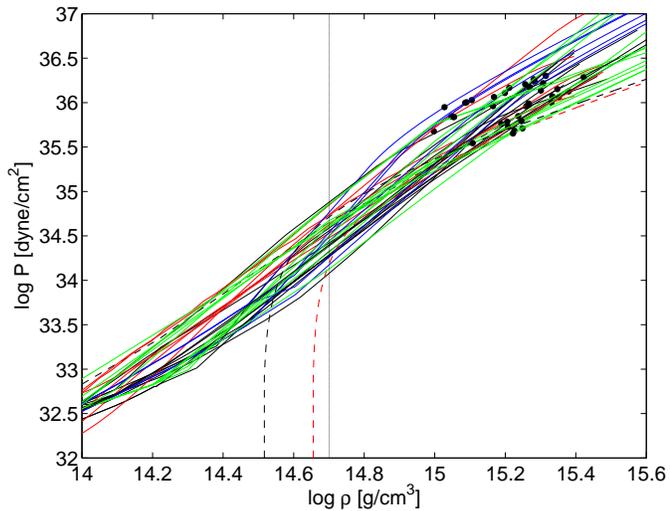}
\caption{\label{fig:eos}Pressure as a function of the rest-mass density for all candidate EoSs at zero temperature and in neutrinoless beta-equilibrium. Microphysical EoSs which provide the full temperature dependence are shown in red (class i), microphysical EoSs that are given at zero temperature and in neutrinoless beta-equilibrium are displayed in black (class ii), green lines correspond to piecewise polytropes (class iii) and EoSs of class iv are depicted in blue. Dashed lines are used for strange quark matter EoSs. The vertical line marks a rest-mass density of $\rho=10^{14.7}~\mathrm{g/cm}^3=1.85 \rho_{0}$. The filled circles indicate the central rest-mass density of the maximum-mass TOV solution for every EoS.}  
\end{figure}

\begin{figure}
\includegraphics[width=8.cm]{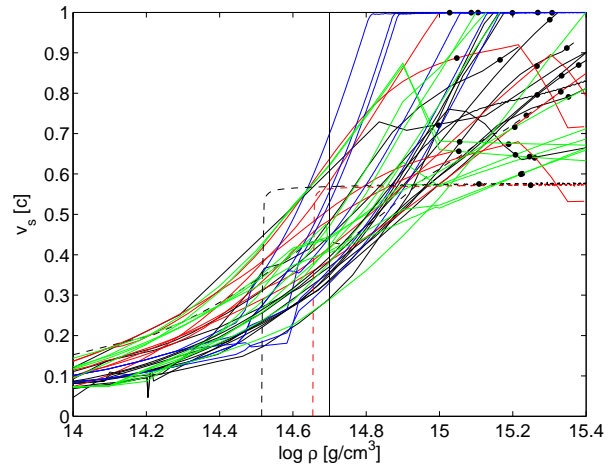}
\caption{\label{fig:vsound}Sound speed for all EoSs used in our survey. Colors and line style have the same meaning as in Fig.~\ref{fig:eos}. The vertical line marks a rest-mass density of $\rho=10^{14.7}~\mathrm{g/cm}^3=1.85 \rho_{0}$. The filled circles indicate the central rest-mass density of the maximum-mass TOV solution for every EoS, if not visible the EoS becomes acausal for stable TOV solutions.}
\end{figure}

\begin{figure}
\includegraphics[width=8.9cm]{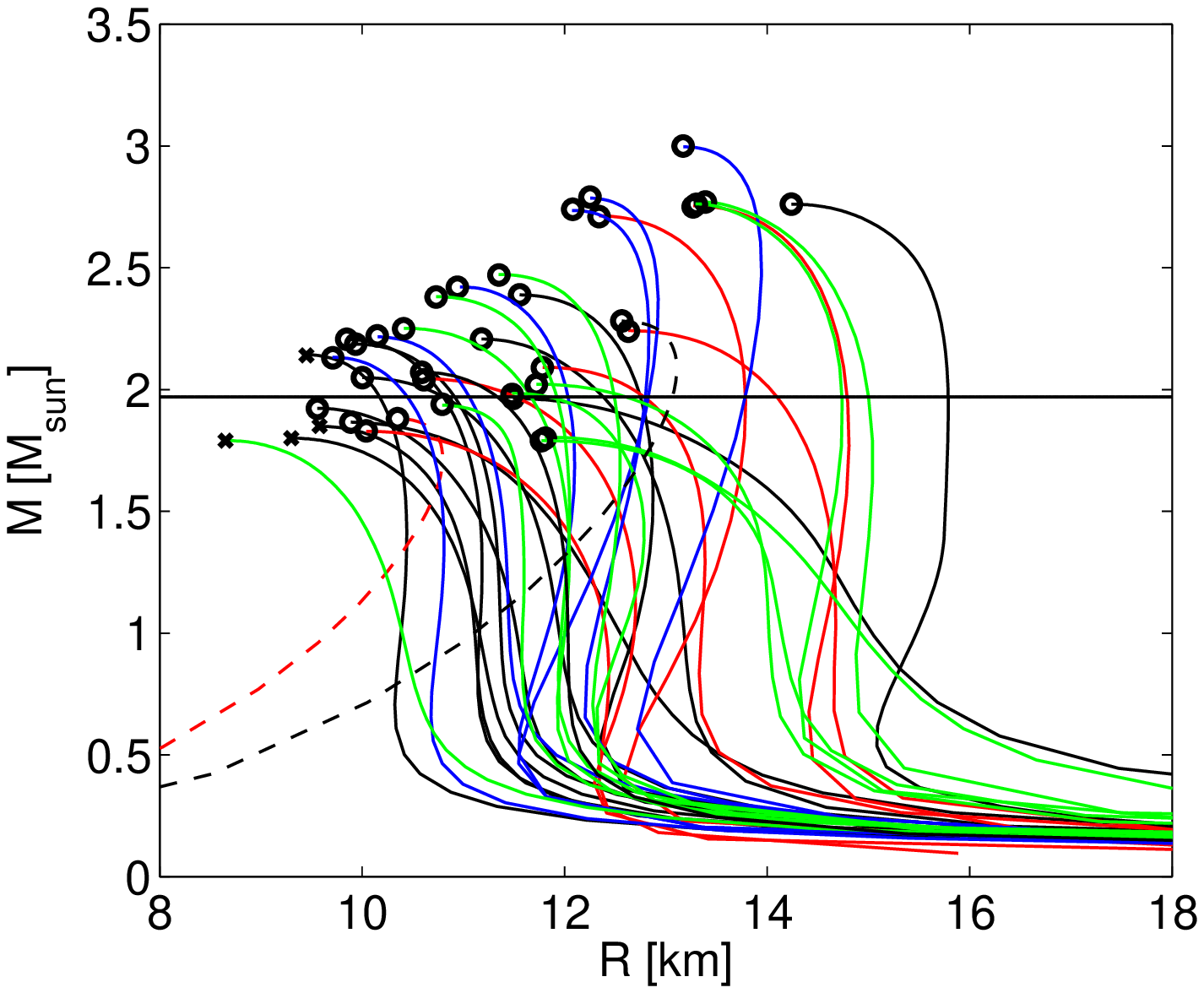}
\caption{\label{fig:tov}Mass-radius relations for all EoSs with the gravitational mass $M$  in isolation and the areal radius $R$. The color scheme is the same as in Fig.~\ref{fig:eos}. The dashed lines denote mass-radius relations for strange quark matter EoSs. The horizontal line corresponds to the observed 1.97~$M_{\odot}$ NS~\cite{2010Natur.467.1081D}. For EoSs where the merger of two stars with 1.35~$M_{\odot}$ leads to the prompt formation of a BH the maximum-mass configuration is indicated by a cross. Maximum-mass configurations depicted by a circle correspond to EoSs where in the simulation of this binary setup the formation of a differentially rotating object is found.}
\end{figure}

The thermodynamic properties of all EoSs available for our study are illustrated in Fig.~\ref{fig:eos} showing the pressure $P$ as a function of the rest-mass density $\rho$. One observes that in the range of one to several times nuclear saturation density ($\rho_0=2.7\cdot 10^{14}~\mathrm{g/cm}^3$) the spread of the pressure between the various models is nearly one order of magnitude. It has been pointed out that the pressure at a density between one and two times $\rho_0$ serves as a characteristic quantity that correlates with the radius of a 1.4~$M_{\odot}$ NS~\cite{2001ApJ...550..426L,2007PhR...442..109L}. We provide the pressure at $1.85 \rho_0$ in Tab.~\ref{tab:models}, which is chosen in order to be consistent with the fiducial rest-mass density for the piecewise polytrope fits presented in~\cite{2009PhRvD..79l4032R}. It ranges from 1.18~$\cdot 10^{34}~\mathrm{dyne/cm}^2$ to 9.56~$\cdot 10^{34}~\mathrm{dyne/cm}^2$. In comparison to the high-density regime the pressure differences decrease towards lower densities (not shown in Fig.~\ref{fig:eos}), see e.g.~\cite{2007ASSL..326.....H,2010PhRvL.105p1102H}.

As another EoS property we display the speed of sound in Fig.~\ref{fig:vsound} for all EoSs. The diversity among the models is obvious, where the differences are larger at higher densities. We extract the sound speed at $1.85 \rho_0$ as a characteristic quantity for all EoSs and give the values in Tab.~\ref{tab:models}. The values vary between 0.29 and 0.70 times the speed of light. Note that for some EoS models the sound velocity exceeds the speed of light at high densities, i.e. the EoSs become acausal (e.g. the eosAU, eosUU and APR EoS).

The variety of the microphysical models employed in our study is also reflected in the mass-radius relations of non-rotating NSs, which are obtained by solving the relativistic equations of hydrostatic equilibrium (Tolman-Oppenheimer-Volkoff (TOV) equations). The TOV equations yield a unique relation between the NS mass and radius for each EoS, which allows to characterize an EoS by the resulting stellar properties. Figure~\ref{fig:tov} displays the TOV solutions for all EoSs. Red curves correspond to EoSs whose temperature dependence is given by the microphysical model, while fully microphysical zero-temperature EoSs (which in the simulations are supplemented with an ideal-gas component to mimic thermal effects) are shown with black lines. The green curves exhibit the mass-radius relations of EoSs of class (iii), while EoSs belonging to class (iv) are shown in blue.

All mass-radius relations feature a maximum-mass configuration. More massive objects collapse to a BH if not additionally supported, e.g. by rotation or thermal effects. The corresponding solutions are marked with symbols in Fig.~\ref{fig:tov}, and their stellar properties are listed in Tab.~\ref{tab:models}. Furthermore, Tab.~\ref{tab:models} gives the central energy density of these maximum-mass configurations, they range from $1.35\cdot 10^{15}~\mathrm{g/cm}^3$ to $3.80\cdot 10^{15}~\mathrm{g/cm}^3$. The corresponding rest-mass densities are indicated by circles in Figs.~\ref{fig:eos} and~\ref{fig:vsound}. Note that EoSs where the dot is not visible in Fig.~\ref{fig:vsound} become acausal (speed of sound larger than the speed of light) at densities below the maximum density.

Our survey considers EoSs that lead to maximum masses in the range of 1.79~$M_{\odot}$ to 3.00~$M_{\odot}$ and to corresponding radii, denoted as $R_{\mathrm{max}}$, spanning from 8.65~km to 14.30~km (Fig.~\ref{fig:tov}, Tab.~\ref{tab:models}). There has not been any special selection procedure for the EoSs, except that we require a maximum mass larger than roughly 1.8~$M_{\odot}$. This being fulfilled  we include every EoS that is available to us. The lower bound of about 1.8~$M_{\odot}$ is motivated by the discovery of a NS with a gravitational mass of $(1.97 \pm 0.04)~M_{\odot}$ \cite{2010Natur.467.1081D}. This measured mass is indicated as horizontal line in Fig.~\ref{fig:tov}. This detection practically rules out some EoSs of our sample with $M_{\mathrm{max}}$ below the limit. We do not dismiss such excluded models because they may still provide a viable model at lower densities (see also Sect.~\ref{ssec:interpretation}). For instance, during the first 5~ms after merging the central density in the merger remnant described by the excluded LS180 EoS remains below the central density of a nonrotating 1.5~$M_{\odot}$ NS modeled by this EoS. For such ``low-mass'' stars the mass-radius relations of excluded EoSs are partially similar to those obtained from EoSs compatible with the observation of~\cite{2010Natur.467.1081D}. Hence, in the corresponding density regimes relevant for the low-mass stars and the merger remnant such EoSs can still yield a viable description of high-density matter. In addition to that, the inclusion of EoSs with relatively low $M_{\mathrm{max}}$ extends (maybe artificially) the range of variations of stellar parameters, and correlations between NS properties and GW characteristics that hold over a wider parameter range can be inferred easier. We note that all of the four technical EoS categories cover a similar range of stellar parameter values. Only the mass-radius relations of class (iv) lie in a more narrow band, which was the main result of~\cite{2010PhRvL.105p1102H}.

A common feature of most EoSs is a relatively small variation of the NS radius between about 0.5~$M_{\odot}$ and about $(M_{\mathrm{max}}-0.5~M_{\odot})$. This suggests to use the radii in this mass range as a characteristic feature of a given EoS.

Finally, the MIT60 and MIT40 EoSs deserve a comment. These models describe absolutely stable strange quark matter within the MIT bag model~\cite{PhysRevD.9.3471,Farhi:1984qu}, i.e. a deconfined quark phase with an energy per baryon lower than the one of nucleonic matter ($E/A=$860~MeV for MIT60 and $E/A=$844~MeV for MIT40). As a consequence of the strange matter hypothesis~\cite{PhysRevD.4.1601,PhysRevD.30.272} underlying these two EoSs, the compact stars observed in the universe, commonly referred to as NSs, would actually be strange quark stars (consisting of strange quark matter). This possibility has not yet been ruled out theoretically or observationally (see e.g.~\cite{1996csnp.book.....G,1999Weber,2007ASSL..326.....H} for details and for observational consequences discriminating this scenario from ordinary NS; see~\cite{PhysRevD.81.024012,2009PhRvL.103a1101B} for the consequences of this hypothetical state of matter in the context of compact binary mergers). As a striking difference to nucleonic NSs, strange quark stars show an inverse mass-radius relation typical of this class of objects because of the self-binding of strange quark matter. The particular model MIT60 with $M_{\mathrm{max}}=1.88~M_{\odot}$ is excluded by the observation of the two-solar-mass pulsar. The MIT40 EoS, however, is compatible with present knowledge. For the MIT40 EoS belonging to class (ii), we adopt $\Gamma_{\mathrm{th}}=1.34$.

Note that throughout this paper we use the more common term NS instead of compact star for all compact stellar objects including strange quark stars. With ``purely'' or ``fully'' microphysical EoSs we refer to models of class (i) or (ii), which do not involve piecewise polytropes (see Sect.~\ref{sec:methods}). Moreover, in this paper ``accepted'' EoSs denote models which are compatible with the detection of the 1.97~$M_{\odot}$ NSs taking into account the error bars of the observation by~\cite{2010Natur.467.1081D}.

\begin{table*}
\caption{\label{tab:models}Models for the EoS with their thermodynamical properties and resulting stellar parameters. $M_{\mathrm{max}}$ is the maximum mass of nonrotating NSs with the radius $R_{\mathrm{max}}$ corresponding to this maximum-mass configuration. The next three columns provide the radii of nonrotating NSs with 1.35~$M_{\odot}$, 1.6~$M_{\odot}$ and 1.8~$M_{\odot}$. $P$ is the pressure at a fiducial rest-mass density of 1.85 times nuclear saturation density. The sound speed at the same reference density is given in the eighth column. The central energy density of the maximum-mass configuration of the TOV solutions is listed as $e_{\mathrm{max}}$. ``Class'' identifies how an EoS is implemented in our code (see Sect.~\ref{sec:eos} for details). References for the EoSs are listed in the last column. For EoSs marked with an ``x'' in parentheses the merging of two NSs with 1.35~$M_{\odot}$ leads to the direct formation of a BH.}
 \begin{ruledtabular}
 \begin{tabular}{|l|l|l|l|l|l|l|l|l|l|l|}
EoS      & $M_{\mathrm{max}}$ & $R_{\mathrm{max}}$ & $R_{1.35}$ & $R_{1.6}$ & $R_{1.8}$ &  $P(1.85\rho_0)$    & $v_{\mathrm{s}}(1.85\rho_0)$ & $e_{\mathrm{max}}$ & class & Ref. \\
         & $[M_{\odot}]$ & [km]   & [km]  & [km]  &  [km]   & $[10^{34}~\mathrm{dyne/cm^2}]$  &  [c]             & $[10^{15}~\mathrm{g/cm^3}]$  &&      \\ \hline
GS1      & 2.75 & 13.27 & 14.72 & 14.79 & 14.81 & 6.9920 &   0.6348   & 1.55  & i    &       \cite{2011PhRvC..83c5802S} \\ \hline
LS375    & 2.71 & 12.34 & 13.56 & 13.71 & 13.78 & 5.3904 &   0.5791   & 1.75  & i    &       \cite{1991NuPhA.535..331L} \\ \hline
Shen     & 2.24 & 12.63 & 14.64 & 14.53 & 14.35 & 5.3480 &   0.5202   & 1.40  & i    &       \cite{1998NuPhA.637..435S} \\ \hline
GS2      & 2.09 & 11.78 & 13.38 & 13.31 & 13.13 & 4.2195 &   0.4847   & 2.04  & i    &       \cite{2011arXiv1103.5174S} \\ \hline
LS220    & 2.04 & 10.61 & 12.64 & 12.43 & 12.10 & 3.3569 &   0.4170   & 2.54  & i    &       \cite{1991NuPhA.535..331L} \\ \hline
MIT60    & 1.88 & 10.35 & 10.43 & 10.74 & 10.75 & 1.4654 &   0.5630   & 2.20  & i    &       \cite{refmit60}            \\ \hline
LS180    & 1.83 & 10.04 & 12.13 & 11.65 & 10.68 & 2.9883 &   0.3840   & 2.70  & i    &       \cite{1991NuPhA.535..331L} \\ \hline
eosL     & 2.76 & 14.30 & 15.74 & 15.77 & 15.79 & 9.5635 &   0.5900   & 1.35  & ii   &       \cite{1975PhLB...59...15P} \\ \hline
eosO     & 2.39 & 11.56 & 12.85 & 12.87 & 12.83 & 3.9106 &   0.4786   & 2.04  & ii   &       \cite{1975PhRvD..12.3043B} \\ \hline
MIT40    & 2.28 & 12.56 & 12.08 & 12.59 & 12.89 & 5.0864 &   0.5687   & 1.49  & ii   &       \cite{2008arXiv0801.4829G,2000AA...359..311Z} \\  \hline
eosUU    & 2.21 & 9.84  & 11.18 & 11.12 & 11.00 & 1.8118 &   0.3477   & 2.80  & ii   &       \cite{1988PhRvC..38.1010W} \\ \hline
SKA      & 2.21 & 11.17 & 13.07 & 12.91 & 12.68 & 3.5929 &   0.4446   & 2.30  & ii\footnote{The full temperature dependence of this EoS is available for $T\le 30$~MeV, which is exceeded in NS merger simulations. Therefore, we use the ideal-gas ansatz for thermal effects combined with the zero-temperature slice of this EoS.}   &       \cite{1991NuPhA.535..331L} \\ \hline
APR      & 2.19 & 9.90  & 11.33 & 11.25 & 11.12 & 1.9152 &   0.3533   & 2.80  & ii   &       \cite{1998PhRvC..58.1804A} \\ \hline
eosAU {\bf (x)}  & 2.14 & 9.45  & 10.44 & 10.42 & 10.33 & 1.1812 &  0.2906   & 3.00  & ii   &       \cite{1988PhRvC..38.1010W} \\ \hline
BurgioNN & 2.07 & 10.58 & 11.99 & 11.87 & 11.67 & 2.4720 &   0.3936    & 2.50  & ii   &       \cite{2003PhLB..562..153B} \\ \hline
Sly4     & 2.05 & 10.01 & 11.79 & 11.59 & 11.29 & 2.4012 &   0.3830   & 2.85  & ii   &       \cite{2001AA...380..151D}  \\ \hline
Glendnh3 & 1.96 & 11.48 & 14.52 & 13.98 & 13.22 & 4.4281 &   0.4326   & 2.35  & ii   &       \cite{1985ApJ...293..470G} \\ \hline
BBB2     & 1.92 & 9.55  & 11.30 & 10.97 & 10.49 & 2.0949 &   0.3549   & 3.15  & ii   &       \cite{1997AA...328..274B}  \\ \hline
eosC     & 1.87 & 9.89  & 12.06 & 11.57 & 10.79 & 2.6024 &   0.3693   & 3.04  & ii   &       \cite{1974NuPhA.230....1B} \\ \hline
eosWS {\bf (x)}  & 1.85 & 9.58  & 10.99 & 10.72 & 10.16 & 1.9070 &  0.3357   & 3.09  & ii   &       \cite{1988PhRvC..38.1010W} \\ \hline
FPS {\bf (x)}    & 1.80 & 9.30  & 10.91 & 10.52 &  9.38 & 1.8720 &  0.3344   & 3.35  & ii   &       \cite{1981NuPhA.361..502F} \\ \hline
MS1      & 2.77 & 13.39 & 14.99 & 15.04 & 15.04 & 7.1495 &   0.6218   & 1.60  & iii  &       \cite{2009PhRvD..79l4032R} \\ \hline
MS1b     & 2.76 & 13.30 & 14.59 & 14.69 & 14.74 & 7.0970 &   0.6194   & 1.60  & iii  &       \cite{2009PhRvD..79l4032R} \\ \hline
MPA1     & 2.47 & 11.35 & 12.49 & 12.54 & 12.54 & 3.0980 &   0.4693   & 2.10  & iii  &       \cite{2009PhRvD..79l4032R} \\ \hline
APR3     & 2.38 & 10.73 & 12.04 & 12.04 & 12.00 & 2.4451 &   0.4202   & 2.35  & iii  &       \cite{2009PhRvD..79l4032R} \\ \hline
ENG      & 2.25 & 10.41 & 12.05 & 11.99 & 11.87 & 2.7104 &   0.4136   & 2.55  & iii  &       \cite{2009PhRvD..79l4032R} \\ \hline
H4       & 2.02 & 11.72 & 13.95 & 13.74 & 13.35 & 4.6293 &   0.4447   & 2.15  & iii  &       \cite{2009PhRvD..79l4032R} \\ \hline
ALF2     & 1.98 & 11.48 & 12.78 & 12.72 & 12.49 & 4.0893 &   0.4396   & 2.10  & iii  &       \cite{2009PhRvD..79l4032R} \\ \hline
ALF4     & 1.94 & 10.79 & 11.60 & 11.55 & 11.43 & 2.0437 &   0.3789   & 2.30  & iii  &       \cite{2009PhRvD..79l4032R} \\ \hline
MS2      & 1.80 & 11.81 & 14.25 & 13.52 & 12.10 & 3.9987 &   0.4088   & 2.20  & iii  &       \cite{2009PhRvD..79l4032R} \\ \hline
APR2 {\bf (x)}   & 1.79 & 8.65  & 10.13 &  9.74 & -     & 1.3266 &   0.2887   & 3.80  & iii  &       \cite{2009PhRvD..79l4032R} \\ \hline
H3       & 1.79 & 11.77 & 13.95 & 13.48 & -     & 4.3852 &   0.4153   & 2.15  & iii  &       \cite{2009PhRvD..79l4032R} \\ \hline
Heb6     & 3.00 & 13.17 & 13.33 & 13.54 & 13.69 & 5.7020 &   0.6967   & 1.50  & iv   &       \cite{2010PhRvL.105p1102H} \\ \hline
Heb5     & 2.79 & 12.25 & 12.38 & 12.59 & 12.73 & 3.3927 &   0.5537   & 1.70  & iv   &       \cite{2010PhRvL.105p1102H} \\ \hline
Heb4     & 2.74 & 12.08 & 12.51 & 12.65 & 12.74 & 3.0665 &   0.5270   & 1.80  & iv   &       \cite{2010PhRvL.105p1102H} \\ \hline
Heb3     & 2.42 & 10.94 & 12.03 & 12.09 & 12.09 & 2.7306 &   0.4414   & 2.25  & iv   &       \cite{2010PhRvL.105p1102H} \\ \hline
Heb2     & 2.22 & 10.15 & 11.42 & 11.35 & 11.24 & 1.8099 &   0.3625   & 2.65  & iv   &       \cite{2010PhRvL.105p1102H} \\ \hline
Heb1     & 2.13 &  9.71 & 10.81 & 10.77 & 10.66 & 1.4685 &   0.3289   & 2.90  & iv   &       \cite{2010PhRvL.105p1102H} \\ \hline

\hline    
\end{tabular}
 \end{ruledtabular} 
\end{table*}

\section{Simulations} \label{sec:simulations}
\subsection{Dynamics}
According to pulsar observations~\cite{1999ApJ...512..288T,2011A&A...527A..83Z} and population synthesis studies~\cite{2008ApJ...680L.129B} binaries of two NSs with a gravitational mass of about 1.35~$M_{\odot}$ each are the most abundant systems in the binary NS population. Therefore, we choose a symmetric binary with $M_1=M_2=1.35~M_{\odot}$ and simulate for all EoSs discussed in Sect.~\ref{sec:eos} the late inspiral phase, the merging, and the early postmerger evolution of this system until an approximately stationary state has formed (10 to 20~ms after merging). The inspiral is driven by the loss of angular momentum and energy due to the GW emission and lasts between some ten and several hundred million years depending on the initial binary separation. Only shortly before the merging takes place, when the orbital period has reached milliseconds, finite-size and post-Newtonian effects become important.

It is known that the remnant formation after the coalescence depends on the total binary mass $M_{\mathrm{tot}}=M_1+M_2$ and the EoS (see e.g.~\cite{2010CQGra..27k4002D}). In this and the following two sections the total binary mass is fixed, thus the outcome depends on the EoS only. We distinguish two scenarios. For four out of the 38 EoSs considered in our study, we observe the direct formation of a BH within less than a ms after the stars come in contact, because the merged object cannot be stabilized against the gravitational collapse. This ``prompt collapse'' scenario is found only for EoSs that lead to relatively small $R_{\mathrm{max}}$ (EoSs labelled with an ``{\bf x}'' in Tab.~\ref{tab:models}). The maximum-mass configurations of these EoSs are marked with a cross in Fig.~\ref{fig:tov}.

In the simulations with the remaining EoSs the colliding stars form a differentially rotating object (DRO), which is supported against the gravitational collapse -- at least for a longer period -- by centrifugal forces. Furthermore, shock heating at the contact interface and compression lead to an increase of the temperature to several 10~MeV (in some cases even more than 100~MeV), which has an additional stabilizing effect. After angular momentum is partially lost by GW emission and redistributed to the outer remnant layers, the merger remnant may collapse to a BH. This ``delayed collapse'' happens typically after several 10--100~ms and only takes place if the mass of the DRO exceeds the upper mass limit of rigidly rotating NSs, which for most EoSs is about~1.2~$M_{\mathrm{max}}$~\cite{1996ApJ...456..300L,2007coaw.book.....C,2007ASSL..326.....H}. We explicitly point out that for some EoSs the total remnant mass may not overstep this threshold and the DRO finally will settle to a uniformly and rapidly rotating NS. In these cases the 1.35-1.35~$M_{\odot}$ NS binaries might be excluded as progenitors of short gamma-ray bursts, at least for models of gamma-ray bursts that rely on the formation of a BH with a hot accretion torus (see~\cite{2007PhR...442..166N} for a review). Rigidly spinning ``supermassive'' NSs (whose gravitational mass is larger than $M_{\mathrm{max}}$) may form BHs on much longer timescales due to the loss of angular momentum e.g. by electromagnetic emission~\cite{2007coaw.book.....C,2007ASSL..326.....H}. Note that the gravitational mass of a remnant resulting from a 1.35-1.35~$M_{\odot}$ binary is of the order of 2.6~$M_{\odot}$ because the gravitational mass of the initial binary is already below 2.7~$M_{\odot}$. Moreover, energy is radiated away by GWs, matter becomes gravitationally unbound during merging, and energy converted into heat is lost by neutrino emission. The latter effect is not modelled here.

Independent of whether a BH is the final outcome, NS mergers evolve dynamically similarly for all systems which form a DRO. As the stars approach each other increasingly faster during the late inspiral phase, the binary components get more strongly deformed and finally collide with a relatively big impact parameter. The bulk matter of the initial stars assembles into a rotating double-core structure, where the dense cores bounce against each other. While matter is shed off from the surface to feed a halo around the central object, the two cores merge into a single core after a few bounces. On a timescale of several milliseconds the oscillations of the initially highly deformed remnant are damped and an approximately stationary, axisymmetric object is left, which is still rotating differentially and ringing with much lower amplitude. 
We simulate all models resulting in a DRO for about 15~ms after the plunge until the nearly stationary phase is reached. In none of these calculations we actually observe the delayed collapse except for the MIT60 EoS, where the merger remnant forms a BH after about 3~ms, and for the H3 EoS, where the DRO is stable for 25~ms.

\subsection{Analysis of the gravitational-wave signal} \label{ssec:gwana}
\begin{figure}
\includegraphics[width=8.9cm]{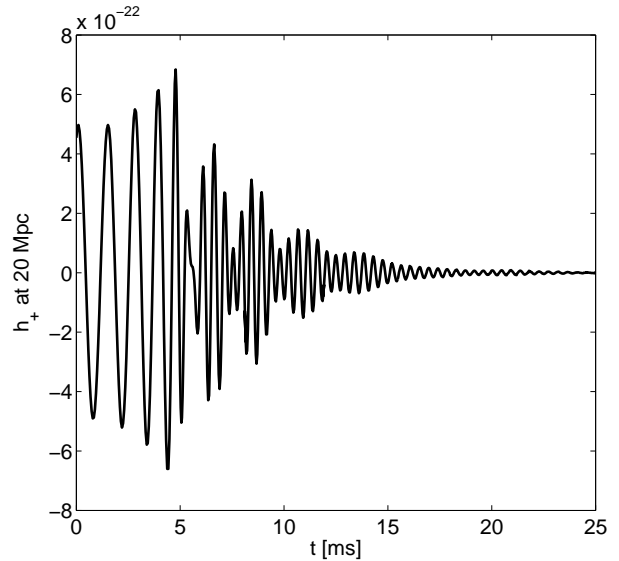}
\caption{\label{fig:amplitude}Gravitational-wave amplitude of the plus polarization measured along the polar axis at a distance of 20~Mpc for the simulation with the Shen EoS.}
\end{figure}
\begin{figure}
\includegraphics[width=8.9cm]{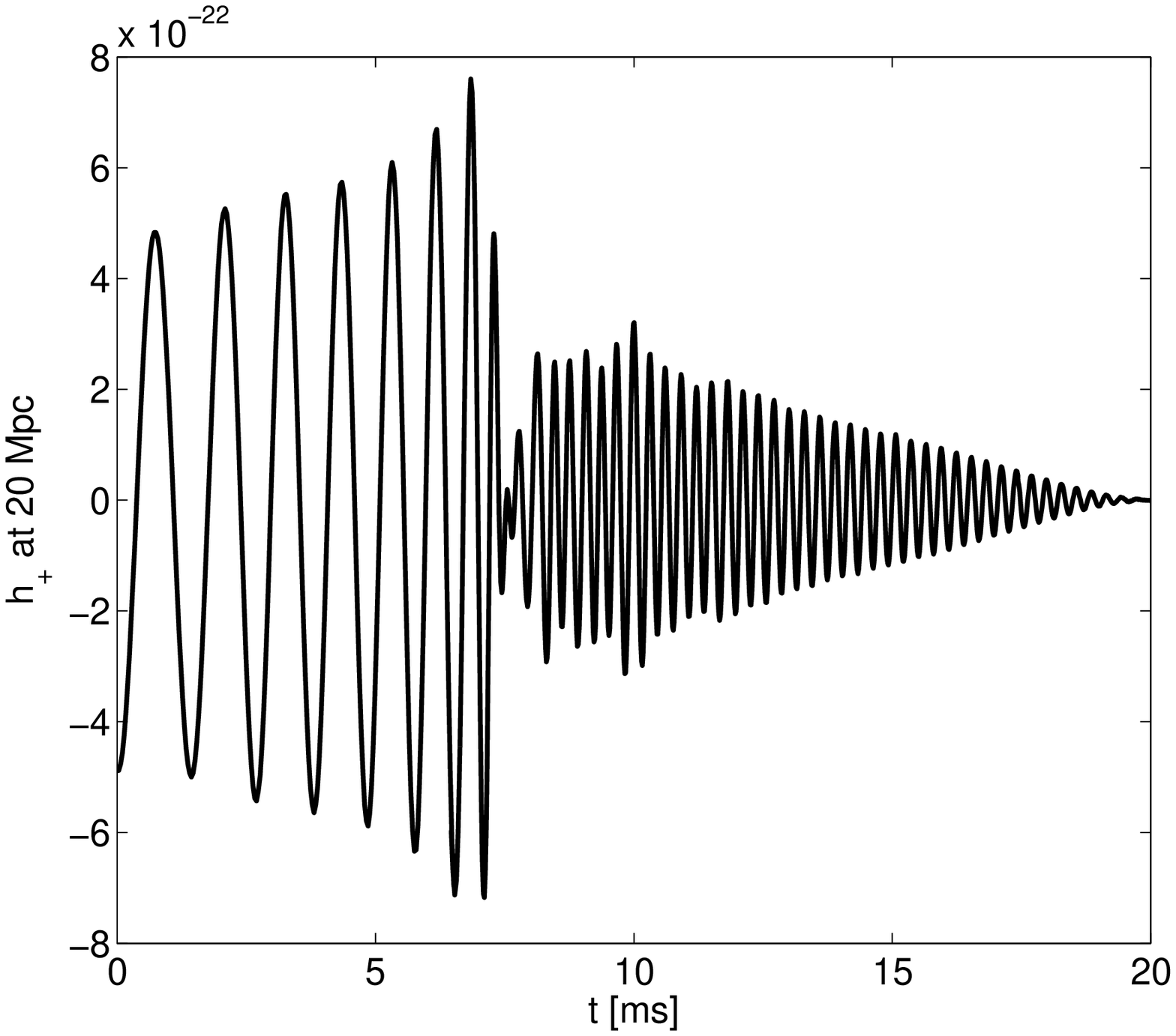}
\caption{\label{fig:amplitude2}Gravitational-wave amplitude of the plus polarization measured along the polar axis at a distance of 20~Mpc for the simulation with the Sly4 EoS.}
\end{figure}
\begin{figure}
\includegraphics[width=8.9cm]{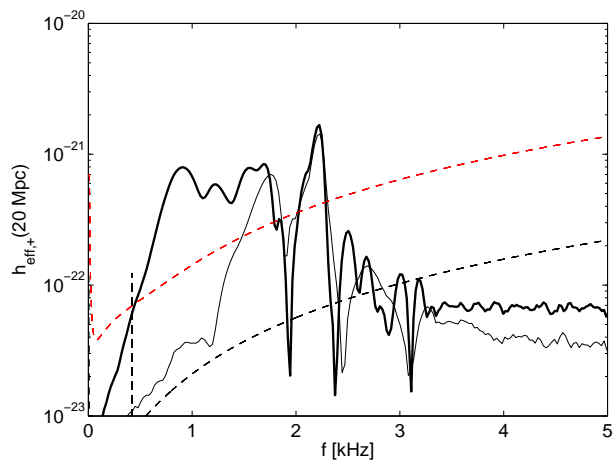}
\caption{\label{fig:heff}Fourier spectrum of the plus polarization of the GW signal at a distance of 20~Mpc for the Shen EoS. The thick line displays the spectrum computed from the signal of the full simulation time, while the thin line shows the spectrum of the postmerger phase only. The dashed lines give the unity SNR sensitivity curve of Advanced LIGO (red) and of the Einstein Telescope (black).}
\end{figure}
\begin{figure}
\includegraphics[width=8.9cm]{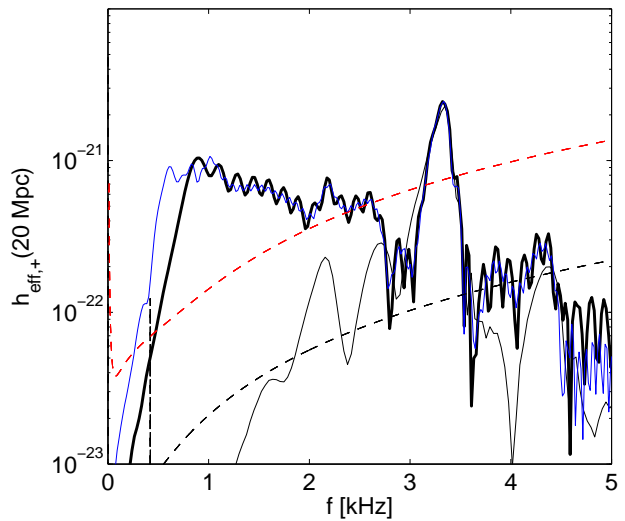}
\caption{\label{fig:heff_sly4}Fourier spectrum of the plus polarization of the GW signal at a distance of 20~Mpc for the Sly4 EoS. The thick line displays the spectrum computed from the signal of the full simulation time, while the thin line shows the spectrum of the postmerger phase only. The spectrum of a simulation starting 5.5 revolutions before merging is given by the blue line. The dashed lines give the unity SNR sensitivity curve of Advanced LIGO (red) and of the Einstein Telescope (black).}
\end{figure}
The dynamical stages of a merger can also be identified in the GW signal, which is computed by means of a modified quadrupole formula that takes into account post-Newtonian effects~\cite{2007A&A...467..395O}. A typical waveform is shown in Fig.~\ref{fig:amplitude} for the Shen EoS as measured perpendicular to the orbital plane at a distance of 20~Mpc. Fig.~\ref{fig:amplitude} displays the plus polarization, the cross polarization looks very similar but is phase-shifted by $\pi /2$. The characteristic inspiral phase with an increasing frequency and an increasing amplitude until 5~ms is clearly visible. Following the plunge the ringdown of the postmerger remnant can be seen by high-frequency oscillations that cease over 10 to 15~ms until the object has reached approximate stationarity. For rather stiff EoSs with $R_{\mathrm{max}}\gtrsim 11$~km we observe a low-frequency modulation of the postmerger signal as in Fig.~\ref{fig:amplitude}. In contrast, for soft EoSs with relatively small $R_{\mathrm{max}}$ such a feature is absent or less pronounced as for instance in the waveform calculated for the model with the Sly4 EoS, which is shown in Fig.~\ref{fig:amplitude2}. The reason for this difference could not be clarified and deserves further investigation, e.g. in terms of an oscillation mode analysis as in~\cite{2011arXiv1105.0368S}. Furthermore, one recognizes in Fig.~\ref{fig:amplitude2} that for this EoS the inspiral phase lasts about 2~ms longer and ends with a larger wave amplitude, which is a consequence of the higher compactness of the inspiraling NSs. Note that all simulations start with the same initial coordinate distance between the stars.

To obtain spectral information of the GW signal we compute the dimensionless quantity $h_{\mathrm{eff,+}}=\tilde{h}_{+}(f)f$ with the Fourier transformed waveform $\tilde{h}_{+}(f)$ of the plus polarization. The results are displayed in Fig.~\ref{fig:heff} and Fig.~\ref{fig:heff_sly4} for the Shen EoS and the Sly4 EoS, respectively. The thick lines show the spectra calculated from the whole signal during the simulation time, while the thin lines correspond to the postmerger phase alone. (Note that slight differences between the spectrum for the postmerger signal and the spectrum for the full signal at the highest frequencies depend on the time chosen to define the beginning of the postmerger phase.) The dashed lines belong to the unity signal-to-noise ratio (SNR) sensitivity curves for Advanced LIGO (red) (broadband configuration)~\cite{2010CQGra..27h4006H} and the planned Einstein Telescope (black)~\cite{2010CQGra..27a5003H}. Note that the low-frequency part of the spectra below 1~kHz computed for the full signal is not reliable because our simulations start only a few revolutions before merging and therefore lower frequencies of the preceding inspiral phase are underrepresented.

A pronounced peak in the spectra at a frequency $f_{\mathrm{peak}}$ between about~2 and 4~kHz is generic to all models resulting in a DRO. It has a frequency of $f_{\mathrm{peak}}=2.19$~kHz for the Shen model and of $f_{\mathrm{peak}}=3.32$~kHz for the Sly4 EoS. This feature clearly originates from the postmerger phase and is produced by the violent oscillations of the deformed merger remnant. As can be seen in the spectra of the postmerger evolution, the structure at $f_{\mathrm{peak}}$ is accompanied by additional, weaker peaks. In a first step towards astroseismology of NS merger remnants it has been shown that the remnant can be considered as an isolated selfgravitating object, where certain oscillation modes are excited by
the formation process of the object~\cite{2011arXiv1105.0368S}. The different structures in the postmerger spectrum are caused by the various oscillation modes in the merger remnant and their couplings. The study of~\cite{2011arXiv1105.0368S} has also identified the mode corresponding to the peak frequency $f_{\mathrm{peak}}$ as the fundamental quadrupolar fluid mode.

In Tab.~\ref{tab:gw} the peak frequencies for all models forming a DRO are given, where the lowest peak frequency is found for the eosL model with $f_{\mathrm{peak}}=1.84$~kHz. The highest peak frequency of 3.73~kHz is obtained in the simulation with the BBB2 EoS. In addition, the full width at half maximum (FWHM) is provided for every model. Typically, the FWHM is in the range of 100 to 200~Hz, with the exception of the MIT60 calculation yielding a broad peak with a FWHM of 460~Hz.

For the Shen EoS we have performed higher resolved simulations with 550,000 and 1,270,000 SPH particles, which confirm that $f_{\mathrm{peak}}$ is determined to an accuracy of about one per cent (see also the discussion and figure 2 in~\cite{2012PhRvL.108a1101B} and~\cite{2010PhRvD..82h4043B}). Moreover, a run starting 3.5 orbits before the merging of the binary components does not yield differences in the peak frequency (see~\cite{2012PhRvL.108a1101B}). Additionally we show in Fig.~\ref{fig:heff_sly4} the resulting spectrum of a calculation starting 5.5 orbits before merging for the Sly4 EoS (blue line). The peak frequency is unaffected by the duration of the simulated inspiral phase (difference in $f_{\mathrm{peak}}$ below 0.5 per cent).

\subsection{Detectability of the gravitational-wave signal} \label{ssec:gwdet}
The prospects to determine $f_{\mathrm{peak}}$ are estimated by computing the signal-to-noise ratio (SNR) via $(\mathrm{SNR})^2=4\int_0^{\infty}\frac{|\tilde{h}_{+}(f)|^2}{S_{\mathrm{h}}(f)}df$ (see e.g.~\cite{2008Maggiore}) adopting the one-sided spectral density $S_{\mathrm{h}}(f)$ of the strain noise of the Advanced LIGO detector, which has a comparable sensitivity to the Advanced VIRGO observatory. In Tab.~\ref{tab:gw} we list the SNRs for the postmerger signal and the emission during the whole simulation time starting from the late inspiral phase. Here we correct the underestimation of the GW amplitude due to the usage of the quadrupole formula (see Appendix~\ref{app:grcomp}). The SNRs are computed for a distance of $D=20$~Mpc assuming an optimal inclination of the binary and an optimal orientation of the detector. Because the GW amplitude scales with $1/D$, also the SNR is proportional to $1/D$ and one can easily obtain the optimal detection horizon for a chosen SNR. Requiring for instance a SNR of 2, the postmerger signal alone can be detected up to a distance of about 20 to 45~Mpc depending on the EoS if we ignore results for EoSs which are incompatible with the observation of the 1.97~$M_{\odot}$ NS~\cite{2010Natur.467.1081D} (marked by an asterisk in Tab.~\ref{tab:gw}) and that yield somewhat lower SNRs. Including also the EoS dependent late inspiral wave train boosts the optimal detection horizon to about 150~Mpc. While a SNR of 2 is rather low, the preceding inspiral signal (not covered by our calculations) provides detailed information on important emission features such as the time of merging, the distance, and the involved masses, justifying a low threshold SNR (in particular, the inspiral signal means a warning that one should search for the postmerger emission in the close temporal vicinity). Interestingly, models with EoSs that result in a high SNR in the postmerger phase, yield in general a lower SNR for the full signal of the simulation and vice versa. This can be understood from the observation that for more compact stars the inspiral phase lasts longer increasing the full SNR. In these cases the frequencies of the late-time GW emission are generally relatively high, which leads to a low SNR in the postmerger phase because of the reduced sensitivity of the GW detector at high frequencies.

For an estimate of the detection rate with Advanced LIGO one considers the Milky Way Equivalent Galaxies (MWEGs) accessible by the detector. The reduction of the rate due to random source and detector orientations can be accounted for by dividing the optimal detection horizon by $1/\sqrt{5}$~\cite{2010CQGra..27q3001A}. For the chosen SNR the number of MWEGs equals 145 to 1190 for the postmerger signal alone, taking advantage of the local over-density of potential host galaxies~\cite{2010CQGra..27q3001A}. The inclusion of the late inspiral signal increases the number of probed MWEGs to more than 14,000. Adopting the ``realistic'' and the ``high'' merger rates listed in~\cite{2010CQGra..27q3001A}, the prospects for the detection of the postmerger signal alone are estimated to be 0.015 to 1.2 events per year. For the combined signal 1.4 to 14 detections per year can be expected. Finally, we emphasize that with the increased sensitivity of the proposed Einstein Telescope~\cite{2011CQGra..28i4013H} a significant number of detections with high SNRs will be possible.

The uncertainties in the determination of the peak frequency by a GW detection can be estimated by evaluating the Fisher information matrix for a single-parameter family of waveforms parametrized by $f_{\mathrm{peak}}$. This approach considers to which extent two different waveforms are distinguishable by a detector with a certain noise level (see e.g.~\cite{2008Maggiore}). Following the procedure of~\cite{2009PhRvD..79l4033R}, we arrange our models in Tab.~\ref{tab:gw} with increasing $f_\mathrm{peak}$ and compute for two subsequent models A and B with corresponding GW amplitudes $h^A$ and $h^B$
\begin{equation}\label{eq:fisher}
\delta f^2=\frac{(f_\mathrm{peak}^A-f_\mathrm{peak}^B)^2}{(h^A-h^B|h^A-h^B)}
\end{equation}
with the inner product between two signal $h_1$ and $h_2$
\begin{equation}
(h_1|h_2)=4 Re \int_0^{\infty}\frac{\tilde{h}_{1}(f)\tilde{h}^*_{2}(f)}{S_{\mathrm{h}}(f)}df,
\end{equation}
where the Fourier transforms of the signal enters (an asterisk marks the complex conjugated). In Tab.~\ref{tab:gw} the entry of $\delta f$ for a given EoS corresponds to the value obtained by~\eqref{eq:fisher} for the waveforms of this EoS and the subsequent EoS in the table (e.g. for the eosL and the MS1 EoS we compute $\delta f_{\mathrm{pm}}=0.062$~kHz). Here, a polar distance of 20~Mpc and optimal detector alignment are assumed. We observe uncertainties in $f_\mathrm{peak}$ of typically better than 50~Hz (on average 40~Hz). Note that these values are smaller than the ones reported in~\cite{2012PhRvL.108a1101B} because a larger number of models is employed leading to a finer coverage of the $f_\mathrm{peak}$-parameter space. It is important to mention that $\delta f$ computed for the postmerger phase serves only as a rough estimate for the uncertainties in the $f_\mathrm{peak}$ determination because the procedure is not fully applicable in the low-SNR limit~\cite{2008PhRvD..77d2001V,2009PhRvD..79l4033R}. However, we conduct the same analysis including the late inspiral phase of the signal, i.e. with a sufficiently high SNR. For a fair comparison the signal is constrained to $f\geq 1$~kHz. In this case smaller uncertainties are found (on average 27~Hz), see column~7 in Tab.~\ref{tab:gw}.

\begin{table*}
\caption{\label{tab:gw}All EoS models for which the merging of two 1.35~$M_{\odot}$ NSs results in the formation of a DRO. $f_{\mathrm{peak}}$ denotes the dominant GW frequency in the postmerger phase, and FWHM characterizes the corresponding width of the peak in the spectrum. SNRs are computed for a distance of 20~Mpc assuming an optimal source and detector orientation considering the postmerger signal only (SNR$_{\mathrm{pm}}$) and the full signal above 1~kHz (SNR$_{\mathrm{tot}}$). Uncertainties in the determination of the peak frequency are estimated by $\delta f_{\mathrm{pm}}$ and $\delta f_\mathrm{tot}$ for the postmerger phase and the signal above 1~kHz, respectively (see main text for details). The last column provides the radius of a sphere enclosing a rest mass of 2.6~$M_{\odot}$ at 8~ms after merging (see Sect.~\ref{ssec:interpretation}). EoSs marked with an asterisk are incompatible with the observation of the 1.97~$M_{\odot}$ pulsar.}
 \begin{ruledtabular}
 \begin{tabular}{|l|l|l|l|l|l|l|l|}
EoS     & $f_{\mathrm{peak}}$~[kHz]  & FWHM~[kHz]  & SNR$_{\mathrm{pm}}$ & SNR$_{\mathrm{tot}}$ & $\delta f_{\mathrm{pm}}$~[kHz] & $\delta f_\mathrm{tot}$~[kHz] & $R_{\mathrm{remnant}}(8~\mathrm{ms})$~[km]\\ \hline
eosL    &  1.84 & 0.102 & 4.59  & 11.74  & 0.062 & 0.060  & 15.20 \\ \hline
MS1     &  2.08 & 0.127 & 3.58  & 12.23  & 0.024 & 0.012  & 13.58 \\ \hline
GS1     &  2.10 & 0.117 & 3.81  & 13.08  & 0.047 & 0.034  & 13.48 \\ \hline
Shen    &  2.19 & 0.151 & 3.62  & 13.07  & 0.0   & 0.0    & 12.92 \\ \hline
MS1b    &  2.19 & 0.104 & 3.56  & 12.52  & 0.054 & 0.040  & 13.14 \\ \hline
Glendnh3&  2.33 & 0.130 & 2.54  & 15.39  & 0.041 & 0.026  & 12.31 \\ \hline
H4      &  2.37 & 0.131 & 2.83  & 13.96  & 0.027 & 0.015  & 11.99 \\ \hline
LS375   &  2.40 & 0.133 & 3.16  & 11.35  & 0.015 & 0.011  & 11.96 \\ \hline
MS2 *   &  2.42 & 0.181 & 2.04  & 16.03  & 0.010 & 0.007  & 12.07 \\ \hline
H3 *    &  2.43 & 0.146 & 2.66  & 14.52  & 0.007 & 0.007  & 11.72 \\ \hline
Heb6    &  2.44 & 0.129 & 3.05  & 14.32  & 0.046 & 0.037  & 11.69 \\ \hline
GS2     &  2.53 & 0.121 & 2.78  & 15.51  & 0.061 & 0.035  & 11.14 \\ \hline
MIT40   &  2.62 & 0.194 & 3.67  & 13.84  & 0.008 & 0.005  &  9.33 \\ \hline
SKA     &  2.64 & 0.131 & 2.83  & 14.86  & 0.015 & 0.013  & 10.73 \\ \hline
eosO    &  2.66 & 0.112 & 2.81  & 15.61  & 0.050 & 0.031  & 10.60 \\ \hline
ALF2    &  2.71 & 0.145 & 2.64  & 14.00  & 0.0   & 0.0    & 10.73 \\ \hline
Heb5    &  2.71 & 0.122 & 2.78  & 14.41  & 0.048 & 0.035  & 10.46 \\ \hline
MPA1    &  2.80 & 0.159 & 2.49  & 13.94  & 0.108 & 0.043  & 10.42 \\ \hline
Heb4    &  2.87 & 0.148 & 2.64  & 14.89  & 0.012 & 0.014  & 10.48 \\ \hline
LS220   &  2.89 & 0.205 & 2.58  & 14.38  & 0.036 & 0.033  &  9.68 \\ \hline
Heb3    &  2.96 & 0.153 & 2.50  & 14.82  & 0.044 & 0.034  &  9.49 \\ \hline
ENG     &  3.02 & 0.180 & 2.19  & 14.25  & 0.0   & 0.0    &  9.39 \\ \hline
APR3    &  3.02 & 0.188 & 2.23  & 14.52  & 0.107 & 0.077  &  9.58 \\ \hline
BurgioNN&  3.20 & 0.183 & 2.29  & 16.29  & 0.047 & 0.031  &  8.53 \\ \hline
LS180 * &  3.26 & 0.250 & 1.66  & 15.35  & 0.028 & 0.018  &  8.34 \\ \hline
ALF4    &  3.29 & 0.172 & 2.13  & 14.52  & 0.047 & 0.031  &  8.90 \\ \hline
Sly4    &  3.32 & 0.194 & 2.16  & 14.82  & 0.011 & 0.006  &  8.63 \\ \hline
eosC *  &  3.33 & 0.220 & 1.52  & 15.87  & 0.051 & 0.032  &  8.37 \\ \hline
Heb2    &  3.39 & 0.154 & 2.10  & 16.59  & 0.027 & 0.008  &  8.69 \\ \hline
MIT60 * &  3.43 & 0.460 & 2.01  &  8.83  & 0.023 & 0.008  &  6.11 \\ \hline
APR     &  3.46 & 0.182 & 2.36  & 15.12  & 0.048 & 0.026  &  8.34 \\ \hline
eosUU   &  3.50 & 0.169 & 2.07  & 14.60  & 0.100 & 0.084  &  8.30 \\ \hline
Heb1    &  3.72 & 0.152 & 2.20  & 15.90  & 0.009 & 0.006  &  7.56 \\ \hline
BBB2 *  &  3.73 & 0.249 & 1.34  & 15.20  &       &        &  7.07 \\ \hline
\hline    
\end{tabular}
 \end{ruledtabular}         
\end{table*}

\section{Dependence on stellar parameters} \label{sec:stellar}

It is the goal of this work to establish the relations between features of the GW signal and properties of the EoS, which can be either thermodynamical properties or stellar parameters. To this end we determine the dominant frequency $f_{\mathrm{peak}}$ of the postmerger GW emission as discussed in the previous section for all EoSs which lead to the formation of a DRO (see Tab.~\ref{tab:gw}). In the following we investigate the dependence of $f_{\mathrm{peak}}$ on various NS and EoS properties, which in turn allows to constrain or measure these quantities when $f_{\mathrm{peak}}$ is determined from a GW observation. We stress that the total binary mass and the initial mass ratio of the binary components can be determined from the GW inspiral signal~\cite{1994PhRvD..49.2658C}. Throughout this and the following section we discuss exclusively binaries with two 1.35~$M_{\odot}$ NSs.

\subsection{Neutron star radii}\label{sssect:nsradii}
\begin{figure}
\includegraphics[width=8.9cm]{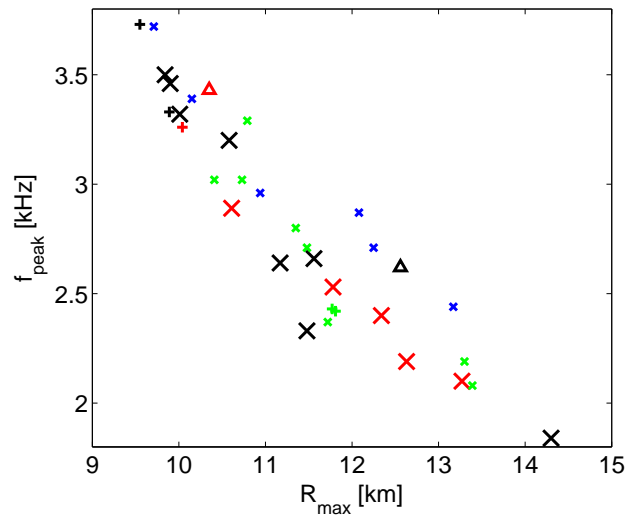}
\caption{\label{fig:frmax}Peak frequency of the postmerger GW emission versus the radius of the maximum-mass configuration of nonrotating NSs for different EoSs. Red symbols correspond to microphysical EoSs with a consistent temperature treatment (class i), black symbols show data points for barotropic microphysical EoSs (without temperature and electron fraction dependence) (class ii), green (smaller) symbols belong to EoSs implemented as piecewise polytropes fitting barotropic microphysical EoSs (class iii), and blue symbols display results for microphysical EoSs at low densities with high-density extensions by piecewise polytropes (class iv). Classes (ii) to (iv) are supplemented with an ideal-gas component for mimicking thermal effects. Plus signs indicate EoSs which are excluded by the observation of a 1.97~$M_{\odot}$ pulsar~\cite{2010Natur.467.1081D}. EoSs describing absolutely stable strange quark matter are denoted by triangles. Note that the MIT60 EoS (red triangle) is ruled out by the 1.97~$M_{\odot}$ mass limit.}
\end{figure}

In~\cite{2012PhRvL.108a1101B} we have identified an anticorrelation between the peak frequency and the minimum radius of nonspinning NSs. Fig.~\ref{fig:frmax} shows this relation for our extended set of EoSs. All data points correspond to simulations of 1.35~$M_{\odot}$-1.35~$M_{\odot}$ binaries. The figure adopts the same color scheme as Fig.~\ref{fig:tov}, i.e. the colors represent the four different possibilities for the technical implementations of the EoSs including the treatment of thermal effects. The crosses denote the results of simulations with EoSs consistent within the error bars with the observation of the 1.97~$M_{\odot}$ pulsar~\cite{2010Natur.467.1081D}, while plus signs belong to calculations employing EoSs which are excluded by this detection. The triangles correspond to the two EoSs describing absolutely stable strange quark matter (MIT60 in red, MIT40 in black). Note that we keep this notation in all figures of this paper except for Fig.~\ref{fig:pcollapse}. In Fig.~\ref{fig:frmax} one observes a clear anticorrelation between $f_{\mathrm{peak}}$ and $R_{\mathrm{max}}$ with a slightly steeper slope at smaller radii. Note that the data points for EoSs implemented as piecewise polytropes (class iii) or with a high-density regime described by piecewise polytropes (class iv) tend to show some scatter from the relation established by the purely microphysical EoSs (black and red symbols). Since these models involve additional approximations or simplifications (see Sect.~\ref{sec:eos} and reasoning below), the corresponding data points are shown with smaller symbols to emphasize the correlation of the different cases. Moreover, excluded models with $M_{\mathrm{max}}<1.93~M_{\odot}$ (lower bound of the error bar given by~\cite{2010Natur.467.1081D}) are plotted with smaller plus signs, although they are not extraordinary outliers in this figure.

The somewhat displaced black cross (microphysical, barotropic EoS) at $R_{\mathrm{max}}=11.48$~km and $f_{\mathrm{peak}}$=2.33~kHz displays the data of the Glendnh3 EoS, which has a strikingly different mass-radius relation in comparison to other microphysical EoSs (see black curve with $R=17$~km and $M=0.5~M_{\odot}$ in Fig.~\ref{fig:tov} or black dashed line in Fig.~1 of~\cite{2012PhRvL.108a1101B}). Note that this EoS leads to a mass-radius relation which seems incompatible with theoretical calculations of the EoS of subnuclear matter~\cite{2010PhRvL.105p1102H}. Furthermore, we stress that among the purely microphysical models only the strange quark matter EoSs behave as outliers. However, if strange quark matter was the true ground state of matter and one of these EoSs was the correct EoS of high-density matter, this would lead to clear observational signatures (e.g. in the form of nuggets of strange quark matter in the cosmic ray flux) which would discriminate this scenario from ordinary NSs composed of nucleonic matter (see~\cite{2007ASSL..326.....H,2009PhRvL.103a1101B,PhysRevD.81.024012}). For all these reasons we classify the results of these three particular EoSs as outliers not spoiling the quality of the relation evident from the black and red crosses.

\begin{figure}
\includegraphics[width=8.9cm]{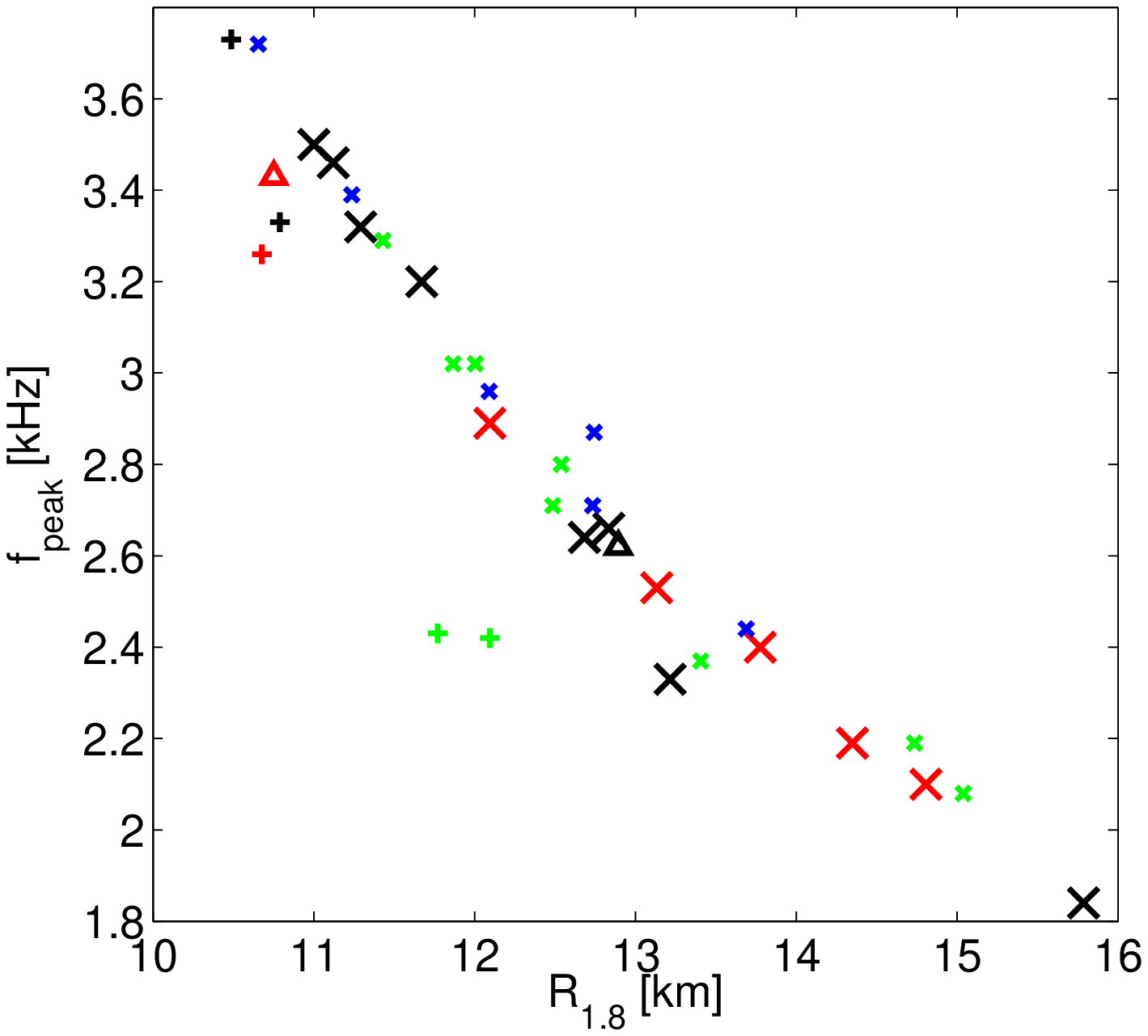}
\caption{\label{fig:fr18}Peak frequency of the postmerger GW emission versus the radius of a nonrotating NS with 1.8~$M_{\odot}$ for different EoSs. Symbols have the same meaning as in Fig.~\ref{fig:frmax}.}
\end{figure}

Figure~\ref{fig:fr18} shows the peak frequencies from our set of models versus the radius $R_{1.8}$ of the corresponding nonrotating NS with 1.8~$M_{\odot}$. Considering in~Fig.~\ref{fig:fr18} only EoSs which are in agreement with the NS mass limit of~\cite{2010Natur.467.1081D}, one recognizes that the relation between the GW frequency and the NS radius is closer in comparison to Fig.~\ref{fig:frmax}. In particular, some of the EoSs modeled as piecewise polytropes (green and blue symbols) better fulfill the relation indicated by the purely microphysical EoSs. As will be discussed in Sect.~\ref{ssec:interpretation} the density regime probed with the merger remnant, i.e. affecting $f_{\mathrm{peak}}$, corresponds roughly to the densities encountered in nonrotating NSs with masses below 1.8~$M_{\odot}$.

It is striking in Fig.~\ref{fig:fr18} that EoSs incompatible with the $(1.97\pm 0.04)~M_{\odot}$ limit (plus signs), appear as outliers at smaller radii. This can be understood from the fact that for these EoSs the radius of a 1.8~$M_{\odot}$ NS is near radius of the maximum-mass configuration because $M_{\mathrm{max}}$ is close to 1.8~$M_{\odot}$. This means that $R_{1.8}$ is outside of the region of the mass-radius relation with approximately constant radius and its location near the limiting mass corresponds to a smaller radius (see Fig.~\ref{fig:tov}). Also in Fig.~\ref{fig:fr18} the data point of the Glendnh3 EoS slightly deviates from the anticorrelation between $f_{\mathrm{peak}}$ and $R_{1.8}$.

\begin{figure}
\includegraphics[width=8.9cm]{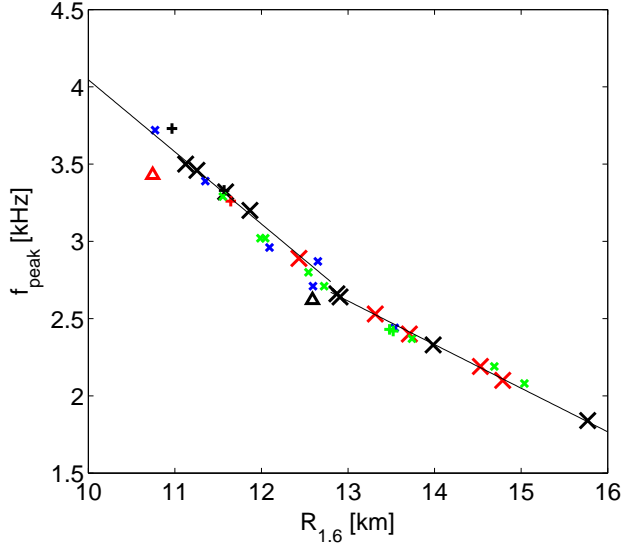}
\caption{\label{fig:fr16}Peak frequency of the postmerger GW emission versus the radius of a nonrotating NS with 1.6~$M_{\odot}$ for different EoSs. Symbols have the same meaning as in Fig.~\ref{fig:frmax}. The solid lines indicate the fits described by Eq.~\eqref{eq:brline}.}
\end{figure}

In contrast to Fig.~\ref{fig:fr18}, a plot using the radius $R_{1.6}$ of a 1.6~$M_{\odot}$ static NS as characteristic EoS property reveals all EoSs fulfilling a very tight anticorrelation (see Fig.~\ref{fig:fr16}). Even the black cross belonging to the Glendnh3 simulation is now located on the relation. Again, one finds a slight steepening of the slope at smaller radii. The EoSs of strange quark matter lie slightly off the relation, whereas in Fig.~\ref{fig:fr18} the MIT40 data point is consistent with the correlation (MIT60 has a relatively low $M_{\mathrm{max}}$ and is therefore expected to appear at a smaller radius as argued above). In this context we stress that MIT40 and MIT60 describe bare strange stars. In principle, strange stars could carry a nuclear crust with a density below the neutron drip density of $4\cdot10^{11}$~g/cm$^3$ (see e.g.~\cite{2007ASSL..326.....H}). This crust would have only a little mass of the order of $10^{-5}~M_{\odot}$ and, thus, would be dynamically unimportant, i.e. it is unlikely to change $f_{\mathrm{peak}}$. But the crust would increase the radius of a nonrotating strange star by a few hundred meters. This means that by considering the nuclear crust the triangles would move to the right, for instance about 500 meters for MIT40 and about 300 meters for MIT60 in the case of stars with 1.6~$M_{\odot}$. Hence, also these EoSs, in particular MIT40, are consistent with the relation that exists between $f_{\mathrm{peak}}$ and $R_{1.6}$ for the nuclear EoSs.

\begin{figure}
\includegraphics[width=8.9cm]{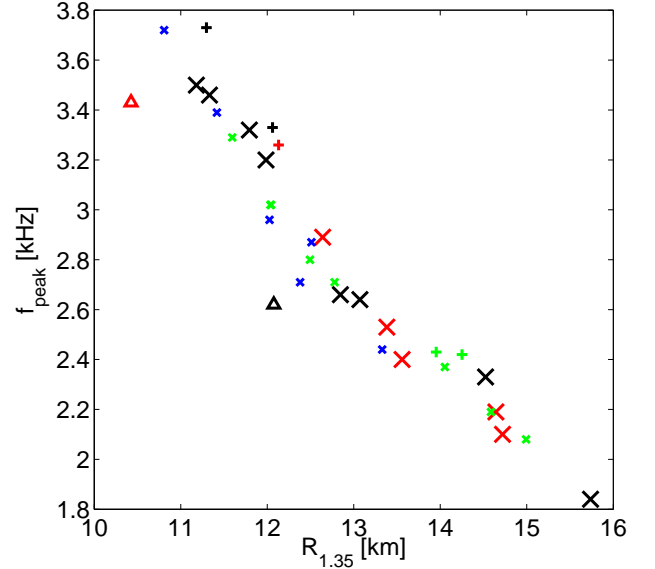}
\caption{\label{fig:fr135}Peak frequency of the postmerger GW emission versus the radius of a nonrotating NS with 1.35~$M_{\odot}$ for different EoSs. Symbols have the same meaning as in Fig.~\ref{fig:frmax}.}
\end{figure}

Finally, Fig.~\ref{fig:fr135} shows the dominant GW frequency as a function of the radius $R_{1.35}$ of a 1.35~$M_{\odot}$ NS. In comparison to~Fig.~\ref{fig:fr16} the scattering is larger, although the relation in particular for the accepted purely microphysical models is still very tight (excluding the data point of the Glendnh3 EoS). Interestingly, in Fig.~\ref{fig:fr135} outliers are on the opposite side of the basic relation as compared to their location in the plot using $R_{\mathrm{max}}$ (Fig.~\ref{fig:frmax}).

Note that the different EoS classes (colors) in Figs.~\ref{fig:frmax}-\ref{fig:fr135} sample in each case nearly the fully parameter range and do not indicate any trend or bias due to the different implementations of the EoSs aside from the increased scatter for implementations using piecewise polytropes. The EoSs of class (iv) (blue) cover a broad range of possible behaviors at intermediate and high densities, which are partially very extreme (e.g., very high pressure and sound speed at high densities, see Figs.~\ref{fig:eos} and~\ref{fig:vsound}). Therefore, it is expected that the resulting variations will also span a broad range, which is however consistent with the chiral effective field theory constraints at saturation densities and below. The models of class (iii) involve a twofold simplification that can explain the larger deviations from the correlations. First, the fits of the EoSs do not perfectly match the underlying microphysical model (e.g. in the sound speed, see~\cite{2009PhRvD..79l4032R}), leading to peak frequencies which may be slightly different from those obtained by the original model. Second, due to the usage of the fit also the stellar parameters of nonrotating NSs differ slightly from those obtained with the original EoSs. Bear in mind that the same reasoning for EoSs of class (iii) and class (iv) EoSs also applies to all following relations discussed in this paper (Fig.~\ref{fig:frrem} to~\ref{fig:fvs185}).


\subsection{Fits and residuals}\label{ssec:fit}
To quantify the discussion above and to introduce a measure for the scatter inherent to the presented relations, we fit power laws of the type $R_{\mathrm{TOV}}=a\cdot (f_{\mathrm{peak}})^b +c$ through the data points of Figs.~\ref{fig:frmax}-\ref{fig:fr135} with $a,b,c$ being parameters to be obtained by a least-square fit. $R_{\mathrm{TOV}}$ denotes the corresponding radius of nonrotating NSs used as characteristic EoS parameter in Figs.~\ref{fig:frmax}-\ref{fig:fr135}. Because of the reasons mentioned in the last paragraph of Sect.~\ref{sssect:nsradii} we restrict the set of EoSs determining the parameters to the accepted purely microphysical models omitting also the EoSs of strange quark matter and the Glendnh3, which is in conflict with radius constraints discussed in~\cite{2010PhRvL.105p1102H}. The derived parameters are given in Tab.~\ref{tab:fit} for the four different characteristic radii of nonrotating NSs. Tab.~\ref{tab:fit} also quantifies the deviations of the data points from the corresponding fit. We list the maximum value and the average value of the residuals for the accepted models only and for all models except for the strange quark matter EoSs. As already apparent from Figs.~\ref{fig:frmax} to~\ref{fig:fr135}, the radius of a star with 1.6~$M_{\odot}$ represents the best choice for characterizing an EoS in the sense that a tight relation between the dominant GW frequency of the postmerger remnant and the radius of the nonrotating NS is desired.

Considering only the 8 purely microphysical models (classes (i) and (ii)) that yield a peak frequency below 2.8~kHz \footnote{Fully microphysical models with $f_{\mathrm{peak}}<2.8$~kHz are eosL, GS1, Shen, Glendnh3, LS375, GS2, SKA, eosO.}, one finds that in~Fig.~\ref{fig:fr16} the maximum residual from a straight line is only 50~meters (average of residuals is 26~meters), allowing for an even better constraint of the NS radius. For the five accepted fully microphysical models (red and black crosses) with $f_{\mathrm{peak}}>2.8$~kHz \footnote{Fully microphysical models with $f_{\mathrm{peak}}>2.8$~kHz are LS220, BurgioNN, Sly4, APR, eosUU.} we observe a maximum deviation from a linear fit of only 58~meters (average of residuals is 35~meters). Hence, in Fig.~\ref{fig:fr16} a broken straight line represents the best approximation to the presented scaling with
\begin{equation} \label{eq:brline}
f_{\mathrm{peak}}=\begin{cases}
-0.2823\cdot R_{1.6}+6.284&\text{ for $f_{\mathrm{peak}}<2.8$~kHz}\\
-0.4667\cdot R_{1.6}+8.713&\text{ for $f_{\mathrm{peak}}>2.8$~kHz}
\end{cases}
\end{equation}
where radii are in km and frequencies are in kHz. The fits are shown with solid lines in Fig.~\ref{fig:fr16}.

In order to quantify to which accuracy the radius of a 1.6~$M_{\odot}$~NS can be obtained from a GW detection of the postmerger phase, we consider the maximum deviation from the derived fits and add the uncertainty in the measurement of the peak frequency: For $f_{\mathrm{peak}}$ we adopt an uncertainty of 40~Hz (see Sect.~\ref{ssec:gwdet}), which via Eq.~\eqref{eq:brline} corresponds to 86~meters for $f_{\mathrm{peak}}>2.8$~kHz and 142~meters for $f_{\mathrm{peak}}<2.8$~kHz. The larger error bar for low peak frequencies is caused by the flatter slope of Eq.~\eqref{eq:brline} in this frequency range. We then add the maximum deviation of 58~meters (50~meters for $f_{\mathrm{peak}}<2.8$~kHz) from the broken line (Eq.~\eqref{eq:brline}) found for the accepted fully microphysical candidate EoSs (ignoring the outliers Glendnh3 and MIT40 for the reasons described in Sect.~\ref{sssect:nsradii}). In total this amounts to 144~meters for $f_{\mathrm{peak}}>2.8$~kHz and to 192~meters for a lower peak frequency. Being conservative one may add an uncertainty of up to 200~meters due to the use of the conformal flatness approximation subsuming also general effects by differences in the details of the implementation of different codes (see Appendix~\ref{app:grcomp}). We do not take into account an error resulting from the approximate treatment of thermal effects (see Appendix~\ref{app:gamma}) because this does not apply to all EoSs. (As estimated in Appendix~\ref{app:gamma} for an ideal-gas index $\Gamma_{\mathrm{th}}=1.75$ instead of 2, the maximum residual of the data points from a power law is 126~meters.) Note that the latter two effects are no fundamental sources of errors and can be overcome by a more elaborate modeling of NS mergers and the high-density EoS. We also mention the possibility that the residuals from a fit might be further reduced by using different radii of nonrotating NSs to characterize the EoSs in different ranges of the peak frequency.

\begin{table*}
\caption{\label{tab:fit}Parameters of power-laws fitted to the data points shown in Figs.~\ref{fig:frmax}-\ref{fig:fr135}. See the main text for definitions, units are such that NS radii are measured in km and $f_{\mathrm{peak}}$ in kHz. $R_{\mathrm{TOV}}$ denotes which radius of nonrotating NSs is used to characterize the different EoSs. The last four columns provide the maximum value of the residuals and the average of the residuals for all accepted models of fully microphysical EoSs (without MIT60, MIT40, Glendnh3) and for all models except for the strange quark matter EoSs.}
 \begin{ruledtabular}
 \begin{tabular}{|l|l|l|l|l|l|l|l|}
$R_{\mathrm{TOV}}$& $a$ & $b$ & $c$ & $\Delta_{\mathrm{max}}^{\mathrm{accepted}}$~[km] & $\Delta_{\mathrm{mean}}^{\mathrm{accepted}}$~[km] & $\Delta_{\mathrm{max}}^{\mathrm{all}}$~[km] & $\Delta_{\mathrm{mean}}^{\mathrm{all}}$~[km] \\ \hline
$R_{\mathrm{max}}$& 17.35 & -1.053  & 5.201  & 0.281 & 0.142 & 1.187 & 0.307 \\ \hline
$R_{1.8}$         & 20.57 & -0.7019 & 2.466  & 0.190 & 0.084 & 1.726 & 0.272 \\ \hline
$R_{1.6}$         & 21.53 & -0.5598 & 0.523  & 0.124 & 0.064 & 0.357 & 0.103 \\ \hline
$R_{1.35}$        & 21.28 & -0.5276 & 0.3394 & 0.232 & 0.109 & 0.565 & 0.211 \\ \hline
\end{tabular}
 \end{ruledtabular}
\end{table*}

\subsection{Interpretation}\label{ssec:interpretation}
\begin{figure}
\includegraphics[width=8.9cm]{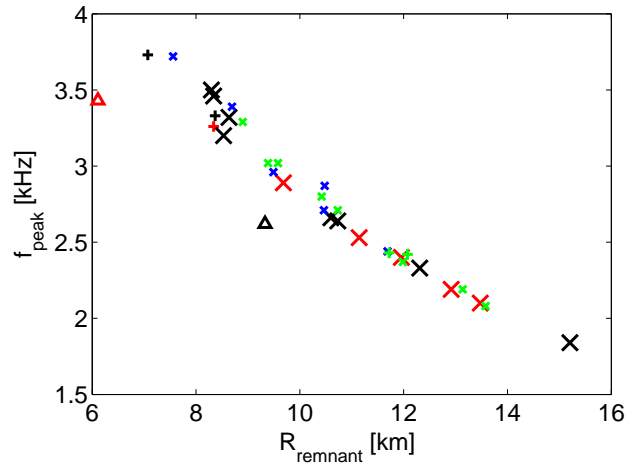}
\caption{\label{fig:frrem}Peak frequency of the postmerger GW emission versus the radius of a sphere enclosing 2.6~$M_{\odot}$ of rest mass of the merger remnant for all fully microphysical EoSs 8~ms after merging. Symbols have the same meaning as in Fig.~\ref{fig:frmax}.}
\end{figure}

In a previous paper we presented arguments why the dominant oscillation frequency of the differentially rotating merger remnant scales with the radius of a nonrotating NS whose mass is generally smaller than the one of the remnant~\cite{2012PhRvL.108a1101B}. As has been shown in~\cite{2011arXiv1105.0368S}, the GW emission at the peak frequency is generated by the fundamental quadrupolar oscillation mode. The frequency of this mode is known to be proportional to the square root of the mean density, $\sqrt{\frac{M}{R^3}}$ with $M$ and $R$ being the mass and the radius of the oscillating object (see~\cite{1998MNRAS.299.1059A,2011PhRvD..83f4031G}). The mass of the merger remnant is approximately given by the total binary mass and therefore it is the same for all models discussed in this section neglecting small amounts of ejecta and differences in the inflated torus surrounding the central object. Hence, the peak frequency is entirely determined by the radius of the DRO. The radius of the merger remnant cannot be defined unambiguously because one cannot identify a well defined surface of the object (see e.g. Fig.~5 in~\cite{PhysRevD.81.024012}). Using arbitrarily the radius of a sphere enclosing 2.6~$M_{\odot}$ of rest mass as the radius of the DRO, Fig.~\ref{fig:frrem} confirms the close relation between $f_{\mathrm{peak}}$ and the so chosen radius of the merger remnant $R_{\mathrm{remnant}}$. Here, $R_{\mathrm{remnant}}$ is measured 8~ms after merging when the oscillations of the DRO are sufficiently damped (see Figs.~\ref{fig:amplitude} and~\ref{fig:amplitude2}). The radii of the merger remnants for different EoSs are also provided in Tab.~\ref{tab:gw}. (The data point in Fig.~\ref{fig:frrem} with $f_{\mathrm{peak}}=3.2$~kHz, which is located slightly below the relation ($R_{\mathrm{remnant}}=8.53$~km), corresponds to the fully microphysical BurgioNN EoS, where our somewhat arbitrary definition of $R_{\mathrm{remnant}}$ fails. In particular the time, when $R_{\mathrm{remnant}}$ is determined, is arbitrarily chosen. It should be sufficiently early to characterize the GW emission, but not too early when the DRO, and thus $R_{\mathrm{remnant}}$, are still strongly oscillating, which is the case for the BurgioNN model.)

\begin{figure}
\includegraphics[width=8.9cm]{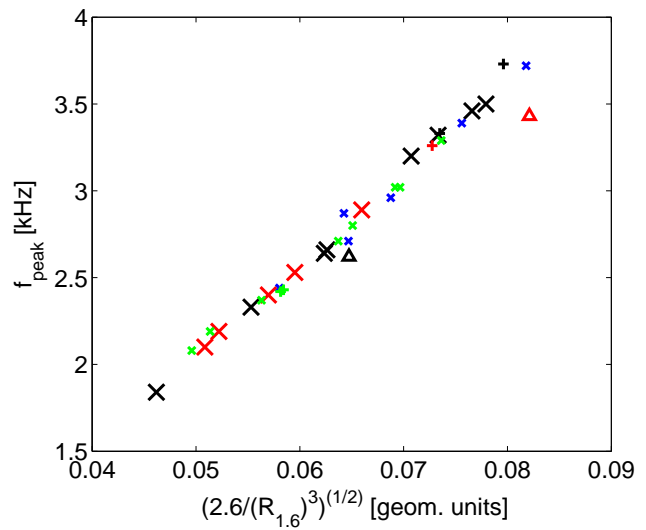}
\caption{\label{fig:frrrhomean}Peak frequency of the postmerger GW emission versus $\sqrt{\frac{2.6}{R_{1.6}^3}}$ in geometrical units for different EoSs. Symbols have the same meaning as in Fig.~\ref{fig:frmax}.}
\end{figure}

To understand the correlations found in Figs.~\ref{fig:frmax} to~\ref{fig:fr135}, where the frequency showed a tight anticorrelation with radii of static TOV configurations, we hypothesize that for a given EoS the radius of the differentially rotating merger remnant of about 2.6~$M_{\odot}$ scales with the radius $R_{\mathrm{TOV}}$ of a nonrotating NS for a chosen mass. This hypothesis is confirmed when considering $R_{\mathrm{remnant}}$ as a function of $R_{\mathrm{TOV}}=R_{1.35}$, $R_{1.6}$, $R_{1.8}$, or $R_{\mathrm{max}}$, of which the relation with $R_{1.6}$ shows the smallest scatter. Neglecting effects due to thermal contributions and differential rotation, a linear relation between $R_{\mathrm{max}}$ and the radius of the most massive, uniformly rotating NS configuration was reported in~\cite{1996ApJ...456..300L}. Adopting therefore a linear dependence between $R_{\mathrm{remnant}}$ and $R_{\mathrm{TOV}}$, one expects that $f_{\mathrm{peak}}$ is proportional to $R_{\mathrm{TOV}}^{-3/2}$. When fitting a power law $f_{\mathrm{peak}}=a'\cdot R_{\mathrm{TOV}}^{-3/2}$ to the data points of Figs.~\ref{fig:frmax} to~\ref{fig:fr135} similar residuals as listed in Tab.~\ref{tab:fit} are found, which implies that in fact there exists a tight relation between $R_{\mathrm{remnant}}$ and $R_{\mathrm{TOV}}$. Additionally, Fig.~\ref{fig:frrrhomean} shows the peak frequency as a function of $\sqrt{\frac{2.6}{R_{1.6}^3}}$, a quantity which according to the above reasoning is proportional to the mean density of the merger remnant. This behavior is confirmed by the linear scaling evident from Fig.~\ref{fig:frrrhomean}, which should be considered as an empirical finding of this work.

\begin{figure}
\includegraphics[width=8.9cm]{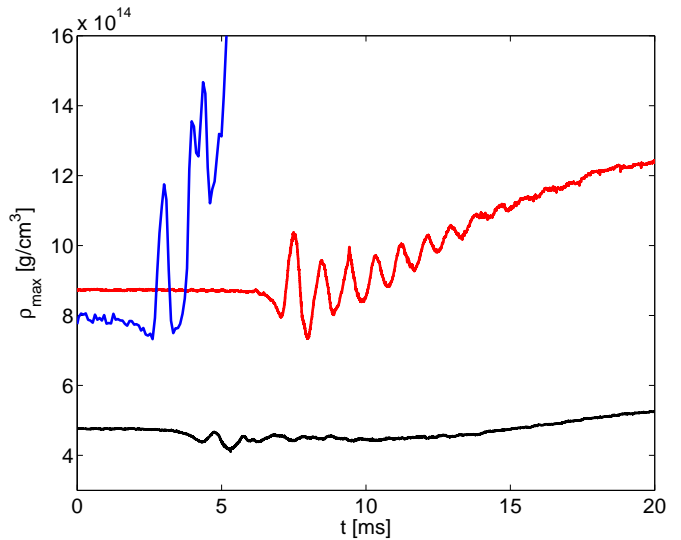}
\caption{\label{fig:rhomax}Evolution of the maximum rest-mass density for the models with the Shen EoS (black), the Sly4 (red) EoS and the MIT60 EoS (blue). At the beginning of the evolution the shown maximum rest-mass density corresponds to the central density in the inspiraling NSs, while after merging the maximum value of the rest-mass density is found in the center of the DRO.}
\end{figure}

The fact that the relation between $f_{\mathrm{peak}}$ and the radius of a NS with 1.6~$M_{\odot}$ shows the best quality, can be understood by investigating the involved density regimes. For a given EoS the central density in the merger remnant is well below the central density $\rho_{\mathrm{max}}$ of the nonrotating maximum-mass configuration as specified in Tab.~\ref{tab:models}, but typically above the central density of the initial binary components, i.e., of 1.35~$M_{\odot}$ NSs in the considered case (exceptions to this relatively general behavior are rather stiff EoSs like Shen or GS1). This can be seen by the density evolution in Fig.~\ref{fig:rhomax}. Hence, the density regime probed by the merger remnant and thus determining the GW emission is comparable to the densities encountered in nonrotating NSs with masses above 1.35~$M_{\odot}$ but below $M_{\mathrm{max}}$. On the one hand this means that the value of $R_{\mathrm{max}}$ as a characteristic EoS property is determined by a density regime of the EoS that does not occur in the merger remnant. This explains the larger scatter in Fig.~\ref{fig:frmax}. On the other hand, the radius $R_{1.35}$ depends only on the EoS at relatively low densities, whereas the structure of a DRO of a merger with $M_{\mathrm{tot}}=2.7~M_{\odot}$ is determined by the EoS at higher densities. Thus, the relatively large deviations in Fig.~\ref{fig:fr135} are a consequence of the inability of a 1.35~$M_{\odot}$ NS to capture the EoS behavior at densities realized in the merger remnant.

\subsection{Additional dependences}
\begin{figure}
\includegraphics[width=8.9cm]{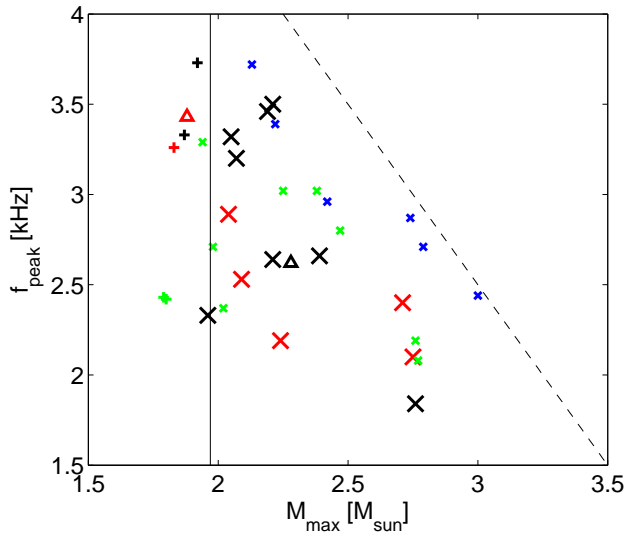}
\caption{\label{fig:fmmax}Peak frequency of the postmerger GW emission versus the maximum mass $M_{\mathrm{max}}$ of nonrotating NSs for different EoSs. Symbols have the same meaning as in Fig.~\ref{fig:frmax}. The vertical solid line marks the 1.97~$M_{\odot}$ mass limit set by~\cite{2010Natur.467.1081D}. The dashed line represents an upper limit of $M_{\mathrm{max}}$ for fully microphysical EoSs (see main text).}
\end{figure}

Though we identify NS radii as the crucial characteristic EoS-dependent quantity that affects the late-time GW emission, additional constraints on NS properties by GW signal features can be derived from our survey of NS mergers for a large set of EoSs. Fig.~\ref{fig:fmmax} shows the peak frequency as a function of the maximum mass $M_{\mathrm{max}}$ of nonrotating NSs. The vertical line indicates the mass limit set by the observation of the 1.97~$M_{\odot}$ pulsar~\cite{2010Natur.467.1081D}. Note that some EoS on the left of the vertical line (crosses) are not considered as excluded because they are compatible with the mass measurement within the error bars. Figure~\ref{fig:fmmax} shows that high peak frequencies tend to exclude high values of $M_{\mathrm{max}}$. Note that the three blue symbols with very high $M_{\mathrm{max}}$ belong to EoSs of class (iv) (with piecewise-polytrope extension at high densities), and that these three particular EoSs are subject to an extreme stiffening of the EoS at high densities (see Figs.~\ref{fig:eos} and~\ref{fig:vsound}). Considering our set of representative EoSs one can conclude that for a given peak frequency $f_{\mathrm{peak}}$ the maximum mass of nonrotating NSs can be constrained to roughly $M_{\mathrm{max}}\le 4.25~M_{\odot} -0.5 (M_{\odot}/\mathrm{kHz})\cdot f_{\mathrm{peak}} $ (see dashed line in~\ref{fig:fmmax}).

\begin{figure}
\includegraphics[width=8.9cm]{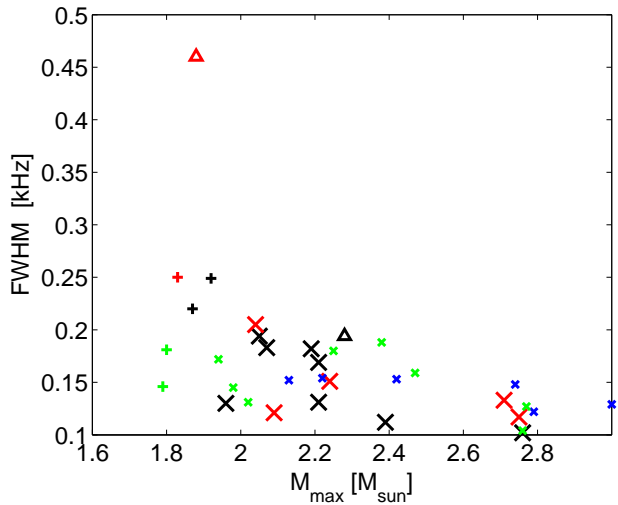}
\caption{\label{fig:fwhmmmax}FWHM of the peak structure of the postmerger GW emission versus the maximum mass $M_{\mathrm{max}}$ of nonrotating NSs for different EoSs. Symbols have the same meaning as in Fig.~\ref{fig:frmax}.}
\end{figure}

\begin{figure}
\includegraphics[width=8.9cm]{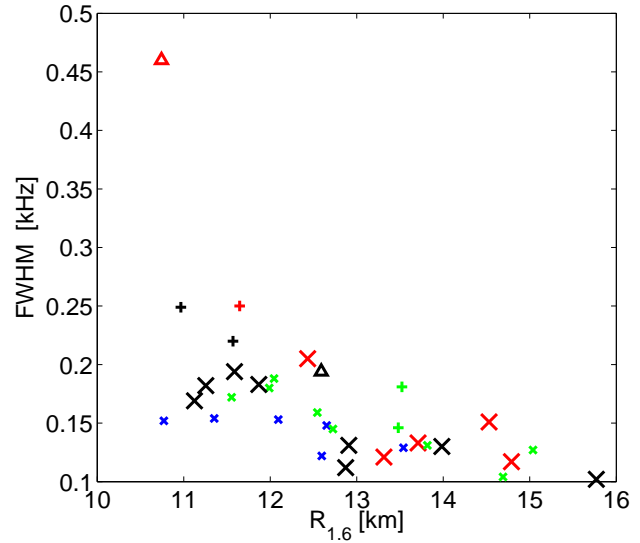}
\caption{\label{fig:fwhmr16}FWHM of the peak structure of the postmerger GW emission versus the radius of nonrotating NSs with 1.6~$M_{\odot}$ for different EoSs. Symbols have the same meaning as in Fig.~\ref{fig:frmax}.}
\end{figure}

In addition to the peak frequency one may also consider to use the width of the peak to characterize the GW emission during the postmerger phase, although it is unlikely to be obtained precisely from a GW observation by Advanced LIGO or Advanced Virgo. However, the situation may change with the increased sensitivity of the Einstein Telescope. Since the peak frequency is determined by the size of the merger remnant, the width of the peak can be considered as a measure for the change of the structure of the DRO and its evolution. In the following we quantify the width of the peak by the FWHM (see Tab.~\ref{tab:gw}). Fig.~\ref{fig:fwhmmmax} illustrates the dependence of the FWHM on $M_{\mathrm{max}}$, whereas in Fig.~\ref{fig:fwhmr16} the peak width is shown as a function of the radius of a nonrotating NS with 1.6~$M_{\odot}$. Using $R_{\mathrm{max}}$ to characterize the different models results in a picture qualitatively similar to Fig.~\ref{fig:fwhmr16}. It is striking that the FWHM does not show a clear dependence on the stellar parameters characterizing a given EoS, instead for a given FWHM a large spread in $M_{\mathrm{max}}$ and $R_{1.6}$ is found. A slight trend is noticeable that EoSs with high $M_{\mathrm{max}}$ or larger $R_{1.6}$ tend to have smaller peak widths. Furthermore, one recognizes that only models that are excluded by the observation of the 1.97~$M_{\odot}$ pulsar yield increased FWHMs, though even in these cases the absolute values are not higher than about 250~Hz (except for the strange quark matter MIT60 EoS). However, we emphasize that not all excluded models show this behavior.

The broadest peak with FWHM=460~Hz is found for the MIT60 EoS describing absolutely stable strange quark matter. It seems unlikely that this finding is a particular consequence of the fact that this EoS models strange quark stars because for MIT40 a narrow peak is found. Rather, what distinguishes MIT60 from the other EoSs is the early occurrence of the delayed BH formation taking place about 3~ms after merging, i.e. during a period when the postmerger remnant is still vividly emitting GWs. This leads to the conclusion that the broadening of the peak is connected to the dramatic structural changes in the DRO prior to collapse, visible for example in the evolution of the central density (see Fig.~\ref{fig:rhomax}). See also the discussion in Sect.~\ref{sec:binpara}.

We stress that MIT60 is the only model out of the 38 discussed here, where the delayed collapse occurs in the period of strong GW emission, i.e. at a time when the peak in the GW spectrum is shaped. In the remaining simulations either the prompt collapse to a BH takes place or the forming DROs are stable for at least 10 to 15~ms until the GW emission has significantly decreased. Only in one other case, namely the H3 EoS, we observe the delayed collapse before the end of the simulation about 25~ms after merging. Note that MIT60 yields an $M_{\mathrm{max}}$ by which this EoS is excluded because of $M_{\mathrm{max}}<1.97~M_{\odot}$. The broadest peak among the accepted models of our survey is only 205~Hz (for the LS220 EoS).

Considering these observations and the large variety in our representative sample of EoSs we conclude that the early occurrence of a delayed collapse of the DRO is very unlikely to happen and, therefore, a broadening of the peak in the postmerger GW spectrum seems improbable to occur at least for the binary setup considered here. Beside this empirical finding, to our knowledge there is no case reported in the literature where the merger of two stars with 1.35~$M_{\odot}$ yields an extraordinary broad peak in the GW spectrum for an EoS which is compatible with the NS mass measurement of~\cite{2010Natur.467.1081D}.

In~\cite{2011PhRvD..83l4008H} the delayed collapse within the first 5~ms of the lifetime of the DRO was found to be generic in a certain range of total binary masses slightly below the threshold for the direct BH formation. In view of this it is unexpected that we do not observe any case of such an early delayed collapse except for the excluded strange quark matter MIT60 EoS when we consider our very large sample of EoSs. For instance, according to~\cite{2011PhRvD..83l4008H} the merger of two NSs of 1.35~$M_{\odot}$ with the Sly4 EoS should have led to a collapse within 5~ms after merging. To explore this difference we perform additional calculations for the Sly4 EoS with slightly increased binary masses. The resulting lifetimes of the remnants are summarized in Tab.~\ref{tab:collapse} for simulations using $\Gamma_{\mathrm{th}}=2$ (our standard choice; see Sect.~\ref{sec:methods}) and $\Gamma_{\mathrm{th}}=1.357$ as employed in~\cite{2011PhRvD..83l4008H}.

For $\Gamma_{\mathrm{th}}=2$ we find either the prompt collapse to a BH or the formation of relatively long-lived merger remnants depending on the total binary mass. In contrast, with a description of thermal effects that corresponds to a very inefficient shock heating ($\Gamma_{\mathrm{th}}=1.357$) we indeed identify a parameter range for which short-lived merger remnants with lifetimes below 5~ms occur. This is in qualitative agreement with the findings of~\cite{2011PhRvD..83l4008H}. Our calculations thus suggest that the frequent occurrence of short DRO lifetimes (below 5~ms) in~\cite{2011PhRvD..83l4008H} is a consequence of the assumption of a very low shock-heating efficiency and it shows that the choice of $\Gamma_{\mathrm{th}}$ has a crucial influence on the stability of the merger remnant. This is understandable since thermal pressure support has generally a stabilizing effect. (In~\cite{2010PhRvD..82h4043B} the thermal ideal-gas index was extracted from microphysical EoSs including a consistent description of temperature effects, showing that $\Gamma_{\mathrm{th}}$ in the range from 1.5 to 2 represents a suitable approximation. See Fig.~2 in~\cite{2010PhRvD..82h4043B}.)

The quantitative differences in Tab.~\ref{tab:collapse} might be explained by differences in the exact implementation of the EoS (original table vs. piecewise polytropic approximation of the EoS) and the different degree of sophistication of the simulations (SPH in combination with the conformal flatness approximation and a GW backreaction scheme vs. grid-based hydrodynamics in full general relativity where the simulations start more than five orbits before merging). See also Appendix~\ref{app:grcomp}.

In the context of DRO lifetimes we also mention that a slight decrease in the total binary mass ($M_{\mathrm{tot}}=2.68~M_{\odot}$ instead of $M_{\mathrm{tot}}=2.7~M_{\odot}$) for the model with the eosAU prevents the prompt collapse and instead leads to the formation of a DRO which is stable for at least 14~ms and generates a narrow peak of 169~Hz at $f_{\mathrm{peak}}=3.94$~kHz. Hence, based on this marginal case and on our survey we conclude that the generic outcome of a NS merger of two stars with 1.35~$M_{\odot}$ is either the prompt collapse or a relatively long-lived DRO (with a lifetime of typically more than 5~ms) where shock heating is crucial for the stability of the DRO. We stress that for a thorough investigation of the stability and lifetime of merger remnants the consideration of a large sample of temperature dependent EoSs is essential.

\begin{table*}
\caption{\label{tab:collapse}Lifetime in ms of the postmerger remnant in simulations for the Sly4 EoS. In the first row the masses of the simulated binary components are given in $M_{\odot}$. The entry ``prompt BH'' indicates direct BH formation. Bars mark cases where no calculations have been performed. Note that we report lower bounds on the lifetime when the collapse to a BH does not occur until the end of the simulation.}
 \begin{ruledtabular}
 \begin{tabular}{|l|l|l|l|l|l|}

                                 & 1.35-1.35  & 1.38-1.38    & 1.4-1.4 & 1.42-1.42   & 1.45-1.45  \\ \hline
$\Gamma_{\mathrm{th}}=2$ (this work)     & $>15$   & $>10$     & $>12$   & $>10$ & prompt BH \\ \hline
$\Gamma_{\mathrm{th}}=1.357$ (this work) & $>11$   & $>9$      & 2.5 & prompt BH    & prompt BH \\ \hline
$\Gamma_{\mathrm{th}}=1.357$ (\cite{2011PhRvD..83l4008H}) & $<5$ & - & prompt BH  & - & - \\ \hline
\end{tabular}
 \end{ruledtabular}
\end{table*}


\section{Dependence on thermodynamical properties} \label{sec:thermo}
\begin{figure}
\includegraphics[width=8.9cm]{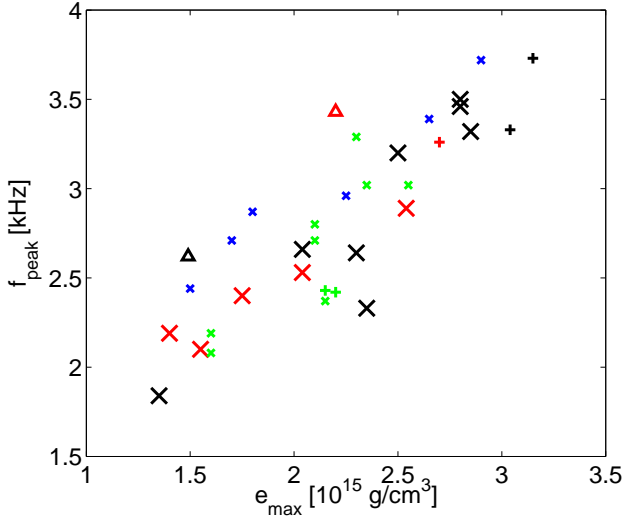}
\caption{\label{fig:frhomax}Peak frequency of the postmerger GW emission versus the maximum central energy density of stable nonrotating NSs for different EoSs. Symbols have the same meaning as in Fig.~\ref{fig:frmax}.}
\end{figure}

Section~\ref{sec:stellar} provides evidence that NS radii play a crucial role for determining the GW emission of NS mergers. Stellar parameters can be considered to be a consequence of the bulk properties of the high-density EoS because they are integral values associated with the solution of the TOV equations taking into account the EoS behavior over a large density regime. In this section we directly relate the dominant GW frequency of the postmerger phase to specific properties of the NS EoS (see Tab.~\ref{tab:models}). Fig.~\ref{fig:frhomax} shows a clear correlation between $f_{\mathrm{peak}}$ and the central energy density of the nonrotating maximum-mass configuration of the TOV solutions. However, in comparison to Figs.~\ref{fig:frmax} to~\ref{fig:fr135} one finds a wider scatter and one cannot infer a clear functional dependence. At most exclusion regions can be identified. As argued in Sect.~\ref{sec:stellar} the deviations are understandable because only a certain density range, which does not include $e_{\mathrm{max}}$, is probed by the postmerger remnant. This effect is most pronounced for hyperon EoSs, where in contrast to nucleonic matter the behavior of the high-density EoS is changed by the appearance of hyperons. Consequently, the hyperon EoSs are found on the lower right of the band established by the correlation of other EoSs.

\begin{figure}
\includegraphics[width=8.9cm]{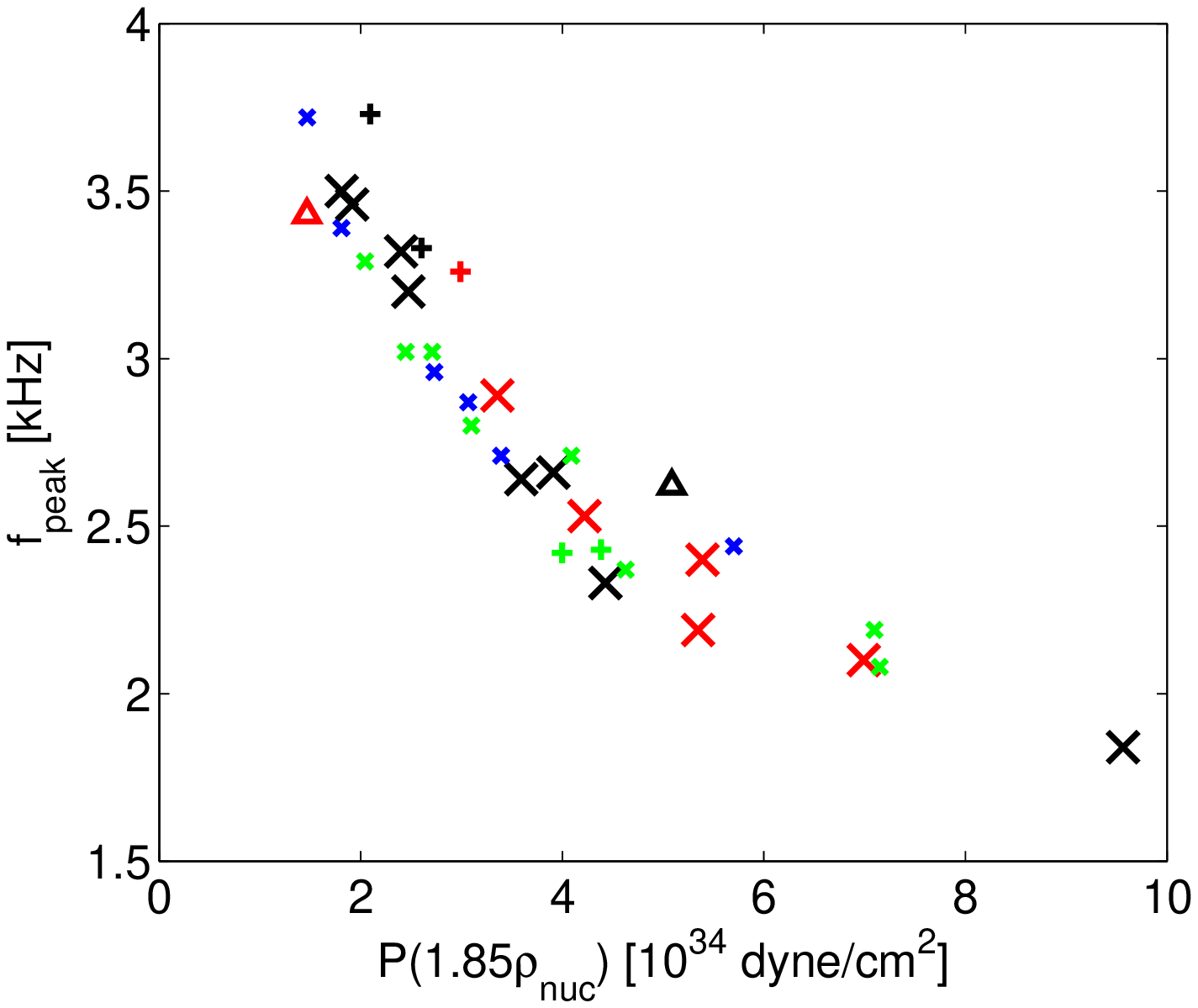}
\caption{\label{fig:fp185}Peak frequency of the postmerger GW emission versus the pressure at 1.85 times nuclear saturation density for different EoSs. Symbols have the same meaning as in Fig.~\ref{fig:frmax}.}
\end{figure}

Figure~\ref{fig:fp185}, which displays $f_{\mathrm{peak}}$ as a function of the pressure at $1.85\rho_0$ confirms these arguments. Here, the intrinsic scatter is smaller because one considers a characteristic quantity at a moderate density which is realized in the merger remnants. One notices that a low pressure at the fiducial rest-mass density of $1.85 \rho_0$ causes a higher peak frequency. This is plausible because a reduced pressure yields more compact stellar objects resulting in higher oscillation frequencies as discussed in Sect.~\ref{sec:stellar}. A direct correlation between $P(1.85 \rho_0)$ and the radius of nonrotating NSs has been pointed out in~\cite{2007PhR...442..109L} and is confirmed by the values given in Tab.~\ref{tab:models}.

\begin{figure}
\includegraphics[width=8.9cm]{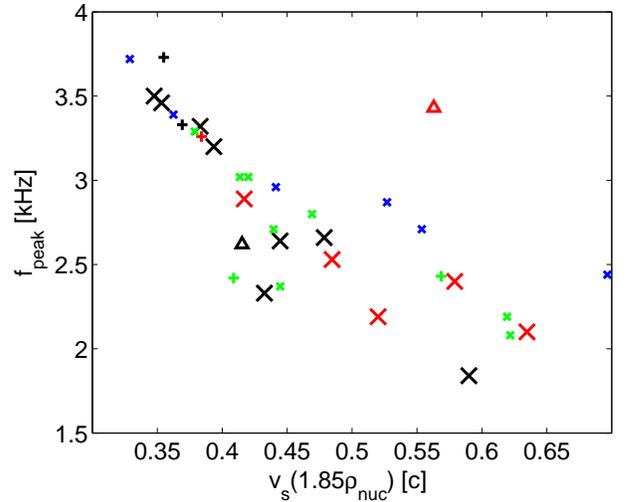}
\caption{\label{fig:fvs185}Peak frequency of the postmerger GW emission versus the sound speed at 1.85 times nuclear saturation density for different EoSs. Symbols have the same meaning as in Fig.~\ref{fig:frmax}.}
\end{figure}

Using the speed of sound as a characteristic EoS property reveals a relation as shown in Fig.~\ref{fig:fvs185}. The sound speed is determined at $1.85 \rho_0$. At high peak frequencies above 3~kHz the anticorrelation is rather tight, whereas the chosen EoS feature cannot be constrained significantly in cases of low peak frequencies. Note that using the speed of sound at saturation density reduces the scatter at frequencies below 2.8~kHz but leads to an increase of the variations at high $f_{\mathrm{peak}}$ (not shown). As a general finding we stress that choosing other fiducial densities for the pressure or the sound speed does not yield tighter correlations over the whole parameter range. We also note that the locus of the strange quark matter EoSs in diagrams like Figs.~\ref{fig:fp185} or~\ref{fig:fvs185} is highly sensitive to the chosen reference density (see Figs.~\ref{fig:eos} and~\ref{fig:vsound} for the determination of the characteristic properties depending on the fiducial density).

\section{Variation of binary parameters} \label{sec:binpara}
\subsection{Dependence on stellar radii}
\begin{figure}
\includegraphics[width=8.9cm]{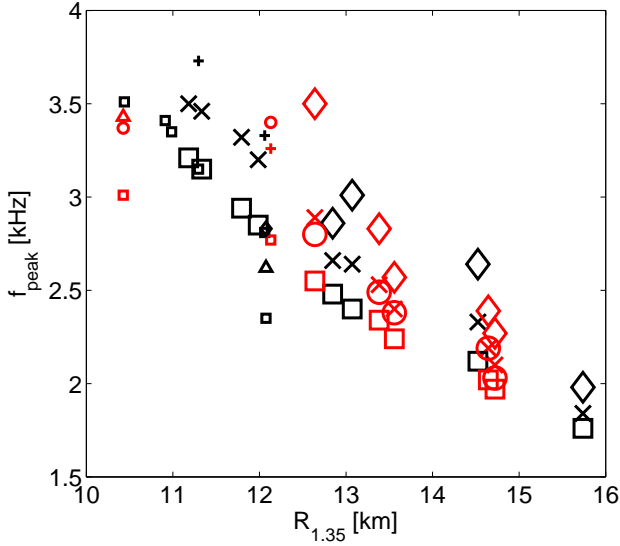}
\caption{\label{fig:fr135mtot}Peak frequency of the postmerger GW emission versus the radius of a nonrotating NS with 1.35~$M_{\odot}$ for different EoSs. The same color scheme as in Fig.~\ref{fig:frmax} is adopted. Squares mark the results of simulations with $M_{\mathrm{tot}}=2.4~M_{\odot}$, crosses and plus signs indicate the peak frequencies of mergers of two stars with 1.35~$M_{\odot}$, circles correspond to calculations of 1.2-1.5~$M_{\odot}$ mergers, and diamonds denote the outcome of symmetric binary coalescences with $M_{\mathrm{tot}}=3.0~M_{\odot}$. Triangles belong to the 1.35-1.35~$M_{\odot}$ simulations with the strange quark matter EoSs (MIT40 and MIT60). Note that the other binary setups for these EoSs are not specially highlighted. Small symbols indicate that the corresponding EoS is excluded by the 1.97~$M_{\odot}$ pulsar observation (except for the strange quark matter EoS MIT40, which is not excluded).}
\end{figure}

\begin{figure}
\includegraphics[width=8.9cm]{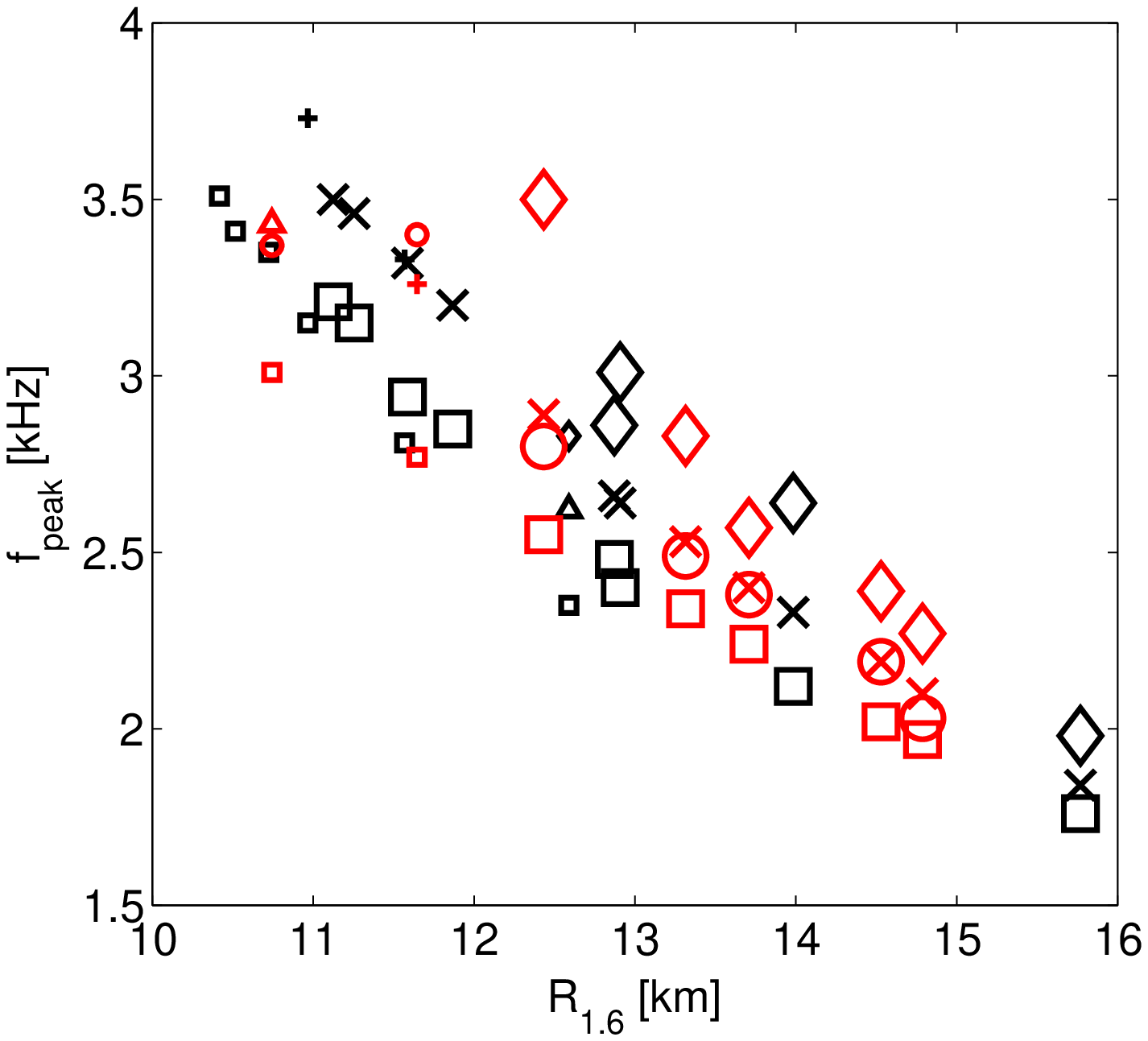}
\caption{\label{fig:fr16mtot}Peak frequency of the postmerger GW emission versus the radius of a nonrotating NS with 1.6~$M_{\odot}$ for different EoSs and varied binary setups. Symbols have the same meaning as in Fig.~\ref{fig:fr135mtot}.}
\end{figure}

\begin{figure}
\includegraphics[width=8.9cm]{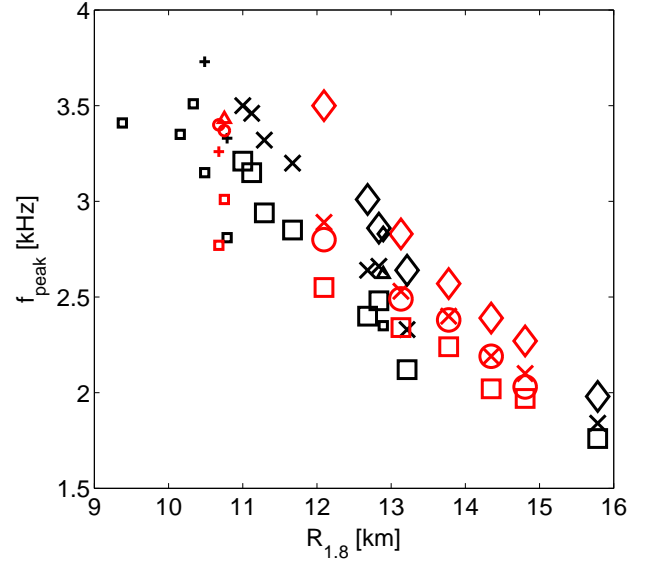}
\caption{\label{fig:fr18mtot}Peak frequency of the postmerger GW emission versus the radius of a nonrotating NS with 1.8~$M_{\odot}$ for different EoSs and varied binary setups. Symbols have the same meaning as in Fig.~\ref{fig:fr135mtot}.}
\end{figure}

In the previous sections NS mergers with a total binary mass of 2.7~$M_{\mathrm{tot}}$ and a mass ratio $q=M_1/M_2=1$ were considered and very tight anticorrelations in particular between the dominant GW frequency in the postmerger phase and the stellar radii of nonrotating NSs were observed. In the present section we explore these relations for different binary setups. In our discussion we include the results of simulations of binaries with total system masses (i.e. the gravitational mass of the two components in isolation added) of $2.4~M_{\odot}$, $2.7~M_{\odot}$ and $3.0~M_{\odot}$ each with a mass ratio $q=1$. Calculations are performed for all fully microphysical EoSs (classes i and ii). For the EoSs with a consistent temperature treatment (class i) we also test the effect of asymmetric initial masses by simulating 1.2-1.5~$M_{\odot}$ mergers. All results are given in Tabs.~\ref{tab:varbin} (symmetric binaries) and~\ref{tab:varbinasym} (asymmetric binaries). Figs.~\ref{fig:fr135mtot}, \ref{fig:fr16mtot} and~\ref{fig:fr18mtot} display the peak frequency of the postmerger stage as a function of the radii of nonrotating NSs with 1.35~$M_{\odot}$, 1.6~$M_{\odot}$ and~1.8~$M_{\odot}$ for the aforementioned binary setups analogous to Figs.~\ref{fig:fr18} to~\ref{fig:fr135}. The simulations with $M_{\mathrm{tot}}=2.7~M_{\odot}$, which were already discussed above, are shown with the same symbols used previously, i.e. crosses and plus signs. The same color scheme as in the previous sections is adopted. Low-mass binary merger are plotted with squares. The results of calculations with $M_{\mathrm{tot}}=3.0~M_{\odot}$ are given by diamonds. Data points corresponding to excluded EoSs are displayed by smaller symbols. Simulations with $q\neq 1$ are indicated by circles. Note that the strange quark-matter cases are highlighted by special symbols only for the 1.35-1.35~$M_{\odot}$ mergers. However, the symbols belonging to a certain EoS can be readily identified by their location at the same NS radius.

The $M_{\mathrm{tot}}=2.7~M_{\odot}$ simulations were extensively discussed in Sect.~\ref{sec:stellar}. Considering the calculations with two 1.2~$M_{\odot}$ stars one finds in Figs.~\ref{fig:fr135mtot} to~\ref{fig:fr18mtot} that also for this binary setup relations between $f_{\mathrm{peak}}$ and the NS radii exist. The anticorrelation found with the radius $R_{1.35}$ is the tightest in comparison to the others. This is fully consistent with the interpretation in Sect.~\ref{ssec:interpretation} that the merger remnant probes mostly a certain density regime which for lower $M_{\mathrm{tot}}$ corresponds to the radii of lower-mass NSs. Consequently, the scatter is enhanced when using the radii of nonrotating high-mass NSs as characteristic properties (Figs.~\ref{fig:fr16mtot} and~\ref{fig:fr18mtot}). The strange quark matter EoSs are off the relations in all three plots. In addition, only the result for the Glendnh3 EoS ($R_{1.35}=14.52$~km) seems to deviate slightly from the very tight anticorrelation established by the other EoSs in Fig.~\ref{fig:fr135mtot}.

Choosing $R_{1.6}$ as characterizing EoS property in Fig.~\ref{fig:fr16mtot}, all excluded models (small symbols) occur as outliers at smaller radii but seem to fulfill their own relation. This behavior is in analogy to the one in Fig.~\ref{fig:fr18} (see discussion in Sect.~\ref{sec:stellar}) and can be explained by the fact that for these EoSs the NSs with 1.6~$M_{\odot}$ are not located on the nearly vertical sequence of the mass-radius relation as it is the case for EoSs with higher $M_{\mathrm{max}}$.

In Fig.~\ref{fig:fr16mtot} the two EoSs appearing at nearly the same radius of about 13~km are the SKA EoS and eosO (black symbols). While they show some difference in $f_{\mathrm{peak}}$ for the $M_{\mathrm{tot}}=2.4~M_{\odot}$ and $M_{\mathrm{tot}}=3.0~M_{\odot}$ runs, the peak frequencies of the 1.35-1.35~$M_{\odot}$ mergers are very close. This illustrates the importance of choosing an appropriate characterizing NS radius dependent on the total binary mass, cf. the location of these EoSs in Figs.~\ref{fig:fr135mtot} and~\ref{fig:fr18mtot}. Accordingly, for mergers of two stars with 1.5~$M_{\odot}$ the radius of a nonrotating 1.8~$M_{\odot}$ NS leads to the most narrow anticorrelation, although compared to the simulations with $M_{\mathrm{tot}}=$2.4~$M_{\odot}$ or 2.7~$M_{\odot}$ still larger scatter is visible. However, note that only the three models based on barotropic EoSs (black diamonds) seem to yield peak frequencies that are systematically on the lower side while the remaining models form a clear relation (that can be described for instance by a power law). The behavior of these three class-ii EoSs at smaller radii may be an artifact connected to the approximate temperature treatment needed for these models. The usage of $\Gamma_{\mathrm{th}}=2$ may underestimate $f_{\mathrm{peak}}$ in comparison to a consistent description of thermal effects (see Appendix~\ref{app:gamma} and~\cite{2010PhRvD..82h4043B}), which becomes in particular important for high-mass binaries and EoSs yielding compact NSs.

Exploring the influence of $M_{\mathrm{tot}}$ on the peak frequency for different EoSs, a stronger effect is found for models with compact NSs, i.e. at higher GW frequencies. While for eosL, for example, $f_{\mathrm{peak}}$ ranges from 1.76~kHz to 1.98~kHz when $M_{\mathrm{tot}}$ varies from 2.4~$M_{\odot}$ to 3.0~$M_{\odot}$, differences in the peak frequency exceeding 500~Hz are observed for EoSs with smaller NS radii for the same variations in the total binary mass. In comparison, the effect of the initial mass ratio of the binaries is more modest. The peak frequencies of 1.2-1.5~$M_{\odot}$ mergers are in general slightly lower than the corresponding $f_{\mathrm{peak}}$ of the symmetric setup with the same total binary mass. This trend can be expected because for an asymmetric binary during the coalescence more matter is spread into the torus surrounding the central object. Hence, the DRO is less massive and thus oscillates at lower frequencies. The increased frequency for the 1.2-1.5~$M_{\odot}$ merger with the excluded LS180 EoS is a consequence of the gravitational collapse only 9~ms after merging, which is accompanied by a compactification of the remnant prior to the collapse. Note that in Figs.~\ref{fig:fr135mtot} to~\ref{fig:fr18mtot} we do not include the result of the asymmetric merger with the MIT60 EoS because this model collapses about 2~ms after merging. For the accepted models the deviations of $f_{\mathrm{peak}}$ from the results of symmetric binaries seem to increase for smaller NS radii, but the GS1 EoS makes an exception to this tendency. The larger differences for smaller radii can be understood from the finding above that changes in the mass of the central DRO have a stronger effect on $f_{\mathrm{peak}}$ for such models. In summary, the influence of the initial mass ratio is relatively small, at most 90~Hz for the accepted models of class i. This demonstrates the insensitivity of the presented relations to small variations of the initial mass ratio $q$ and shows the robustness of our method for constraining NS radii with respect to uncertainties in the accurate determination of $q$ from the GW inspiral signal.

\subsection{Prompt collapse}
\begin{figure}
\includegraphics[width=8.9cm]{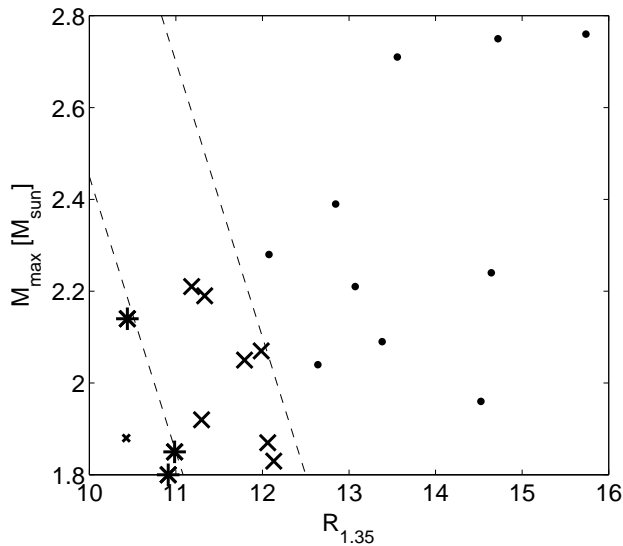}
\caption{\label{fig:pcollapse}Equation of state properties for direct BH formation are characterized by the corresponding maximum mass of nonrotating NS and the radius of a nonrotating NS with 1.35~$M_{\odot}$. Models for which the merger of a symmetric binary with $M_{\mathrm{tot}}=2.7~M_{\odot}$ leads to the prompt formation of a BH are displayed with asterisks. Crosses indicate cases where the coalescence of two stars with 1.5~$M_{\odot}$ each results in a prompt collapse. Dots denote simulations where the formation of a DRO is found for binaries with $M_{\mathrm{tot}}=3.0~M_{\odot}$. Note that we only consider EoSs of classes i and ii. The small cross marks the MIT60 EoS describing absolutely stable strange quark matter. Models collapsing promptly for $M_{\mathrm{tot}}=2.7~M_{\odot}$ also form directly a BH in a 1.5-1.5~$M_{\odot}$ merger. The dashed lines indicate the approximate boundary between prompt and delayed collapse for $M_{\mathrm{tot}}=2.7~M_{\odot}$ and $M_{\mathrm{tot}}=3.0~M_{\odot}$ (see text).}
\end{figure}

It can be noticed in Figs.~\ref{fig:fr135mtot} to~\ref{fig:fr18mtot} that in comparison to the 1.35-1.35~$M_{\odot}$ simulations there are more data points for 1.2-1.2~$M_{\odot}$ binaries and less for 1.5-1.5~$M_{\odot}$ mergers because with higher total binary masses for more EoSs the prompt collapse to a BH occurs. Characterizing a given EoS by $M_{\mathrm{max}}$ and $R_{1.35}$, Fig.~\ref{fig:pcollapse} illustrates the situation: Models are marked with asterisks there when the direct BH formation is observed for 1.35-1.35~$M_{\odot}$ mergers. Crosses indicate EoSs which lead to the prompt collapse for a total binary mass of $M_{\mathrm{tot}}=3.0~M_{\odot}$. The remaining fully microphysical models of 1.5-1.5~$M_{\odot}$ mergers with DRO formation are denoted by dots. Note that we do not find cases where the coalescence of two NSs with 1.2~$M_{\odot}$ results in the direct occurrence of a BH. The data point at $R_{1.35}=10.43$~km and $M_{\mathrm{max}}=1.88~M_{\odot}$ marks the MIT60 EoS, which as a strange quark matter EoS should be disregarded in this discussion. Despite the coarse sampling in the $R_{1.35}-M_{\mathrm{max}}$ space one can identify the region in the parameter space that allows for the direct BH formation for a given $M_{\mathrm{tot}}$. For a fixed total binary mass the boundary between this scenario and the formation of a DRO may be roughly approximated by
\begin{equation}
M_{\mathrm{max}} = - 0.6 (\frac{M_{\odot}}{\mathrm{km}}) R_{1.35} + M_{\mathrm{offset}}(M_{\mathrm{tot}}),
\end{equation}
where only the offset depends on $M_{\mathrm{tot}}$. We determine $M_{\mathrm{offset}}(2.7~M_{\odot})=8.45~M_{\odot}$ and $M_{\mathrm{offset}}(3.0~M_{\odot})=9.3~M_{\odot}$ (see dashed lines in Fig.~\ref{fig:pcollapse}). These findings are qualitatively consistent with the analysis of~\cite{2011PhRvD..83l4008H}, who proposed for the threshold binary mass distinguishing the two scenarios of a prompt and a delayed collapse a dependence $M_{\mathrm{threshold}}=k M_{\mathrm{max}}$. The parameter $k$ was found to depend on the radius of a 1.4~$M_{\odot}$~NS, which is confirmed by our calculations. A quantitative comparison is not meaningful because of the coarse sampling of our models in $M_{\mathrm{tot}}$ and because of the usage of $\Gamma_{\mathrm{th}}=1.357$ in~\cite{2011PhRvD..83l4008H} which implies a very low pressure increase associated with shock heating. But we stress that the relation between $k$ and $R_{1.4}$ is not very tight, which can be also seen in the data of~\cite{2011PhRvD..83l4008H}. This impedes the determination of $M_{\mathrm{max}}$ by means of this scaling. Besides, there are indications for $k$ being also sensitive to $M_{\mathrm{tot}}$. Note that in this subsection we focus entirely on symmetric mergers ($q=1$). The picture may change for asymmetric setups.

\subsection{Width of the peak}
\begin{figure}
\includegraphics[width=8.9cm]{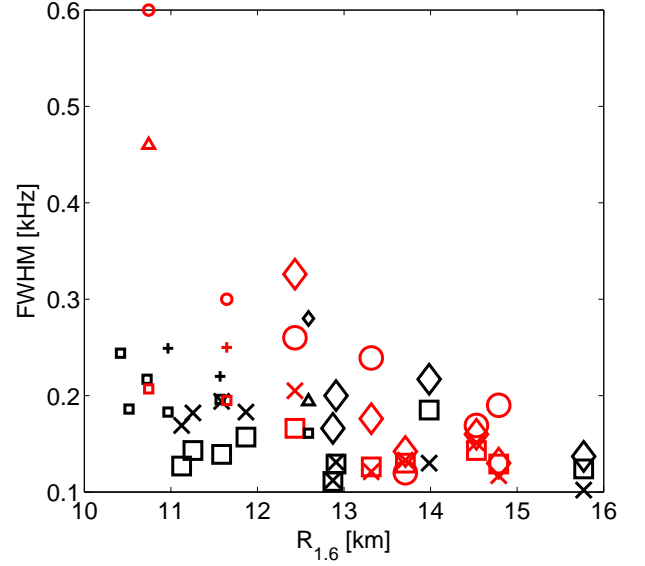}
\caption{\label{fig:fwhmmtot}FWHM of the peak structure of the postmerger GW emission versus the radius of nonrotating NSs with 1.6~$M_{\odot}$ for different EoSs and varied binary setups. Symbols have the same meaning as in Fig.~\ref{fig:fr135mtot}.}
\end{figure}

Tables~\ref{tab:varbin} and~\ref{tab:varbinasym} list the results of all simulations discussed in this section. In particular, the tables provide also the widths of the peaks. As a general trend one finds that the higher the total binary mass is, the broader is the peak in the GW spectrum. This is demonstrated in Fig.~\ref{fig:fwhmmtot} showing the FWHM as a function of the radius $R_{1.6}$ of a NS with 1.6~$M_{\odot}$ adopting the same notation as in Fig.~\ref{fig:fr135mtot}. As already indicated in Fig.~\ref{fig:fwhmr16} the simulations with EoSs yielding more compact NSs result in broader peaks, which is confirmed by Fig.~\ref{fig:fwhmmtot} also for other total binary masses. It is apparent that in calculations with excluded EoSs (small symbols) generally larger FWHMs are observed. Asymmetric binary setups also tend to generate broader peaks (circles), which for many EoSs is a stronger effect than varying $M_{\mathrm{tot}}$ by 0.3~$M_{\odot}$. Note that extraordinarily broadened peaks occur only in simulations where the DRO collapses in the phase that still produces a substantial emission of GWs. This is the case for the models LS180-1.2-1.5, MIT60-1.2-1.5, LS180-1.4-1.4, and LS220-1.5-1.5. From these general observations we conclude that the width of the peak in general is a measure of the proximity of the DRO to collapse, prior to which strong structural changes of the remnant occur. This can be seen, for instance, by the outcome and the FWHM for all simulations with the LS180 EoS when we include in addition the calculation for a symmetric binary with $M_{\mathrm{tot}}=2.8~M_{\odot}$ (not considered in the plots but listed in Tab.~\ref{tab:varbin}). This also means that our study cannot confirm that a broadened peak is a particular and generic consequence of the appearance of hyperons as it has been concluded in~\cite{2011PhRvL.107u1101S} based on the investigation of only two EoSs. These differ not only with respect to their hyperon content but in particular also in the resulting stellar parameters, both of which can have crucial influence on the dynamical behavior. For instance we find broad peaks also for purely nucleonic EoSs which lead to a collapse a few ms after merging, while not every EoS including hyperons causes an enlarged width of the peak (cf. models Glendnh3-1.2-1.2, Glendnh3-1.35-1.35, H3-1.35-1.35, H4-1.35-1.35, Glendnh3-1.5-1.5). We stress that for the considered binary configurations broadened peaks are rather rare, in particular with EoSs compatible with the observation of a 1.97~$M_{\odot}$ pulsar. Hence, it is unlikely that the broadening of the dominant peak in the postmerger GW spectrum jeopardizes the determination of NS properties by the relations presented in this work, because wide peaks occur only in the seldom cases when the gravitational collapse of the DRO sets in during the phase of strong GW emission. Moreover, for the models examined here the peak frequency of broadened peaks still fulfills the discussed relations.

\begin{table} 
\caption{\label{tab:varbin}Models with varied total binary masses and an initial mass ratio of $q=1$. $M_1$ and $M_2$ refer to the masses of the binary components. In the second column the ideal-gas index for mimicking thermal effects is given. ``full'' denotes a consistent temperature treatment. The third column lists the dominant GW frequency $f_{\mathrm{peak}}$ in the postmerger phase, ``prompt BH'' indicates the prompt BH formation for the given model. The width of the peak is provided in the last column. EoSs which are excluded by the pulsar mass measurement of~\cite{2010Natur.467.1081D} are marked by an asterisk.}
 \begin{ruledtabular}
 \begin{tabular}{|l|l|l|l|}
EoS $M_1-M_2$   & $\Gamma_{\mathrm{th}}$ & $f_{\mathrm{peak}}$ [kHz] & FWHM [kHz] \\   \hline
Sly4 1.2-1.2    & 2                      & 2.94                      & 0.139      \\   \hline
APR 1.2-1.2     & 2                      & 3.15                      & 0.143      \\   \hline
FPS 1.2-1.2 {\bf*}   & 2                      & 3.41                      & 0.186      \\   \hline
BBB2 1.2-1.2 {\bf*}  & 2                      & 3.15                      & 0.183      \\   \hline
Glendnh3 1.2-1.2& 2                      & 2.12                      & 0.185      \\   \hline
eosAU 1.2-1.2   & 2                      & 3.51                      & 0.244      \\   \hline
eosC 1.2-1.2 {\bf*}  & 2                      & 2.81                      & 0.196      \\   \hline
eosL 1.2-1.2    & 2                      & 1.76                      & 0.124      \\   \hline
eosO 1.2-1.2    & 2                      & 2.48                      & 0.111      \\   \hline
eosUU 1.2-1.2   & 2                      & 3.21                      & 0.127      \\   \hline
eosWS 1.2-1.2 {\bf*} & 2                      & 3.35                      & 0.170      \\   \hline
SKA 1.2-1.2     & 2                      & 2.40                      & 0.129      \\   \hline
Shen 1.2-1.2    & full                   & 2.02                      & 0.143      \\   \hline
LS180 1.2-1.2 {\bf*} & full                   & 2.77                      & 0.195      \\   \hline
LS220 1.2-1.2   & full                   & 2.55                      & 0.166      \\   \hline
LS375 1.2-1.2   & full                   & 2.24                      & 0.130      \\   \hline
GS1 1.2-1.2     & full                   & 1.97                      & 0.129      \\   \hline
GS2 1.2-1.2     & full                   & 2.34                      & 0.126      \\   \hline
MIT60 1.2-1.2 {\bf*} & full                   & 3.01                      & 0.207      \\   \hline
BurgioNN 1.2-1.2& 2                      & 2.85                      & 0.157      \\   \hline
MIT40 1.2-1.2   & 1.34                   & 2.35                      & 0.161      \\   \hline
LS180 1.4-1.4 {\bf*} & full                   & 3.87\footnote{collapse after 7~ms} & 0.475           \\ \hline
Sly4 1.5-1.5    & 2                      & prompt BH                 &            \\  \hline
APR 1.5-1.5     & 2                      & prompt BH                 &            \\  \hline
FPS 1.5-1.5 {\bf*}   & 2                      & prompt BH                 &            \\  \hline
BBB2 1.5-1.5 {\bf*}  & 2                      & prompt BH                 &            \\  \hline
Glendnh3 1.5-1.5& 2                      & 2.64                      & 0.217      \\  \hline
eosAU 1.5-1.5   & 2                      & prompt BH                 &            \\  \hline
eosC 1.5-1.5 {\bf*}  & 2                      & prompt BH                 &            \\  \hline
eosL 1.5-1.5    & 2                      & 1.98                      & 0.137      \\  \hline
eosO 1.5-1.5    & 2                      & 2.86                      & 0.166      \\  \hline
eosUU 1.5-1.5   & 2                      & prompt BH                 &            \\  \hline
eosWS 1.5-1.5 {\bf*} & 2                      & prompt BH                 &            \\  \hline
SKA 1.5-1.5     & 2                      & 3.01                      & 0.200      \\  \hline
Shen 1.5-1.5    & full                   & 2.39                      & 0.160      \\  \hline
LS180 1.5-1.5 {\bf*} & full                   & prompt BH                 &            \\  \hline
LS220 1.5-1.5   & full                   & 3.50\footnote{collapse after 9~ms} & 0.326\\ \hline
LS375 1.5-1.5   & full                   & 2.57                      & 0.142      \\  \hline
GS1 1.5-1.5     & full                   & 2.27                      & 0.130      \\  \hline
GS2 1.5-1.5     & full                   & 2.83                      & 0.176      \\  \hline
MIT60 1.5-1.5 {\bf*} & full                   & prompt BH                 &            \\  \hline
BurgioNN 1.5-1.5& 2                      & prompt BH                 &            \\  \hline
MIT40 1.5-1.5   & 1.34                   & 2.83                      & 0.280      \\  \hline
\end{tabular}
 \end{ruledtabular}
\end{table}
\begin{table} 
\caption{\label{tab:varbinasym}Models with varied initial mass ratio and $M_{\mathrm{tot}}=2.7~M_{\odot}$. $M_1$ and $M_2$ refer to the masses of the binary components. In the second column the ideal-gas index for mimicking thermal effects is given. ``full'' denotes a consistent temperature treatment. The third column lists the dominant GW frequency $f_{\mathrm{peak}}$ in the postmerger phase. The width of the peak is provided in the last column. EoSs which are excluded by the pulsar mass measurement of~\cite{2010Natur.467.1081D} are marked by an asterisk.}
 \begin{ruledtabular}
 \begin{tabular}{|l|l|l|l|}
Shen 1.2-1.5    & full                   & 2.19             & 0.169    \\ \hline
LS180 1.2-1.5 {\bf*} & full                   & $\sim$3.4\footnote{collapse after 9~ms}        & $\sim$0.3         \\  \hline
LS220 1.2-1.5   & full                   & 2.80             & 0.260    \\ \hline
LS375 1.2-1.5   & full                   & 2.38             & 0.120    \\ \hline
GS1 1.2-1.5     & full                   & 2.03             & 0.190    \\ \hline
GS2 1.2-1.5     & full                   & 2.49           & 0.239    \\ \hline
MIT60 1.2-1.5 {\bf*} & full                   & $\sim$3.4\footnote{collapse after 2~ms}       & $\sim$0.6\\ \hline
\end{tabular}
 \end{ruledtabular}
\end{table}
\section{Conclusions} \label{sec:con}
In this work we have investigated an extended set of NS merger simulations considering in total 38 microphysical EoSs. This representative sample of high-density EoS candidates shows a large diversity in its properties, e.g. concerning the resulting stellar parameters, where the maximum NS mass varies from 1.79 to 3.00~$M_{\odot}$ with corresponding radii ranging from 8.65 to 14.30~km (Fig.~\ref{fig:tov}). We emphasize that no selection procedure is applied to the choice of the EoSs (except for requiring $M_{\mathrm{max}}\ge 1.97~M_{\odot}$ because of the pulsar mass measurement of~\cite{2010Natur.467.1081D}). Our models also include two EoSs describing absolutely stable strange quark matter.

For 34 EoSs of our sample the coalescence of two stars with 1.35~$M_{\odot}$ leads to the formation of a DRO, whose oscillations generate GW emission mainly in the kHz range. In particular, the fundamental quadrupolar fluid mode produces a pronounced peak between 1.84 and 3.73~kHz depending on the employed EoS. Except for one strange quark matter EoS, which is incompatible with observations, the peaks are very narrow with FWHMs of typically below 200~Hz. This pronounced feature of the postmerger GW emission is measurable up to a distance of 20 to 45~Mpc with a SNR of 2. Including also the late inspiral signal shifts the detection horizon up to 150~Mpc. Assuming a reasonably optimistic merger rate (1000 MWEG$^{-1}$~Myr$^{-1}$) this implies a detection rate for Advanced LIGO of about 1 event per year (up to 14 with the inclusion of the late inspiral phase).

The frequency of the peak in the GW spectrum turns out to be a direct measure for the size of the merger remnant. Furthermore, the peak frequency is found to anticorrelate very well with the radii of nonrotating NSs (Figs.~\ref{fig:frmax} to~\ref{fig:fr135}). The relation is in particular tight when choosing the radius of a 1.6~$M_{\odot}$ NS as a property characterizing an EoS (Fig.~\ref{fig:fr16}). For such a configuration the density regimes encountered in the massive DRO and in the less massive, nonrotating NS are similar. In this case the maximum offset from a broken linear fit describing the relation between the peak frequency and the radius is 60 meters for all fully microphysical EoSs of our sample which are in agreement with current knowledge of NSs (mass limit set by the observation of the 1.97~$M_{\odot}$ pulsar). Considering in addition that the peak frequency can be measured with an uncertainty of about 40~Hz, NS radii can be determined by a GW detection with an accuracy of about 144--200~meters depending on the peak frequency. This should be compared with the 1~km precision found for the radius determination from the GW inspiral signal with an optimal detection horizon of 100~Mpc~\cite{2009PhRvD..79l4033R}. Our method as presented in this paper therefore promises more accurate results but with a lower detection rate. In addition, the robustness concerning small uncertainties in the initial mass ratio of the binary has been shown (Figs.~\ref{fig:fr135mtot} to~\ref{fig:fr18mtot}). It is also worth mentioning that inspiral signals will provide NS radius information of the inspiraling stars, thus typically for NSs with 1.35~$M_{\odot}$, which are expected to be the most abundant cases in NS binaries. Using the postmerger signal of the same binary will constrain the radius of a more massive star ($R_{1.6}$ for $M_{\mathrm{tot}}=2.7~M_{\odot}$), which seems advantageous because it probes higher densities of the supranuclear EoS.

Examining the correlations between the dominant GW frequency of the postmerger signal and NS radii, we find that EoSs relying on piecewise polytropes show a larger scatter in these relations, which is probably a consequence of the involved simplifications. In any case it is assuring that also with these models the scatter in the observed relations is still small (maximum deviation of 357~meters).

Concerning the strange quark matter models we note that they approximately fulfill the relations established for the nucleonic EoSs, in particular, when including an additional nuclear crust of strange stars. However, the scenario of absolutely stable strange quark matter is observationally distinguishable from ordinary neutron stars, e.g. by strangelets in the cosmic ray flux. Whether strange stars actually obey their ``own'' peak frequency-radius relation should be explored if more evidence arose for the existence of these objects.

As an advantage compared to the analysis of the GW inspiral signal, the GW peak frequencies extracted from our simulations can also be set into relation to other NS and EoS properties. By this, exclusion regions for the maximum mass of nonrotating NSs are identified for certain values of the peak frequency. High frequencies tend to exclude high maximum NS masses and a single determination of the peak frequency may significantly constrain the maximum NS mass (Fig.~\ref{fig:fmmax}). This provides an alternative to estimating the maximum mass via the determination of the binary mass at the boundary between prompt and delayed collapse, which requires several detections and an a priori knowledge of NS radii~\cite{2005PhRvL..94t1101S,2011PhRvD..83l4008H}.

Moreover, we found that the dominant GW frequency of the DRO scales with the maximum central energy density of nonrotating NSs and with the pressure and the speed of sound at 1.85 times the nuclear saturation density (Figs.~\ref{fig:frhomax} to~\ref{fig:fvs185}). All these dependences can be used to constrain the high-density EoS, although in comparison to stellar radii as consequence of the bulk properties of the high-density EoS the relations with local EoS properties are less tight. Finally, we note that the widths of the peaks in the GW spectra show the general trend of being smaller in simulations with EoSs which yield higher maximum masses or less compact NSs (Figs.~\ref{fig:fwhmmmax} and~\ref{fig:fwhmr16}).

Variations with the binary parameters were investigated and scalings between the peak frequency and NS radii were determined for simulations with total binary masses of 2.4, 2.7 and 3.0~$M_{\odot}$. Bear in mind that the binary masses are measurable from the GW inspiral signal. It was noticed that for mergers with $M_{\mathrm{tot}}=2.4~M_{\odot}$ the radius of a nonrotating NS with 1.35~$M_{\odot}$ represents a better quantity to characterize EoSs in the sense that a tighter correlation is found (Fig.~\ref{fig:fr135mtot}). Correspondingly, for $M_{\mathrm{tot}}=3.0~M_{\odot}$ the radius of a 1.8~$M_{\odot}$ NS is preferable as EoS characterizing feature (Fig.~\ref{fig:fr18mtot}). These trends are understandable from the density regimes probed by the merger remnant.

The effects of the initial mass ratio of the binaries are only partially addressed in this paper, and a full exploration is left for the future. Moreover, further studies should involve an even larger set of microphysical EoSs especially models with a consistent description of thermal effects. Also, a treatment in full general relativity is desirable, preferentially with the inclusion of magnetic fields and neutrino effects, to consolidate the scalings presented in this work. Finally, a detailed examination of the possibility to measure the peak frequency is needed to establish the practical use of the described concept based on the discovered correlations. For instance, it should be explored to which extent the simultaneous observation of electromagnetic emission (a gamma-ray burst and/or optical transient) and the knowledge of the preceding inspiral signal can support the GW analysis. Moreover, the prospects of detector networks should be investigated.

\appendix
\section{Comparison to fully relativistic calculations}\label{app:grcomp}
In this appendix we explore the reliability of using the conformal flatness approximation  (see Sect.~\ref{sec:methods}) instead of solving the full Einstein equations. To this end we compare the GW emission characterized by the amplitude and the frequency extracted from our calculations to results which are available in the literature focusing on the postmerger phase as the main subject of this work. From earlier works it is known that the quadrupole formula underestimates the GW amplitude by about 30--40~\%, whereas the frequency is determined in agreement with more sophisticated GW extraction methods~\cite{2005PhRvD..71h4021S}. In order to judge on the quality of the conformal flatness approach we compare the peak frequencies obtained for the APR, SLy4, FPS, H3, H4 and Shen EoSs to the values found in fully relativistic simulations~\cite{2005PhRvD..71h4021S,2006PhRvD..73f4027S,2011PhRvD..83l4008H,2011PhRvL.107e1102S}. Note that in the calculations of~\cite{2011PhRvD..83l4008H} an ideal-gas index $\Gamma_{\mathrm{th}}=1.357$ was used, for which reason we perform additional runs with this value for a better comparison. The values of $f_{\mathrm{peak}}$ are listed in Tab.~\ref{tab:grcomp} together with the results of fully relativistic studies. If the peak frequencies are not explicitly given in the mentioned references, we extract them from the provided spectra.

Note that it is important to take into account differences in the exact implementation of the EoSs: In~\cite{2005PhRvD..71h4021S,2006PhRvD..73f4027S} analytical fits to the microphysical models have been used, while piecewise polytropes have been employed in~\cite{2011PhRvD..83l4008H}. For the FPS, Sly4 and APR models we utilize tables provided for these EoSs. The corresponding differences in the implementation of the EoSs lead to differences in the stellar structure and thus affect also the peak frequency. For instance, for a NS with 1.6~$M_{\odot}$ we find that the fit to the Sly4 EoS yields a stellar radius which is 160~m smaller than the one obtained with our EoS table. Therefore, in Tab.~\ref{tab:grcomp} we recalibrate our values for $f_{\mathrm{peak}}$ according to the differences found in the TOV solutions for 1.6~$M_{\odot}$ NSs for the various implementations of the EoSs. A difference in the radius is translated into a difference in the frequency by using the fit formula derived in Sect.~\ref{ssec:fit}. When such a correction is applied, the original value is also given in parentheses. One observes a very good quantitative agreement with maximum deviations in $f_{\mathrm{peak}}$ of at most a few per cent. Note that one cannot identify a systematic over- or underestimation of the peak frequency by our simulations based on the conformal flatness approximation.

The comparison of the H3 EoS deserves a further comment. In the fully relativistic simulation of~\cite{2011PhRvD..83l4008H} the DRO collapses about 5~ms after the merging, whereas in our calculation the collapse occurs only after~12~ms. In line with the reasoning in Sects.~\ref{sec:stellar} and~\ref{sec:binpara} that an early collapse broadens the peak structure, one observes a double-peak in the spectrum of~\cite{2011PhRvD..83l4008H} while we obtain a somewhat sharper peak (FWHM=255~Hz). The peak at the lower frequency in~\cite{2011PhRvD..83l4008H} coincides very well with $f_{\mathrm{peak}}$ found in our simulation, which is why we conclude that the broadening of the peak is connected to the onset of the collapse during the GW emission phase. We stress that the determination of the exact collapse timescale is sensitive to the numerical treatment (and e.g., corresponding numerical viscosity leading to angular momentum redistribution)~\cite{2010PhRvD..82h4043B,2011PhRvD..83l4008H}. Therefore differences in the particular case of the H3 EoS are not surprising. (Note that the H3 EoS is excluded because of its low $M_{\mathrm{max}}$. The latter is likely to be responsible for the early collapse.) A similar reasoning applies to the 1.35-1.35~$M_{\odot}$ merger with the Sly4 Eos, for which in~\cite{2005PhRvD..71h4021S} the gravitational collapse occurs within the first 10~ms after merging whereas in our simulation the DRO remains stable until the end of the simulation (14~ms after merging).

Notice also that the uncertainties in $f_{\mathrm{peak}}$ due to different resolutions and initial data can be up to a few 10~Hz (see~\cite{2010PhRvD..82h4043B} and Sect.~\ref{ssec:gwana}). Furthermore, we note that in the calculations of~\cite{2011PhRvL.107e1102S} an approximate treatment of neutrino effects has been taken into account. These may affect the exact value of $f_{\mathrm{peak}}$, but the detailed effect of neutrinos on $f_{\mathrm{peak}}$ is not clear from the results of~\cite{2011PhRvL.107e1102S}.

Our approach predicts a prompt collapse in close agreement with fully relativistic simulations, e.g. for the binary with two 1.5~$M_{\odot}$ NSs described by the APR EoS (see Tab.~\ref{tab:grcomp}). In~\cite{2005PhRvD..71h4021S} a prompt collapse for the models FPS1313 and Sly1414 has been reported, whereas we observe for these models the formation of a DRO. However, a slight increase in the total binary mass ($M_{\mathrm{tot}}=2.64~M_{\odot}$ for FPS, $M_{\mathrm{tot}}=2.9~M_{\odot}$ for Sly4) is already sufficient to lead to a prompt collapse. Note that the transition between prompt and delayed collapse is crucially determined by the EoS. Thus, the observed discrepancies in the collapse behavior may be explained by the slight differences in the implementations of the EoSs, where larger deviations are found for FPS and Sly4 in comparison to the APR EoS (compare original and recalibrated values of $f_{\mathrm{peak}}$ in Tab.~\ref{tab:grcomp}).

\begin{table}
\caption{\label{tab:grcomp}Models for a comparison between our work and fully relativistic calculations. The different setups are characterized by the employed EoS and the binary components $M_1$ and $M_2$. The second column provides the ideal-gas index used for an approximate treatment of thermal effects, ``full'' refers to a consistent description of temperature effects. In the third column peak frequencies obtained in our simulations are listed where we correct for slight differences in the EoS implementation (see text). Values in parentheses are the uncorrected peak frequencies. $f_{\mathrm{peak}}^{\mathrm{literature}}$ denotes the peak frequencies found in the literature with the corresponding reference given in the last column. Models marked by ``X'' lead to a delayed collapse within the first 10 milliseconds after merging. ``prompt BH'' indicates the prompt formation of a BH in the given model.}
 \begin{ruledtabular}
 \begin{tabular}{|l|l|l|l|l|}
EoS $M_1-M_2$ & $\Gamma_{\mathrm{th}}$ & $f_{\mathrm{peak}}$ [kHz] & $f_{\mathrm{peak}}^{\mathrm{literature}}$ [kHz] & reference \\ \hline
APR 1.35-1.35 & 2     & 3.45 (3.46) & 3.35          & \cite{2006PhRvD..73f4027S}\\ \hline
APR 1.35-1.35 & 1.357 & 3.69 (3.68) & 3.58          & \cite{2011PhRvD..83l4008H}\\ \hline
H3 1.35-1.35  & 1.357 & 2.67 & 2.69--3.00 X         & \cite{2011PhRvD..83l4008H}\\ \hline
H4 1.35-1.35  & 1.357 & 2.56 & 2.57                 & \cite{2011PhRvD..83l4008H}\\ \hline
APR 1.5-1.5   & 2     & prompt BH & prompt BH & \cite{2006PhRvD..73f4027S}\\ \hline
Sly4 1.2-1.2  & 2     & 3.01 (2.94) & 3.1           & \cite{2005PhRvD..71h4021S}\\ \hline
Sly4 1.35-1.35& 2     & 3.39 (3.32) & 3.6 X & \cite{2005PhRvD..71h4021S}\\ \hline
FPS 1.2-1.2   & 2     & 3.47 (3.41) & 3.5 X & \cite{2005PhRvD..71h4021S}\\ \hline
Shen 1.35-1.35& full  & 2.19 & 2.18          & \cite{2011PhRvL.107e1102S}\\ \hline
Shen 1.5-1.5  & full  & 2.39 & 2.30          & \cite{2011PhRvL.107e1102S}\\ \hline
Shen 1.6-1.6  & full  & 2.57 & 2.49 X        & \cite{2011PhRvL.107e1102S}\\ \hline
\end{tabular}
 \end{ruledtabular}
\end{table}

Despite the differences and the remaining uncertainties it is reassuring that codes with different degrees of sophistication yield quantitatively very similar results. Note that not only the gravity treatment (conformal flatness approximation versus full general relativity) differs between our implementation and the codes used in~\cite{2005PhRvD..71h4021S,2006PhRvD..73f4027S,2011PhRvD..83l4008H,2011PhRvL.107e1102S}, but also different hydrodynamics schemes are employed (SPH versus grid-based methods), and GWs are extracted by different methods. Without more detailed exploration it is therefore unclear to which extent on the one hand the different treatment of gravity and on the other hand the different hydrodynamics schemes, the different numerical implementations and the different resolutions cause the observed differences in the peak frequency. After all, the good match between fully relativistic calculations and the conformal flatness approximation is not really astonishing in the light of the findings of~\cite{2011MNRAS.416L...1T}, where agreement of both methods in determining the frequency of the fundamental quadrupole mode has been reported in the case of isolated NSs. Moreover, the conformal flatness approximation is known to yield exact results in spherical symmetry.

For assessing the reliability of the GW amplitudes extracted in our calculations we directly compare the amplitudes for 1.35-1.35~$M_{\odot}$ mergers with the APR, the H3, and the H4 EoSs with $\Gamma_{\mathrm{th}}=1.357$ to the results shown in~\cite{2011PhRvD..83l4008H}, where the GW signal has been computed by means of the complex Weyl scalar. This comparison confirms earlier findings that GW amplitudes are underestimated by the quadrupole formula by about 30--40 percent~\cite{2005PhRvD..71h4021S}. In the ringdown phase we find even slightly larger differences.

\section{Effect of the thermal ideal-gas component}\label{app:gamma}
\begin{table}
\caption{\label{tab:gammath}Simulations for different EoSs with varied ideal-gas index $\Gamma_{\mathrm{th}}$ for mimicking thermal effects. ``full'' in the second column provides results with a consistent temperature treatment. All models refer to a binary setup with two stars of 1.35~$M_{\odot}$. EoSs providing the full temperature dependence (class i) are restricted to zero temperature and neutrinoless beta-equilibrium. $f_{\mathrm{peak}}$ denotes the dominant GW frequency in the postmerger phase. Note that models which lead to a prompt collapse with $\Gamma_{\mathrm{th}}=2$ are not simulated and not listed here. ``prompt BH'' indicates the prompt formation of a BH in the given model.}
 \begin{ruledtabular}
 \begin{tabular}{|l|l|l|}
EoS      & $\Gamma_{\mathrm{th}}$ & $f_{\mathrm{peak}}$~[kHz]\\ \hline
GS1      & full                   & 2.10                     \\ \hline
GS1      & 2                      & 2.09                     \\ \hline
GS1      & 1.5                    & 2.18                     \\ \hline
LS375    & full                   & 2.40                     \\ \hline
LS375    & 2                      & 2.34                     \\ \hline
LS375    & 1.5                    & 2.41                     \\ \hline
Shen     & full                   & 2.19                     \\ \hline
Shen     & 2                      & 2.16                     \\ \hline
Shen     & 1.5                    & 2.26                     \\ \hline
GS2      & full                   & 2.53                     \\ \hline
GS2      & 2                      & 2.52                     \\ \hline
GS2      & 1.5                    & 2.65                     \\ \hline
LS220    & full                   & 2.89                     \\ \hline
LS220    & 2                      & 2.73                     \\ \hline
LS220    & 1.5                    & 2.92                     \\ \hline
LS180    & full                   & 3.26                     \\ \hline
LS180    & 2                      & 3.08                     \\ \hline
LS180    & 1.5                    & 3.34                     \\ \hline
eosL     & 2                      & 1.84                     \\ \hline
eosL     & 1.5                    & 1.95                     \\ \hline
eosO     & 2                      & 2.66                     \\ \hline
eosO     & 1.5                    & 2.77                     \\ \hline
eosUU    & 2                      & 3.50                     \\ \hline
eosUU    & 1.5                    & 3.70                     \\ \hline
SKA      & 2                      & 2.64                     \\ \hline
SKA      & 1.5                    & 2.81                     \\ \hline
APR      & 2                      & 3.46                     \\ \hline
APR      & 1.5                    & 3.63                     \\ \hline
BurgioNN & 2                      & 3.20                     \\ \hline
BurgioNN & 1.5                    & 3.33                     \\ \hline
Sly4     & 2                      & 3.32                     \\ \hline
Sly4     & 1.5                    & 3.53                     \\ \hline
Glendnh3 & 2                      & 2.33                     \\ \hline
Glendnh3 & 1.5                    & 2.51                     \\ \hline
BBB2     & 2                      & 3.73                     \\ \hline
BBB2     & 1.5                    & prompt BH                \\ \hline
eosC     & 2                      & 3.33                     \\ \hline
eosC     & 1.5                    & 3.83                     \\ \hline
\end{tabular}
 \end{ruledtabular}
\end{table}
As mentioned in Sects.~\ref{sec:methods} and~\ref{sec:eos} not all EoSs include temperature effects consistently, for which reason these models are supplemented with an ideal-gas component to mimic thermal effects. This description requires the specification of the corresponding ideal-gas index $\Gamma_{\mathrm{th}}$. A choice in the range from 1.5 to 2 seems reasonable~\cite{2010PhRvD..82h4043B}. Throughout this paper $\Gamma_{\mathrm{th}}$ is fixed to 2, which can be considered as suitable approximation but may be slightly too high for EoSs yielding compact NSs~\cite{2010PhRvD..82h4043B}.

\begin{figure}
\includegraphics[width=8.9cm]{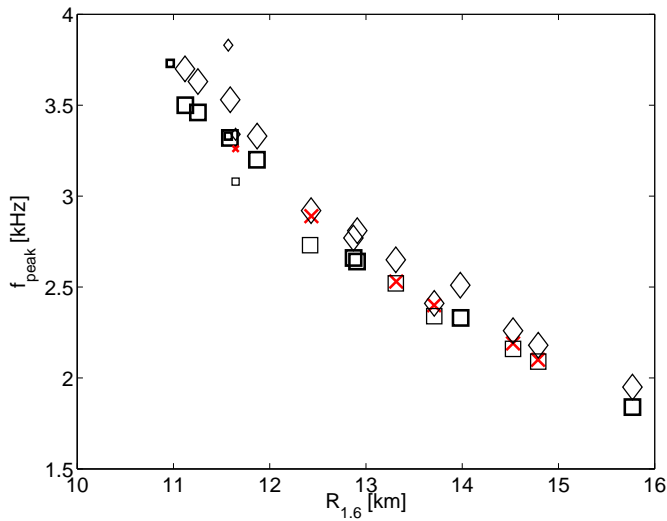}
\caption{\label{fig:gammath}Peak frequency of the postmerger GW emission versus the radius of a nonrotating NS with 1.6~$M_{\odot}$ for fully microphysical EoSs (classes (i) and (ii)). Red symbols show the results for simulations including thermal effects consistently. Squares mark the frequencies for computations with an approximate treatment of thermal effects using an ideal-gas index of $\Gamma_{\mathrm{th}}=2$. Diamonds correspond to runs with $\Gamma_{\mathrm{th}}=1.5$. Small symbols indicate EoSs which are excluded by the observation of the 1.97~$M_{\odot}$ pulsar.}
\end{figure}

To illustrate the uncertainties connected with the approximate treatment of thermal effects, Fig.~\ref{fig:gammath} shows the results of additional simulations of 1.35-1.35~$M_{\odot}$ mergers with $\Gamma_{\mathrm{th}}=1.5$ (see Tab.~\ref{tab:gammath}). The plot displays the peak frequency as a function of the radius of a nonrotating NS with 1.6~$M_{\odot}$ for all purely microphysical EoSs. For all temperature dependent EoSs of our sample (Shen, LS180, LS220, LS375, GS1 and GS2) we include additional calculations, where the EoSs are restricted to the zero-temperature slice and are supplemented with the ideal-gas ansatz for thermal effects. For these EoSs we construct barotropic relations by assuming in addition neutrinoless beta-equilibrium and run simulations with $\Gamma_{\mathrm{th}}=2$ and $\Gamma_{\mathrm{th}}=1.5$. Peak frequencies obtained from these simulations using $\Gamma_{\mathrm{th}}=2$ are shown with squares, while diamonds denote the results of $\Gamma_{\mathrm{th}}=1.5$ runs. Red crosses indicate the peak frequency when the full temperature dependence of the EoS is considered. Small symbols correspond to EoSs which are excluded by the 1.97~$M_{\odot}$ pulsar. For these models based on class-(i) EoSs the $\Gamma_{\mathrm{th}}=1.5$ results appear at higher frequency compared to the fully consistent model, and the $\Gamma_{\mathrm{th}}=2$ data points are located at lower frequencies. An ideal-gas index $\Gamma_{\mathrm{th}}=2$ appears to be a good choice except for the LS EoSs, for which a slightly lower $\Gamma_{\mathrm{th}}$ seems more suitable. Note that the LS180 EoS is excluded.
\begin{figure}
\includegraphics[width=8.9cm]{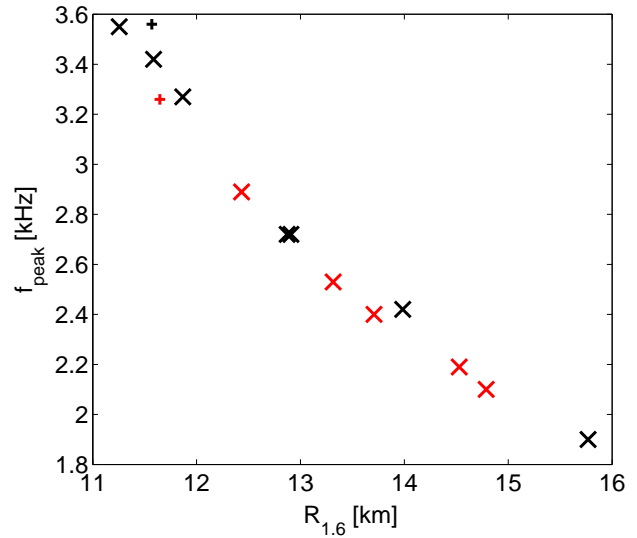}
\caption{\label{fig:fr16gth175}Peak frequency of the postmerger GW emission versus the radius of a nonrotating NS with 1.6~$M_{\odot}$ for fully microphysical EoSs (classes (i) and (ii)). Symbols have the same meaning as in Fig.~\ref{fig:frmax}. The data points for EoSs of class (ii) (black symbols) are an average of the results for calculations with $\Gamma_{\mathrm{th}}=1.5$ and $\Gamma_{\mathrm{th}}=2$.}
\end{figure}

In addition, we simulate for all EoSs of class (ii) (fully microphysical EoSs at zero temperature) the symmetric merger with $M_{\mathrm{tot}}=2.7~M_{\odot}$ using $\Gamma_{\mathrm{th}}=1.5$, which should be considered a firm lower bound for $\Gamma_{\mathrm{th}}$~\cite{2010PhRvD..82h4043B}. The corresponding peak frequencies (diamonds in Fig.~\ref{fig:gammath}) occur at higher values, but the differences to the $\Gamma_{\mathrm{th}}=2$ runs (squares) seem to increase slightly with smaller $R_{1.6}$. Note that for the BBB2 EoS a prompt collapse to a BH occurs for $\Gamma_{\mathrm{th}}=1.5$. The data points at $R_{1.6}=13.98$~km belong to the Glendnh3 EoS whose behavior one should take with a grain of salt because of its qualitatively different mass-radius relation as argued in Sect.~\ref{sec:stellar}. For the accepted models the largest difference between peak frequencies extracted from $\Gamma_{\mathrm{th}}=2$ and $\Gamma_{\mathrm{th}}=1.5$ simulations is 210~Hz for the Sly4 EoS. However, the $\Gamma_{\mathrm{th}}=1.5$ data points should be considered as safe upper limits for the peak frequencies. Because of the lack of better alternatives we favor the use of $\Gamma_{\mathrm{th}}=2$ in this work until more EoSs including a consistent description of temperature effects become available.

In Fig.~\ref{fig:fr16gth175} we display the relation between $f_{\mathrm{peak}}$ and $R_{1.6}$ for the case that the peak frequencies for class (ii) EoSs (black symbols) are averaged between the runs with $\Gamma_{\mathrm{th}}=1.5$ and $\Gamma_{\mathrm{th}}=2$ approximately estimating the outcome of a choice of $\Gamma_{\mathrm{th}}=1.75$. Also in this case a very tight correlation is found, but note that in comparison to Fig.~\ref{fig:fr16} the EoSs excluded by the pulsar mass measurement of~\cite{2010Natur.467.1081D} scatter more strongly from the accepted models. Ignoring the result with the Glendnh3 EoS one finds that the relation for the accepted models obeys approximately a power law with a maximum deviation from this functional relation of 126~meters. The scatter may be reduced by choosing a slightly different radius of nonrotating NSs for characterizing the EoSs.

Moreover, we note that in comparison to nucleonic EoSs the exact choice of the thermal ideal-gas index $\Gamma_{\mathrm{th}}$ for strange quark matter is less relevant. Neglecting thermal effects for example in a merger of two stars with 1.35~$M_{\odot}$ described by the MIT60 EoS, decreases the peak frequency by only 80~Hz~\cite{PhysRevD.81.024012}.

\begin{acknowledgments}
We thank F.~Burgio, M.~Oertel, and N.~Stergioulas for helpful discussions.

We thank S.~Hild for providing the sensitivity curve of the Einstein Telescope and D.~Shoemaker for the Advanced LIGO sensitivity curve.

This work was supported by the Deutsche Forschungsgemeinschaft through Sonderforschungsbereich Transregio 7 ``Gravitational Wave Astronomy", Sonderforschungsbereich Transregio 27 ``Neutrinos and Beyond", and the Cluster of Excellence EXC 153 ``Origin and Structure of the Universe", by CompStar, a research networking programme of the European Science Foundation, and in part by the NSF under Grant PHY--1002478,
the UNEDF SciDAC Collaboration under DOE Grant DE-FC02-07ER41457,
by the Helmholtz Alliance Program of the Helmholtz Association,
contract HA216/EMMI ``Extremes of Density and Temperature:
Cosmic Matter in the Laboratory'', and by the Deutsche Forschungsgemeinschaft through Sonderforschungsbereich 634. The computations were performed at the Rechenzentrum Garching of the Max-Planck-Gesellschaft, the Leibniz-Rechenzentrum Garching and the Max Planck Institute for Astrophysics.
\end{acknowledgments}

\bibliography{references}

\begin{thebibliography}{88}
\expandafter\ifx\csname natexlab\endcsname\relax\def\natexlab#1{#1}\fi
\expandafter\ifx\csname bibnamefont\endcsname\relax
  \def\bibnamefont#1{#1}\fi
\expandafter\ifx\csname bibfnamefont\endcsname\relax
  \def\bibfnamefont#1{#1}\fi
\expandafter\ifx\csname citenamefont\endcsname\relax
  \def\citenamefont#1{#1}\fi
\expandafter\ifx\csname url\endcsname\relax
  \def\url#1{\texttt{#1}}\fi
\expandafter\ifx\csname urlprefix\endcsname\relax\def\urlprefix{URL }\fi
\providecommand{\bibinfo}[2]{#2}
\providecommand{\eprint}[2][]{\url{#2}}

\bibitem[{\citenamefont{{Lattimer} and {Prakash}}(2007)}]{2007PhR...442..109L}
\bibinfo{author}{\bibfnamefont{J.~M.} \bibnamefont{{Lattimer}}}
  \bibnamefont{and}
  \bibinfo{author}{\bibfnamefont{M.}~\bibnamefont{{Prakash}}},
  \bibinfo{journal}{Phys. Rep.} \textbf{\bibinfo{volume}{442}},
  \bibinfo{pages}{109} (\bibinfo{year}{2007}).

\bibitem[{\citenamefont{{Haensel} et~al.}(2007)\citenamefont{{Haensel},
  {Potekhin}, and {Yakovlev}}}]{2007ASSL..326.....H}
\bibinfo{author}{\bibfnamefont{P.}~\bibnamefont{{Haensel}}},
  \bibinfo{author}{\bibfnamefont{A.~Y.} \bibnamefont{{Potekhin}}},
  \bibnamefont{and} \bibinfo{author}{\bibfnamefont{D.~G.}
  \bibnamefont{{Yakovlev}}}, \emph{\bibinfo{title}{{Neutron Stars 1}}}
  (\bibinfo{publisher}{Springer-Verlag, New York}, \bibinfo{year}{2007}).

\bibitem[{\citenamefont{{Glendenning}}(1996)}]{1996csnp.book.....G}
\bibinfo{author}{\bibfnamefont{N.}~\bibnamefont{{Glendenning}}},
  \emph{\bibinfo{title}{{Compact Stars}}} (\bibinfo{publisher}{Springer-Verlag,
  New York}, \bibinfo{year}{1996}).

\bibitem[{\citenamefont{{Horowitz}}(2011)}]{2011IJMPE..20.2077H}
\bibinfo{author}{\bibfnamefont{C.~J.} \bibnamefont{{Horowitz}}},
  \bibinfo{journal}{Int. J. Mod. Phys. E} \textbf{\bibinfo{volume}{20}},
  \bibinfo{pages}{2077} (\bibinfo{year}{2011}).

\bibitem[{\citenamefont{{Camenzind}}(2007)}]{2007coaw.book.....C}
\bibinfo{author}{\bibfnamefont{M.}~\bibnamefont{{Camenzind}}},
  \emph{\bibinfo{title}{{Compact objects in astrophysics : white dwarfs,
  neutron stars, and black holes}}} (\bibinfo{publisher}{Springer-Verlag,
  Berlin Heidelberg}, \bibinfo{year}{2007}).

\bibitem[{\citenamefont{{Thorsett} and
  {Chakrabarty}}(1999)}]{1999ApJ...512..288T}
\bibinfo{author}{\bibfnamefont{S.~E.} \bibnamefont{{Thorsett}}}
  \bibnamefont{and}
  \bibinfo{author}{\bibfnamefont{D.}~\bibnamefont{{Chakrabarty}}},
  \bibinfo{journal}{Astrophys. J.} \textbf{\bibinfo{volume}{512}},
  \bibinfo{pages}{288} (\bibinfo{year}{1999}).

\bibitem[{\citenamefont{{Zhang} et~al.}(2011)\citenamefont{{Zhang}, {Wang},
  {Zhao}, {Yin}, {Song}, {Menezes}, {Wickramasinghe}, {Ferrario}, and
  {Chardonnet}}}]{2011A&A...527A..83Z}
\bibinfo{author}{\bibfnamefont{C.~M.} \bibnamefont{{Zhang}}},
  \bibinfo{author}{\bibfnamefont{J.}~\bibnamefont{{Wang}}},
  \bibinfo{author}{\bibfnamefont{Y.~H.} \bibnamefont{{Zhao}}},
  \bibinfo{author}{\bibfnamefont{H.~X.} \bibnamefont{{Yin}}},
  \bibinfo{author}{\bibfnamefont{L.~M.} \bibnamefont{{Song}}},
  \bibinfo{author}{\bibfnamefont{D.~P.} \bibnamefont{{Menezes}}},
  \bibinfo{author}{\bibfnamefont{D.~T.} \bibnamefont{{Wickramasinghe}}},
  \bibinfo{author}{\bibfnamefont{L.}~\bibnamefont{{Ferrario}}},
  \bibnamefont{and}
  \bibinfo{author}{\bibfnamefont{P.}~\bibnamefont{{Chardonnet}}},
  \bibinfo{journal}{Astron. Astrophys.} \textbf{\bibinfo{volume}{527}},
  \bibinfo{pages}{A83} (\bibinfo{year}{2011}).

\bibitem[{\citenamefont{{Steiner} et~al.}(2010)\citenamefont{{Steiner},
  {Lattimer}, and {Brown}}}]{2010ApJ...722...33S}
\bibinfo{author}{\bibfnamefont{A.~W.} \bibnamefont{{Steiner}}},
  \bibinfo{author}{\bibfnamefont{J.~M.} \bibnamefont{{Lattimer}}},
  \bibnamefont{and} \bibinfo{author}{\bibfnamefont{E.~F.}
  \bibnamefont{{Brown}}}, \bibinfo{journal}{\apj}
  \textbf{\bibinfo{volume}{722}}, \bibinfo{pages}{33} (\bibinfo{year}{2010}).

\bibitem[{\citenamefont{{Galloway} and {Lampe}}(2012)}]{2012arXiv1201.1680G}
\bibinfo{author}{\bibfnamefont{D.~K.} \bibnamefont{{Galloway}}}
  \bibnamefont{and} \bibinfo{author}{\bibfnamefont{N.}~\bibnamefont{{Lampe}}},
  \bibinfo{journal}{Astrophys. J.} \textbf{\bibinfo{volume}{747}},
  \bibinfo{eid}{75} (\bibinfo{year}{2012}).

\bibitem[{\citenamefont{{Demorest} et~al.}(2010)\citenamefont{{Demorest},
  {Pennucci}, {Ransom}, {Roberts}, and {Hessels}}}]{2010Natur.467.1081D}
\bibinfo{author}{\bibfnamefont{P.~B.} \bibnamefont{{Demorest}}},
  \bibinfo{author}{\bibfnamefont{T.}~\bibnamefont{{Pennucci}}},
  \bibinfo{author}{\bibfnamefont{S.~M.} \bibnamefont{{Ransom}}},
  \bibinfo{author}{\bibfnamefont{M.~S.~E.} \bibnamefont{{Roberts}}},
  \bibnamefont{and} \bibinfo{author}{\bibfnamefont{J.~W.~T.}
  \bibnamefont{{Hessels}}}, \bibinfo{journal}{\nat}
  \textbf{\bibinfo{volume}{467}}, \bibinfo{pages}{1081} (\bibinfo{year}{2010}).

\bibitem[{\citenamefont{{Belczynski} et~al.}(2008)\citenamefont{{Belczynski},
  {O'Shaughnessy}, {Kalogera}, {Rasio}, {Taam}, and
  {Bulik}}}]{2008ApJ...680L.129B}
\bibinfo{author}{\bibfnamefont{K.}~\bibnamefont{{Belczynski}}},
  \bibinfo{author}{\bibfnamefont{R.}~\bibnamefont{{O'Shaughnessy}}},
  \bibinfo{author}{\bibfnamefont{V.}~\bibnamefont{{Kalogera}}},
  \bibinfo{author}{\bibfnamefont{F.}~\bibnamefont{{Rasio}}},
  \bibinfo{author}{\bibfnamefont{R.~E.} \bibnamefont{{Taam}}},
  \bibnamefont{and} \bibinfo{author}{\bibfnamefont{T.}~\bibnamefont{{Bulik}}},
  \bibinfo{journal}{Astrophys. J. Lett.} \textbf{\bibinfo{volume}{680}},
  \bibinfo{pages}{L129} (\bibinfo{year}{2008}).

\bibitem[{\citenamefont{{Kalogera} et~al.}(2004)\citenamefont{{Kalogera},
  {Kim}, {Lorimer}, {Burgay}, {D'Amico}, {Possenti}, {Manchester}, {Lyne},
  {Joshi}, {McLaughlin} et~al.}}]{2004ApJ...601L.179K}
\bibinfo{author}{\bibfnamefont{V.}~\bibnamefont{{Kalogera}}},
  \bibinfo{author}{\bibfnamefont{C.}~\bibnamefont{{Kim}}},
  \bibinfo{author}{\bibfnamefont{D.~R.} \bibnamefont{{Lorimer}}},
  \bibinfo{author}{\bibfnamefont{M.}~\bibnamefont{{Burgay}}},
  \bibinfo{author}{\bibfnamefont{N.}~\bibnamefont{{D'Amico}}},
  \bibinfo{author}{\bibfnamefont{A.}~\bibnamefont{{Possenti}}},
  \bibinfo{author}{\bibfnamefont{R.~N.} \bibnamefont{{Manchester}}},
  \bibinfo{author}{\bibfnamefont{A.~G.} \bibnamefont{{Lyne}}},
  \bibinfo{author}{\bibfnamefont{B.~C.} \bibnamefont{{Joshi}}},
  \bibinfo{author}{\bibfnamefont{M.~A.} \bibnamefont{{McLaughlin}}},
  \bibnamefont{et~al.}, \bibinfo{journal}{Astrophys. J. Lett.}
  \textbf{\bibinfo{volume}{601}}, \bibinfo{pages}{L179} (\bibinfo{year}{2004}).

\bibitem[{\citenamefont{{Faber}}(2009)}]{2009CQGra..26k4004F}
\bibinfo{author}{\bibfnamefont{J.}~\bibnamefont{{Faber}}},
  \bibinfo{journal}{Class. Quantum Grav.} \textbf{\bibinfo{volume}{26}},
  \bibinfo{pages}{114004} (\bibinfo{year}{2009}).

\bibitem[{\citenamefont{{Duez}}(2010)}]{2010CQGra..27k4002D}
\bibinfo{author}{\bibfnamefont{M.~D.} \bibnamefont{{Duez}}},
  \bibinfo{journal}{Class. Quantum Grav.} \textbf{\bibinfo{volume}{27}},
  \bibinfo{pages}{114002} (\bibinfo{year}{2010}).

\bibitem[{\citenamefont{{Rosswog}}(2010)}]{2010arXiv1012.0912R}
\bibinfo{author}{\bibfnamefont{S.}~\bibnamefont{{Rosswog}}},
  \bibinfo{journal}{ArXiv e-prints}  (\bibinfo{year}{2010}),
  \eprint{1012.0912}.

\bibitem[{\citenamefont{{Nakar}}(2007)}]{2007PhR...442..166N}
\bibinfo{author}{\bibfnamefont{E.}~\bibnamefont{{Nakar}}},
  \bibinfo{journal}{Phys. Rep.} \textbf{\bibinfo{volume}{442}},
  \bibinfo{pages}{166} (\bibinfo{year}{2007}).

\bibitem[{\citenamefont{{Oechslin} and {Janka}}(2006)}]{2006MNRAS.368.1489O}
\bibinfo{author}{\bibfnamefont{R.}~\bibnamefont{{Oechslin}}} \bibnamefont{and}
  \bibinfo{author}{\bibfnamefont{H.-T.} \bibnamefont{{Janka}}},
  \bibinfo{journal}{Mon. Not. R. Astron. Soc.} \textbf{\bibinfo{volume}{368}},
  \bibinfo{pages}{1489} (\bibinfo{year}{2006}).

\bibitem[{\citenamefont{{Shibata} and {Taniguchi}}(2006)}]{2006PhRvD..73f4027S}
\bibinfo{author}{\bibfnamefont{M.}~\bibnamefont{{Shibata}}} \bibnamefont{and}
  \bibinfo{author}{\bibfnamefont{K.}~\bibnamefont{{Taniguchi}}},
  \bibinfo{journal}{\prd} \textbf{\bibinfo{volume}{73}},
  \bibinfo{pages}{064027} (\bibinfo{year}{2006}).

\bibitem[{\citenamefont{{Bauswein} et~al.}(2010)\citenamefont{{Bauswein},
  {Janka}, and {Oechslin}}}]{2010PhRvD..82h4043B}
\bibinfo{author}{\bibfnamefont{A.}~\bibnamefont{{Bauswein}}},
  \bibinfo{author}{\bibfnamefont{H.-T.} \bibnamefont{{Janka}}},
  \bibnamefont{and}
  \bibinfo{author}{\bibfnamefont{R.}~\bibnamefont{{Oechslin}}},
  \bibinfo{journal}{\prd} \textbf{\bibinfo{volume}{82}},
  \bibinfo{pages}{084043} (\bibinfo{year}{2010}).

\bibitem[{\citenamefont{{Oechslin} et~al.}(2007)\citenamefont{{Oechslin},
  {Janka}, and {Marek}}}]{2007A&A...467..395O}
\bibinfo{author}{\bibfnamefont{R.}~\bibnamefont{{Oechslin}}},
  \bibinfo{author}{\bibfnamefont{H.-T.} \bibnamefont{{Janka}}},
  \bibnamefont{and} \bibinfo{author}{\bibfnamefont{A.}~\bibnamefont{{Marek}}},
  \bibinfo{journal}{Astron. Astrophys.} \textbf{\bibinfo{volume}{467}},
  \bibinfo{pages}{395} (\bibinfo{year}{2007}).

\bibitem[{\citenamefont{{Goriely} et~al.}(2011)\citenamefont{{Goriely},
  {Bauswein}, and {Janka}}}]{2011ApJ...738L..32G}
\bibinfo{author}{\bibfnamefont{S.}~\bibnamefont{{Goriely}}},
  \bibinfo{author}{\bibfnamefont{A.}~\bibnamefont{{Bauswein}}},
  \bibnamefont{and} \bibinfo{author}{\bibfnamefont{H.-T.}
  \bibnamefont{{Janka}}}, \bibinfo{journal}{Astrophys. J. Lett.}
  \textbf{\bibinfo{volume}{738}}, \bibinfo{eid}{L32} (\bibinfo{year}{2011}).

\bibitem[{\citenamefont{{Li} and {Paczy{\'n}ski}}(1998)}]{1998ApJ...507L..59L}
\bibinfo{author}{\bibfnamefont{L.-X.} \bibnamefont{{Li}}} \bibnamefont{and}
  \bibinfo{author}{\bibfnamefont{B.}~\bibnamefont{{Paczy{\'n}ski}}},
  \bibinfo{journal}{Astrophys. J. Lett.} \textbf{\bibinfo{volume}{507}},
  \bibinfo{pages}{L59} (\bibinfo{year}{1998}).

\bibitem[{\citenamefont{{Kulkarni}}(2005)}]{2005astro.ph.10256K}
\bibinfo{author}{\bibfnamefont{S.~R.} \bibnamefont{{Kulkarni}}},
  \bibinfo{journal}{ArXiv Astrophysics e-prints}  (\bibinfo{year}{2005}),
  \eprint{arXiv:astro-ph/0510256}.

\bibitem[{\citenamefont{{Metzger} et~al.}(2010)\citenamefont{{Metzger},
  {Mart{\'{\i}}nez-Pinedo}, {Darbha}, {Quataert}, {Arcones}, {Kasen}, {Thomas},
  {Nugent}, {Panov}, and {Zinner}}}]{2010MNRAS.406.2650M}
\bibinfo{author}{\bibfnamefont{B.~D.} \bibnamefont{{Metzger}}},
  \bibinfo{author}{\bibfnamefont{G.}~\bibnamefont{{Mart{\'{\i}}nez-Pinedo}}},
  \bibinfo{author}{\bibfnamefont{S.}~\bibnamefont{{Darbha}}},
  \bibinfo{author}{\bibfnamefont{E.}~\bibnamefont{{Quataert}}},
  \bibinfo{author}{\bibfnamefont{A.}~\bibnamefont{{Arcones}}},
  \bibinfo{author}{\bibfnamefont{D.}~\bibnamefont{{Kasen}}},
  \bibinfo{author}{\bibfnamefont{R.}~\bibnamefont{{Thomas}}},
  \bibinfo{author}{\bibfnamefont{P.}~\bibnamefont{{Nugent}}},
  \bibinfo{author}{\bibfnamefont{I.~V.} \bibnamefont{{Panov}}},
  \bibnamefont{and} \bibinfo{author}{\bibfnamefont{N.~T.}
  \bibnamefont{{Zinner}}}, \bibinfo{journal}{Mon. Not. R. Astron. Soc.}
  \textbf{\bibinfo{volume}{406}}, \bibinfo{pages}{2650} (\bibinfo{year}{2010}).

\bibitem[{\citenamefont{{Roberts} et~al.}(2011)\citenamefont{{Roberts},
  {Kasen}, {Lee}, and {Ramirez-Ruiz}}}]{2011ApJ...736L..21R}
\bibinfo{author}{\bibfnamefont{L.~F.} \bibnamefont{{Roberts}}},
  \bibinfo{author}{\bibfnamefont{D.}~\bibnamefont{{Kasen}}},
  \bibinfo{author}{\bibfnamefont{W.~H.} \bibnamefont{{Lee}}}, \bibnamefont{and}
  \bibinfo{author}{\bibfnamefont{E.}~\bibnamefont{{Ramirez-Ruiz}}},
  \bibinfo{journal}{Astrophys. J. Lett.} \textbf{\bibinfo{volume}{736}},
  \bibinfo{eid}{L21} (\bibinfo{year}{2011}).

\bibitem[{\citenamefont{{Harry} and {the LIGO Scientific
  Collaboration}}(2010)}]{2010CQGra..27h4006H}
\bibinfo{author}{\bibfnamefont{G.~M.} \bibnamefont{{Harry}}} \bibnamefont{and}
  \bibinfo{author}{\bibnamefont{{the LIGO Scientific Collaboration}}},
  \bibinfo{journal}{Class. Quantum Grav.} \textbf{\bibinfo{volume}{27}},
  \bibinfo{pages}{084006} (\bibinfo{year}{2010}).

\bibitem[{\citenamefont{Acernese et~al.}(2006)}]{Acernese:2006bj}
\bibinfo{author}{\bibfnamefont{F.}~\bibnamefont{Acernese}}
  \bibnamefont{et~al.}, \bibinfo{journal}{Class. Quantum Grav.}
  \textbf{\bibinfo{volume}{23}}, \bibinfo{pages}{S635} (\bibinfo{year}{2006}).

\bibitem[{\citenamefont{{Abadie} et~al.}(2010)}]{2010CQGra..27q3001A}
\bibinfo{author}{\bibfnamefont{J.}~\bibnamefont{{Abadie}}}
  \bibnamefont{et~al.}, \bibinfo{journal}{Class. Quantum Grav.}
  \textbf{\bibinfo{volume}{27}}, \bibinfo{pages}{173001}
  (\bibinfo{year}{2010}).

\bibitem[{\citenamefont{{Cutler} and {Flanagan}}(1994)}]{1994PhRvD..49.2658C}
\bibinfo{author}{\bibfnamefont{C.}~\bibnamefont{{Cutler}}} \bibnamefont{and}
  \bibinfo{author}{\bibfnamefont{{\'E}.~E.} \bibnamefont{{Flanagan}}},
  \bibinfo{journal}{\prd} \textbf{\bibinfo{volume}{49}}, \bibinfo{pages}{2658}
  (\bibinfo{year}{1994}).

\bibitem[{\citenamefont{{Faber} et~al.}(2002)\citenamefont{{Faber},
  {Grandcl{\'e}ment}, {Rasio}, and {Taniguchi}}}]{2002PhRvL..89w1102F}
\bibinfo{author}{\bibfnamefont{J.~A.} \bibnamefont{{Faber}}},
  \bibinfo{author}{\bibfnamefont{P.}~\bibnamefont{{Grandcl{\'e}ment}}},
  \bibinfo{author}{\bibfnamefont{F.~A.} \bibnamefont{{Rasio}}},
  \bibnamefont{and}
  \bibinfo{author}{\bibfnamefont{K.}~\bibnamefont{{Taniguchi}}},
  \bibinfo{journal}{\prl} \textbf{\bibinfo{volume}{89}},
  \bibinfo{pages}{231102} (\bibinfo{year}{2002}).

\bibitem[{\citenamefont{{Read} et~al.}(2009{\natexlab{a}})\citenamefont{{Read},
  {Markakis}, {Shibata}, {Ury{\= u}}, {Creighton}, and
  {Friedman}}}]{2009PhRvD..79l4033R}
\bibinfo{author}{\bibfnamefont{J.~S.} \bibnamefont{{Read}}},
  \bibinfo{author}{\bibfnamefont{C.}~\bibnamefont{{Markakis}}},
  \bibinfo{author}{\bibfnamefont{M.}~\bibnamefont{{Shibata}}},
  \bibinfo{author}{\bibfnamefont{K.}~\bibnamefont{{Ury{\= u}}}},
  \bibinfo{author}{\bibfnamefont{J.~D.~E.} \bibnamefont{{Creighton}}},
  \bibnamefont{and} \bibinfo{author}{\bibfnamefont{J.~L.}
  \bibnamefont{{Friedman}}}, \bibinfo{journal}{\prd}
  \textbf{\bibinfo{volume}{79}}, \bibinfo{pages}{124033}
  (\bibinfo{year}{2009}{\natexlab{a}}).

\bibitem[{\citenamefont{{Baiotti} et~al.}(2010)\citenamefont{{Baiotti},
  {Damour}, {Giacomazzo}, {Nagar}, and {Rezzolla}}}]{2010PhRvL.105z1101B}
\bibinfo{author}{\bibfnamefont{L.}~\bibnamefont{{Baiotti}}},
  \bibinfo{author}{\bibfnamefont{T.}~\bibnamefont{{Damour}}},
  \bibinfo{author}{\bibfnamefont{B.}~\bibnamefont{{Giacomazzo}}},
  \bibinfo{author}{\bibfnamefont{A.}~\bibnamefont{{Nagar}}}, \bibnamefont{and}
  \bibinfo{author}{\bibfnamefont{L.}~\bibnamefont{{Rezzolla}}},
  \bibinfo{journal}{\prl} \textbf{\bibinfo{volume}{105}},
  \bibinfo{pages}{261101} (\bibinfo{year}{2010}).

\bibitem[{\citenamefont{{Zhuge} et~al.}(1994)\citenamefont{{Zhuge},
  {Centrella}, and {McMillan}}}]{1994PhRvD..50.6247Z}
\bibinfo{author}{\bibfnamefont{X.}~\bibnamefont{{Zhuge}}},
  \bibinfo{author}{\bibfnamefont{J.~M.} \bibnamefont{{Centrella}}},
  \bibnamefont{and} \bibinfo{author}{\bibfnamefont{S.~L.~W.}
  \bibnamefont{{McMillan}}}, \bibinfo{journal}{\prd}
  \textbf{\bibinfo{volume}{50}}, \bibinfo{pages}{6247} (\bibinfo{year}{1994}).

\bibitem[{\citenamefont{{Shibata}}(2005)}]{2005PhRvL..94t1101S}
\bibinfo{author}{\bibfnamefont{M.}~\bibnamefont{{Shibata}}},
  \bibinfo{journal}{\prl} \textbf{\bibinfo{volume}{94}},
  \bibinfo{pages}{201101} (\bibinfo{year}{2005}).

\bibitem[{\citenamefont{{Shibata} et~al.}(2005)\citenamefont{{Shibata},
  {Taniguchi}, and {Ury{\= u}}}}]{2005PhRvD..71h4021S}
\bibinfo{author}{\bibfnamefont{M.}~\bibnamefont{{Shibata}}},
  \bibinfo{author}{\bibfnamefont{K.}~\bibnamefont{{Taniguchi}}},
  \bibnamefont{and} \bibinfo{author}{\bibfnamefont{K.}~\bibnamefont{{Ury{\=
  u}}}}, \bibinfo{journal}{\prd} \textbf{\bibinfo{volume}{71}},
  \bibinfo{pages}{084021} (\bibinfo{year}{2005}).

\bibitem[{\citenamefont{{Oechslin} and {Janka}}(2007)}]{2007PhRvL..99l1102O}
\bibinfo{author}{\bibfnamefont{R.}~\bibnamefont{{Oechslin}}} \bibnamefont{and}
  \bibinfo{author}{\bibfnamefont{H.-T.} \bibnamefont{{Janka}}},
  \bibinfo{journal}{\prl} \textbf{\bibinfo{volume}{99}},
  \bibinfo{pages}{121102} (\bibinfo{year}{2007}).

\bibitem[{\citenamefont{Bauswein et~al.}(2010)\citenamefont{Bauswein, Oechslin,
  and Janka}}]{PhysRevD.81.024012}
\bibinfo{author}{\bibfnamefont{A.}~\bibnamefont{Bauswein}},
  \bibinfo{author}{\bibfnamefont{R.}~\bibnamefont{Oechslin}}, \bibnamefont{and}
  \bibinfo{author}{\bibfnamefont{H.-T.} \bibnamefont{Janka}},
  \bibinfo{journal}{\prd} \textbf{\bibinfo{volume}{81}},
  \bibinfo{pages}{024012} (\bibinfo{year}{2010}).

\bibitem[{\citenamefont{{Hotokezaka} et~al.}(2011)\citenamefont{{Hotokezaka},
  {Kyutoku}, {Okawa}, {Shibata}, and {Kiuchi}}}]{2011PhRvD..83l4008H}
\bibinfo{author}{\bibfnamefont{K.}~\bibnamefont{{Hotokezaka}}},
  \bibinfo{author}{\bibfnamefont{K.}~\bibnamefont{{Kyutoku}}},
  \bibinfo{author}{\bibfnamefont{H.}~\bibnamefont{{Okawa}}},
  \bibinfo{author}{\bibfnamefont{M.}~\bibnamefont{{Shibata}}},
  \bibnamefont{and} \bibinfo{author}{\bibfnamefont{K.}~\bibnamefont{{Kiuchi}}},
  \bibinfo{journal}{\prd} \textbf{\bibinfo{volume}{83}},
  \bibinfo{pages}{124008} (\bibinfo{year}{2011}).

\bibitem[{\citenamefont{{Sekiguchi}
  et~al.}(2011{\natexlab{a}})\citenamefont{{Sekiguchi}, {Kiuchi}, {Kyutoku},
  and {Shibata}}}]{2011PhRvL.107u1101S}
\bibinfo{author}{\bibfnamefont{Y.}~\bibnamefont{{Sekiguchi}}},
  \bibinfo{author}{\bibfnamefont{K.}~\bibnamefont{{Kiuchi}}},
  \bibinfo{author}{\bibfnamefont{K.}~\bibnamefont{{Kyutoku}}},
  \bibnamefont{and}
  \bibinfo{author}{\bibfnamefont{M.}~\bibnamefont{{Shibata}}},
  \bibinfo{journal}{\prl} \textbf{\bibinfo{volume}{107}}, \bibinfo{eid}{211101}
  (\bibinfo{year}{2011}{\natexlab{a}}).

\bibitem[{\citenamefont{{Bauswein} and {Janka}}(2012)}]{2012PhRvL.108a1101B}
\bibinfo{author}{\bibfnamefont{A.}~\bibnamefont{{Bauswein}}} \bibnamefont{and}
  \bibinfo{author}{\bibfnamefont{H.-T.} \bibnamefont{{Janka}}},
  \bibinfo{journal}{\prl} \textbf{\bibinfo{volume}{108}}, \bibinfo{eid}{011101}
  (\bibinfo{year}{2012}).

\bibitem[{\citenamefont{{Isenberg} and {Nester}}(1980)}]{1980grg..conf...23I}
\bibinfo{author}{\bibfnamefont{J.}~\bibnamefont{{Isenberg}}} \bibnamefont{and}
  \bibinfo{author}{\bibfnamefont{J.}~\bibnamefont{{Nester}}}, in
  \emph{\bibinfo{booktitle}{General Relativity and Gravitation}}
  (\bibinfo{publisher}{Plenum Press, New York}, \bibinfo{year}{1980}),
  p.~\bibinfo{pages}{23}.

\bibitem[{\citenamefont{{Wilson} et~al.}(1996)\citenamefont{{Wilson},
  {Mathews}, and {Marronetti}}}]{1996PhRvD..54.1317W}
\bibinfo{author}{\bibfnamefont{J.~R.} \bibnamefont{{Wilson}}},
  \bibinfo{author}{\bibfnamefont{G.~J.} \bibnamefont{{Mathews}}},
  \bibnamefont{and}
  \bibinfo{author}{\bibfnamefont{P.}~\bibnamefont{{Marronetti}}},
  \bibinfo{journal}{\prd} \textbf{\bibinfo{volume}{54}}, \bibinfo{pages}{1317}
  (\bibinfo{year}{1996}).

\bibitem[{\citenamefont{{Oechslin} et~al.}(2002)\citenamefont{{Oechslin},
  {Rosswog}, and {Thielemann}}}]{2002PhRvD..65j3005O}
\bibinfo{author}{\bibfnamefont{R.}~\bibnamefont{{Oechslin}}},
  \bibinfo{author}{\bibfnamefont{S.}~\bibnamefont{{Rosswog}}},
  \bibnamefont{and} \bibinfo{author}{\bibfnamefont{F.-K.}
  \bibnamefont{{Thielemann}}}, \bibinfo{journal}{\prd}
  \textbf{\bibinfo{volume}{65}}, \bibinfo{pages}{103005}
  (\bibinfo{year}{2002}).

\bibitem[{\citenamefont{{Read} et~al.}(2009{\natexlab{b}})\citenamefont{{Read},
  {Lackey}, {Owen}, and {Friedman}}}]{2009PhRvD..79l4032R}
\bibinfo{author}{\bibfnamefont{J.~S.} \bibnamefont{{Read}}},
  \bibinfo{author}{\bibfnamefont{B.~D.} \bibnamefont{{Lackey}}},
  \bibinfo{author}{\bibfnamefont{B.~J.} \bibnamefont{{Owen}}},
  \bibnamefont{and} \bibinfo{author}{\bibfnamefont{J.~L.}
  \bibnamefont{{Friedman}}}, \bibinfo{journal}{\prd}
  \textbf{\bibinfo{volume}{79}}, \bibinfo{pages}{124032}
  (\bibinfo{year}{2009}{\natexlab{b}}).

\bibitem[{\citenamefont{{Hebeler} et~al.}(2010)\citenamefont{{Hebeler},
  {Lattimer}, {Pethick}, and {Schwenk}}}]{2010PhRvL.105p1102H}
\bibinfo{author}{\bibfnamefont{K.}~\bibnamefont{{Hebeler}}},
  \bibinfo{author}{\bibfnamefont{J.~M.} \bibnamefont{{Lattimer}}},
  \bibinfo{author}{\bibfnamefont{C.~J.} \bibnamefont{{Pethick}}},
  \bibnamefont{and}
  \bibinfo{author}{\bibfnamefont{A.}~\bibnamefont{{Schwenk}}},
  \bibinfo{journal}{\prl} \textbf{\bibinfo{volume}{105}},
  \bibinfo{pages}{161102} (\bibinfo{year}{2010}).

\bibitem[{\citenamefont{{Liu} et~al.}(2008)\citenamefont{{Liu}, {Shapiro},
  {Etienne}, and {Taniguchi}}}]{2008PhRvD..78b4012L}
\bibinfo{author}{\bibfnamefont{Y.~T.} \bibnamefont{{Liu}}},
  \bibinfo{author}{\bibfnamefont{S.~L.} \bibnamefont{{Shapiro}}},
  \bibinfo{author}{\bibfnamefont{Z.~B.} \bibnamefont{{Etienne}}},
  \bibnamefont{and}
  \bibinfo{author}{\bibfnamefont{K.}~\bibnamefont{{Taniguchi}}},
  \bibinfo{journal}{\prd} \textbf{\bibinfo{volume}{78}},
  \bibinfo{pages}{024012} (\bibinfo{year}{2008}).

\bibitem[{\citenamefont{{Giacomazzo} et~al.}(2009)\citenamefont{{Giacomazzo},
  {Rezzolla}, and {Baiotti}}}]{2009MNRAS.399L.164G}
\bibinfo{author}{\bibfnamefont{B.}~\bibnamefont{{Giacomazzo}}},
  \bibinfo{author}{\bibfnamefont{L.}~\bibnamefont{{Rezzolla}}},
  \bibnamefont{and}
  \bibinfo{author}{\bibfnamefont{L.}~\bibnamefont{{Baiotti}}},
  \bibinfo{journal}{Mon. Not. R. Astron. Soc.} \textbf{\bibinfo{volume}{399}},
  \bibinfo{pages}{L164} (\bibinfo{year}{2009}).

\bibitem[{\citenamefont{{Giacomazzo} et~al.}(2011)\citenamefont{{Giacomazzo},
  {Rezzolla}, and {Baiotti}}}]{2011PhRvD..83d4014G}
\bibinfo{author}{\bibfnamefont{B.}~\bibnamefont{{Giacomazzo}}},
  \bibinfo{author}{\bibfnamefont{L.}~\bibnamefont{{Rezzolla}}},
  \bibnamefont{and}
  \bibinfo{author}{\bibfnamefont{L.}~\bibnamefont{{Baiotti}}},
  \bibinfo{journal}{\prd} \textbf{\bibinfo{volume}{83}},
  \bibinfo{pages}{044014} (\bibinfo{year}{2011}).

\bibitem[{\citenamefont{{Ruffert} et~al.}(1996)\citenamefont{{Ruffert},
  {Janka}, and {Schaefer}}}]{1996A&A...311..532R}
\bibinfo{author}{\bibfnamefont{M.}~\bibnamefont{{Ruffert}}},
  \bibinfo{author}{\bibfnamefont{H.-T.} \bibnamefont{{Janka}}},
  \bibnamefont{and}
  \bibinfo{author}{\bibfnamefont{G.}~\bibnamefont{{Schaefer}}},
  \bibinfo{journal}{Astron. Astrophys.} \textbf{\bibinfo{volume}{311}},
  \bibinfo{pages}{532} (\bibinfo{year}{1996}).

\bibitem[{\citenamefont{{Ruffert} et~al.}(1997)\citenamefont{{Ruffert},
  {Janka}, {Takahashi}, and {Schaefer}}}]{1997A&A...319..122R}
\bibinfo{author}{\bibfnamefont{M.}~\bibnamefont{{Ruffert}}},
  \bibinfo{author}{\bibfnamefont{H.-T.} \bibnamefont{{Janka}}},
  \bibinfo{author}{\bibfnamefont{K.}~\bibnamefont{{Takahashi}}},
  \bibnamefont{and}
  \bibinfo{author}{\bibfnamefont{G.}~\bibnamefont{{Schaefer}}},
  \bibinfo{journal}{Astron. Astrophys.} \textbf{\bibinfo{volume}{319}},
  \bibinfo{pages}{122} (\bibinfo{year}{1997}).

\bibitem[{\citenamefont{{Bildsten} and {Cutler}}(1992)}]{1992ApJ...400..175B}
\bibinfo{author}{\bibfnamefont{L.}~\bibnamefont{{Bildsten}}} \bibnamefont{and}
  \bibinfo{author}{\bibfnamefont{C.}~\bibnamefont{{Cutler}}},
  \bibinfo{journal}{Astrophys. J.} \textbf{\bibinfo{volume}{400}},
  \bibinfo{pages}{175} (\bibinfo{year}{1992}).

\bibitem[{\citenamefont{{Kochanek}}(1992)}]{1992ApJ...398..234K}
\bibinfo{author}{\bibfnamefont{C.~S.} \bibnamefont{{Kochanek}}},
  \bibinfo{journal}{Astrophys. J.} \textbf{\bibinfo{volume}{398}},
  \bibinfo{pages}{234} (\bibinfo{year}{1992}).

\bibitem[{\citenamefont{{Weber}}(1999)}]{1999Weber}
\bibinfo{author}{\bibfnamefont{F.}~\bibnamefont{{Weber}}},
  \emph{\bibinfo{title}{{Pulsars as Astrophysical Laboratories for Nuclear and
  Particle Physics}}} (\bibinfo{publisher}{IOP Publishing, Bristol},
  \bibinfo{year}{1999}).

\bibitem[{\citenamefont{{Hebeler} and {Schwenk}}(2010)}]{2010PhRvC..82a4314H}
\bibinfo{author}{\bibfnamefont{K.}~\bibnamefont{{Hebeler}}} \bibnamefont{and}
  \bibinfo{author}{\bibfnamefont{A.}~\bibnamefont{{Schwenk}}},
  \bibinfo{journal}{\prc} \textbf{\bibinfo{volume}{82}}, \bibinfo{eid}{014314}
  (\bibinfo{year}{2010}).

\bibitem[{\citenamefont{{Hebeler} et~al.}(2012)\citenamefont{{Hebeler},
  {Pethick}, {Lattimer}, and {Schwenk}}}]{Hebelerprep}
\bibinfo{author}{\bibfnamefont{K.}~\bibnamefont{{Hebeler}}},
  \bibinfo{author}{\bibfnamefont{C.~J.} \bibnamefont{{Pethick}}},
  \bibinfo{author}{\bibfnamefont{J.~M.} \bibnamefont{{Lattimer}}},
  \bibnamefont{and}
  \bibinfo{author}{\bibfnamefont{A.}~\bibnamefont{{Schwenk}}},
  \emph{\bibinfo{title}{in preparation}} (\bibinfo{year}{2012}).

\bibitem[{\citenamefont{{Lattimer} and {Prakash}}(2001)}]{2001ApJ...550..426L}
\bibinfo{author}{\bibfnamefont{J.~M.} \bibnamefont{{Lattimer}}}
  \bibnamefont{and}
  \bibinfo{author}{\bibfnamefont{M.}~\bibnamefont{{Prakash}}},
  \bibinfo{journal}{\apj} \textbf{\bibinfo{volume}{550}}, \bibinfo{pages}{426}
  (\bibinfo{year}{2001}).

\bibitem[{\citenamefont{Chodos et~al.}(1974)\citenamefont{Chodos, Jaffe,
  Johnson, Thorn, and Weisskopf}}]{PhysRevD.9.3471}
\bibinfo{author}{\bibfnamefont{A.}~\bibnamefont{Chodos}},
  \bibinfo{author}{\bibfnamefont{R.~L.} \bibnamefont{Jaffe}},
  \bibinfo{author}{\bibfnamefont{K.}~\bibnamefont{Johnson}},
  \bibinfo{author}{\bibfnamefont{C.~B.} \bibnamefont{Thorn}}, \bibnamefont{and}
  \bibinfo{author}{\bibfnamefont{V.~F.} \bibnamefont{Weisskopf}},
  \bibinfo{journal}{Phys. Rev. D} \textbf{\bibinfo{volume}{9}},
  \bibinfo{pages}{3471} (\bibinfo{year}{1974}).

\bibitem[{\citenamefont{Farhi and Jaffe}(1984)}]{Farhi:1984qu}
\bibinfo{author}{\bibfnamefont{E.}~\bibnamefont{Farhi}} \bibnamefont{and}
  \bibinfo{author}{\bibfnamefont{R.~L.} \bibnamefont{Jaffe}},
  \bibinfo{journal}{Phys. Rev.} \textbf{\bibinfo{volume}{D30}},
  \bibinfo{pages}{2379} (\bibinfo{year}{1984}).

\bibitem[{\citenamefont{Bodmer}(1971)}]{PhysRevD.4.1601}
\bibinfo{author}{\bibfnamefont{A.~R.} \bibnamefont{Bodmer}},
  \bibinfo{journal}{Phys. Rev. D} \textbf{\bibinfo{volume}{4}},
  \bibinfo{pages}{1601} (\bibinfo{year}{1971}).

\bibitem[{\citenamefont{Witten}(1984)}]{PhysRevD.30.272}
\bibinfo{author}{\bibfnamefont{E.}~\bibnamefont{Witten}},
  \bibinfo{journal}{Phys. Rev. D} \textbf{\bibinfo{volume}{30}},
  \bibinfo{pages}{272} (\bibinfo{year}{1984}).

\bibitem[{\citenamefont{{Bauswein} et~al.}(2009)\citenamefont{{Bauswein},
  {Janka}, {Oechslin}, {Pagliara}, {Sagert}, {Schaffner-Bielich}, {Hohle}, and
  {Neuh{\"a}user}}}]{2009PhRvL.103a1101B}
\bibinfo{author}{\bibfnamefont{A.}~\bibnamefont{{Bauswein}}},
  \bibinfo{author}{\bibfnamefont{H.-T.} \bibnamefont{{Janka}}},
  \bibinfo{author}{\bibfnamefont{R.}~\bibnamefont{{Oechslin}}},
  \bibinfo{author}{\bibfnamefont{G.}~\bibnamefont{{Pagliara}}},
  \bibinfo{author}{\bibfnamefont{I.}~\bibnamefont{{Sagert}}},
  \bibinfo{author}{\bibfnamefont{J.}~\bibnamefont{{Schaffner-Bielich}}},
  \bibinfo{author}{\bibfnamefont{M.~M.} \bibnamefont{{Hohle}}},
  \bibnamefont{and}
  \bibinfo{author}{\bibfnamefont{R.}~\bibnamefont{{Neuh{\"a}user}}},
  \bibinfo{journal}{\prl} \textbf{\bibinfo{volume}{103}},
  \bibinfo{pages}{011101} (\bibinfo{year}{2009}).

\bibitem[{\citenamefont{{Shen} et~al.}(2011{\natexlab{a}})\citenamefont{{Shen},
  {Horowitz}, and {Teige}}}]{2011PhRvC..83c5802S}
\bibinfo{author}{\bibfnamefont{G.}~\bibnamefont{{Shen}}},
  \bibinfo{author}{\bibfnamefont{C.~J.} \bibnamefont{{Horowitz}}},
  \bibnamefont{and} \bibinfo{author}{\bibfnamefont{S.}~\bibnamefont{{Teige}}},
  \bibinfo{journal}{\prc} \textbf{\bibinfo{volume}{83}},
  \bibinfo{pages}{035802} (\bibinfo{year}{2011}{\natexlab{a}}).

\bibitem[{\citenamefont{{Lattimer} and {Swesty}}(1991)}]{1991NuPhA.535..331L}
\bibinfo{author}{\bibfnamefont{J.~M.} \bibnamefont{{Lattimer}}}
  \bibnamefont{and} \bibinfo{author}{\bibfnamefont{F.~D.}
  \bibnamefont{{Swesty}}}, \bibinfo{journal}{Nucl. Phys. A}
  \textbf{\bibinfo{volume}{535}}, \bibinfo{pages}{331} (\bibinfo{year}{1991}).

\bibitem[{\citenamefont{{Shen} et~al.}(1998)\citenamefont{{Shen}, {Toki},
  {Oyamatsu}, and {Sumiyoshi}}}]{1998NuPhA.637..435S}
\bibinfo{author}{\bibfnamefont{H.}~\bibnamefont{{Shen}}},
  \bibinfo{author}{\bibfnamefont{H.}~\bibnamefont{{Toki}}},
  \bibinfo{author}{\bibfnamefont{K.}~\bibnamefont{{Oyamatsu}}},
  \bibnamefont{and}
  \bibinfo{author}{\bibfnamefont{K.}~\bibnamefont{{Sumiyoshi}}},
  \bibinfo{journal}{Nucl. Phys. A} \textbf{\bibinfo{volume}{637}},
  \bibinfo{pages}{435} (\bibinfo{year}{1998}).

\bibitem[{\citenamefont{{Shen} et~al.}(2011{\natexlab{b}})\citenamefont{{Shen},
  {Horowitz}, and {O'Connor}}}]{2011arXiv1103.5174S}
\bibinfo{author}{\bibfnamefont{G.}~\bibnamefont{{Shen}}},
  \bibinfo{author}{\bibfnamefont{C.~J.} \bibnamefont{{Horowitz}}},
  \bibnamefont{and}
  \bibinfo{author}{\bibfnamefont{E.}~\bibnamefont{{O'Connor}}},
  \bibinfo{journal}{\prc} \textbf{\bibinfo{volume}{83}}, \bibinfo{eid}{065808}
  (\bibinfo{year}{2011}{\natexlab{b}}).

\bibitem[{\citenamefont{{The EoS table has been provided by G.~Pagliara,
  I.~Sagert (details can be found in~\cite{PhysRevD.81.024012})}}()}]{refmit60}
\bibinfo{author}{\bibnamefont{{The EoS table has been provided by G.~Pagliara,
  I.~Sagert (details can be found in~\cite{PhysRevD.81.024012})}}}.

\bibitem[{\citenamefont{{Pandharipande} and
  {Smith}}(1975)}]{1975PhLB...59...15P}
\bibinfo{author}{\bibfnamefont{V.~R.} \bibnamefont{{Pandharipande}}}
  \bibnamefont{and} \bibinfo{author}{\bibfnamefont{R.~A.}
  \bibnamefont{{Smith}}}, \bibinfo{journal}{Phys. Lett. B}
  \textbf{\bibinfo{volume}{59}}, \bibinfo{pages}{15} (\bibinfo{year}{1975}).

\bibitem[{\citenamefont{{Bowers} et~al.}(1975)\citenamefont{{Bowers},
  {Gleeson}, and {Daryl Pedigo}}}]{1975PhRvD..12.3043B}
\bibinfo{author}{\bibfnamefont{R.~L.} \bibnamefont{{Bowers}}},
  \bibinfo{author}{\bibfnamefont{A.~M.} \bibnamefont{{Gleeson}}},
  \bibnamefont{and} \bibinfo{author}{\bibfnamefont{R.}~\bibnamefont{{Daryl
  Pedigo}}}, \bibinfo{journal}{\prd} \textbf{\bibinfo{volume}{12}},
  \bibinfo{pages}{3043} (\bibinfo{year}{1975}).

\bibitem[{\citenamefont{{Gondek-Rosinska} and
  {Limousin}}(2008)}]{2008arXiv0801.4829G}
\bibinfo{author}{\bibfnamefont{D.}~\bibnamefont{{Gondek-Rosinska}}}
  \bibnamefont{and}
  \bibinfo{author}{\bibfnamefont{F.}~\bibnamefont{{Limousin}}},
  \bibinfo{journal}{ArXiv e-prints}  (\bibinfo{year}{2008}),
  \eprint{0801.4829}.

\bibitem[{\citenamefont{{Zdunik}}(2000)}]{2000AA...359..311Z}
\bibinfo{author}{\bibfnamefont{J.~L.} \bibnamefont{{Zdunik}}},
  \bibinfo{journal}{Astron. Astrophys.} \textbf{\bibinfo{volume}{359}},
  \bibinfo{pages}{311} (\bibinfo{year}{2000}).

\bibitem[{\citenamefont{{Wiringa} et~al.}(1988)\citenamefont{{Wiringa}, {Fiks},
  and {Fabrocini}}}]{1988PhRvC..38.1010W}
\bibinfo{author}{\bibfnamefont{R.~B.} \bibnamefont{{Wiringa}}},
  \bibinfo{author}{\bibfnamefont{V.}~\bibnamefont{{Fiks}}}, \bibnamefont{and}
  \bibinfo{author}{\bibfnamefont{A.}~\bibnamefont{{Fabrocini}}},
  \bibinfo{journal}{\prc} \textbf{\bibinfo{volume}{38}}, \bibinfo{pages}{1010}
  (\bibinfo{year}{1988}).

\bibitem[{\citenamefont{{Akmal} et~al.}(1998)\citenamefont{{Akmal},
  {Pandharipande}, and {Ravenhall}}}]{1998PhRvC..58.1804A}
\bibinfo{author}{\bibfnamefont{A.}~\bibnamefont{{Akmal}}},
  \bibinfo{author}{\bibfnamefont{V.~R.} \bibnamefont{{Pandharipande}}},
  \bibnamefont{and} \bibinfo{author}{\bibfnamefont{D.~G.}
  \bibnamefont{{Ravenhall}}}, \bibinfo{journal}{\prc}
  \textbf{\bibinfo{volume}{58}}, \bibinfo{pages}{1804} (\bibinfo{year}{1998}).

\bibitem[{\citenamefont{{Baldo} et~al.}(2003)\citenamefont{{Baldo}, {Buballa},
  {Burgio}, {Neumann}, {Oertel}, and {Schulze}}}]{2003PhLB..562..153B}
\bibinfo{author}{\bibfnamefont{M.}~\bibnamefont{{Baldo}}},
  \bibinfo{author}{\bibfnamefont{M.}~\bibnamefont{{Buballa}}},
  \bibinfo{author}{\bibfnamefont{G.~F.} \bibnamefont{{Burgio}}},
  \bibinfo{author}{\bibfnamefont{F.}~\bibnamefont{{Neumann}}},
  \bibinfo{author}{\bibfnamefont{M.}~\bibnamefont{{Oertel}}}, \bibnamefont{and}
  \bibinfo{author}{\bibfnamefont{H.-J.} \bibnamefont{{Schulze}}},
  \bibinfo{journal}{Phys. Lett. B} \textbf{\bibinfo{volume}{562}},
  \bibinfo{pages}{153} (\bibinfo{year}{2003}).

\bibitem[{\citenamefont{{Douchin} and {Haensel}}(2001)}]{2001AA...380..151D}
\bibinfo{author}{\bibfnamefont{F.}~\bibnamefont{{Douchin}}} \bibnamefont{and}
  \bibinfo{author}{\bibfnamefont{P.}~\bibnamefont{{Haensel}}},
  \bibinfo{journal}{Astron. Astrophys.} \textbf{\bibinfo{volume}{380}},
  \bibinfo{pages}{151} (\bibinfo{year}{2001}).

\bibitem[{\citenamefont{{Glendenning}}(1985)}]{1985ApJ...293..470G}
\bibinfo{author}{\bibfnamefont{N.~K.} \bibnamefont{{Glendenning}}},
  \bibinfo{journal}{\apj} \textbf{\bibinfo{volume}{293}}, \bibinfo{pages}{470}
  (\bibinfo{year}{1985}).

\bibitem[{\citenamefont{{Baldo} et~al.}(1997)\citenamefont{{Baldo}, {Bombaci},
  and {Burgio}}}]{1997AA...328..274B}
\bibinfo{author}{\bibfnamefont{M.}~\bibnamefont{{Baldo}}},
  \bibinfo{author}{\bibfnamefont{I.}~\bibnamefont{{Bombaci}}},
  \bibnamefont{and} \bibinfo{author}{\bibfnamefont{G.~F.}
  \bibnamefont{{Burgio}}}, \bibinfo{journal}{Astron. Astrophys.}
  \textbf{\bibinfo{volume}{328}}, \bibinfo{pages}{274} (\bibinfo{year}{1997}).

\bibitem[{\citenamefont{{Bethe} and {Johnson}}(1974)}]{1974NuPhA.230....1B}
\bibinfo{author}{\bibfnamefont{H.~A.} \bibnamefont{{Bethe}}} \bibnamefont{and}
  \bibinfo{author}{\bibfnamefont{M.~B.} \bibnamefont{{Johnson}}},
  \bibinfo{journal}{Nucl. Phys. A} \textbf{\bibinfo{volume}{230}},
  \bibinfo{pages}{1} (\bibinfo{year}{1974}).

\bibitem[{\citenamefont{{Friedman} and
  {Pandharipande}}(1981)}]{1981NuPhA.361..502F}
\bibinfo{author}{\bibfnamefont{B.}~\bibnamefont{{Friedman}}} \bibnamefont{and}
  \bibinfo{author}{\bibfnamefont{V.~R.} \bibnamefont{{Pandharipande}}},
  \bibinfo{journal}{Nucl. Phys. A} \textbf{\bibinfo{volume}{361}},
  \bibinfo{pages}{502} (\bibinfo{year}{1981}).

\bibitem[{\citenamefont{{Lasota} et~al.}(1996)\citenamefont{{Lasota},
  {Haensel}, and {Abramowicz}}}]{1996ApJ...456..300L}
\bibinfo{author}{\bibfnamefont{J.-P.} \bibnamefont{{Lasota}}},
  \bibinfo{author}{\bibfnamefont{P.}~\bibnamefont{{Haensel}}},
  \bibnamefont{and} \bibinfo{author}{\bibfnamefont{M.~A.}
  \bibnamefont{{Abramowicz}}}, \bibinfo{journal}{\apj}
  \textbf{\bibinfo{volume}{456}}, \bibinfo{pages}{300} (\bibinfo{year}{1996}).

\bibitem[{\citenamefont{{Stergioulas} et~al.}(2011)\citenamefont{{Stergioulas},
  {Bauswein}, {Zagkouris}, and {Janka}}}]{2011arXiv1105.0368S}
\bibinfo{author}{\bibfnamefont{N.}~\bibnamefont{{Stergioulas}}},
  \bibinfo{author}{\bibfnamefont{A.}~\bibnamefont{{Bauswein}}},
  \bibinfo{author}{\bibfnamefont{K.}~\bibnamefont{{Zagkouris}}},
  \bibnamefont{and} \bibinfo{author}{\bibfnamefont{H.-T.}
  \bibnamefont{{Janka}}}, \bibinfo{journal}{Mon. Not. R. Astron. Soc.}
  \textbf{\bibinfo{volume}{418}}, \bibinfo{pages}{427} (\bibinfo{year}{2011}).

\bibitem[{\citenamefont{{Hild} et~al.}(2010)\citenamefont{{Hild}, {Chelkowski},
  {Freise}, {Franc}, {Morgado}, {Flaminio}, and
  {DeSalvo}}}]{2010CQGra..27a5003H}
\bibinfo{author}{\bibfnamefont{S.}~\bibnamefont{{Hild}}},
  \bibinfo{author}{\bibfnamefont{S.}~\bibnamefont{{Chelkowski}}},
  \bibinfo{author}{\bibfnamefont{A.}~\bibnamefont{{Freise}}},
  \bibinfo{author}{\bibfnamefont{J.}~\bibnamefont{{Franc}}},
  \bibinfo{author}{\bibfnamefont{N.}~\bibnamefont{{Morgado}}},
  \bibinfo{author}{\bibfnamefont{R.}~\bibnamefont{{Flaminio}}},
  \bibnamefont{and}
  \bibinfo{author}{\bibfnamefont{R.}~\bibnamefont{{DeSalvo}}},
  \bibinfo{journal}{Class. Quantum Grav.} \textbf{\bibinfo{volume}{27}},
  \bibinfo{pages}{015003} (\bibinfo{year}{2010}).

\bibitem[{\citenamefont{{Maggiore}}(2008)}]{2008Maggiore}
\bibinfo{author}{\bibfnamefont{M.}~\bibnamefont{{Maggiore}}},
  \emph{\bibinfo{title}{{Gravitational Waves, Volume 1: Theory and
  Experiments}}} (\bibinfo{publisher}{Oxford University Press, Oxford},
  \bibinfo{year}{2008}).

\bibitem[{\citenamefont{{Hild} et~al.}(2011)\citenamefont{{Hild}, {Abernathy},
  {Acernese}, {Amaro-Seoane}, {Andersson}, {Arun}, {Barone}, {Barr},
  {Barsuglia}, {Beker} et~al.}}]{2011CQGra..28i4013H}
\bibinfo{author}{\bibfnamefont{S.}~\bibnamefont{{Hild}}},
  \bibinfo{author}{\bibfnamefont{M.}~\bibnamefont{{Abernathy}}},
  \bibinfo{author}{\bibfnamefont{F.}~\bibnamefont{{Acernese}}},
  \bibinfo{author}{\bibfnamefont{P.}~\bibnamefont{{Amaro-Seoane}}},
  \bibinfo{author}{\bibfnamefont{N.}~\bibnamefont{{Andersson}}},
  \bibinfo{author}{\bibfnamefont{K.}~\bibnamefont{{Arun}}},
  \bibinfo{author}{\bibfnamefont{F.}~\bibnamefont{{Barone}}},
  \bibinfo{author}{\bibfnamefont{B.}~\bibnamefont{{Barr}}},
  \bibinfo{author}{\bibfnamefont{M.}~\bibnamefont{{Barsuglia}}},
  \bibinfo{author}{\bibfnamefont{M.}~\bibnamefont{{Beker}}},
  \bibnamefont{et~al.}, \bibinfo{journal}{Class. Quantum Grav.}
  \textbf{\bibinfo{volume}{28}}, \bibinfo{pages}{094013}
  (\bibinfo{year}{2011}).

\bibitem[{\citenamefont{{Vallisneri}}(2008)}]{2008PhRvD..77d2001V}
\bibinfo{author}{\bibfnamefont{M.}~\bibnamefont{{Vallisneri}}},
  \bibinfo{journal}{\prd} \textbf{\bibinfo{volume}{77}}, \bibinfo{eid}{042001}
  (\bibinfo{year}{2008}).

\bibitem[{\citenamefont{{Andersson} and
  {Kokkotas}}(1998)}]{1998MNRAS.299.1059A}
\bibinfo{author}{\bibfnamefont{N.}~\bibnamefont{{Andersson}}} \bibnamefont{and}
  \bibinfo{author}{\bibfnamefont{K.~D.} \bibnamefont{{Kokkotas}}},
  \bibinfo{journal}{Mon. Not. R. Astron. Soc.} \textbf{\bibinfo{volume}{299}},
  \bibinfo{pages}{1059} (\bibinfo{year}{1998}).

\bibitem[{\citenamefont{{Gaertig} and {Kokkotas}}(2011)}]{2011PhRvD..83f4031G}
\bibinfo{author}{\bibfnamefont{E.}~\bibnamefont{{Gaertig}}} \bibnamefont{and}
  \bibinfo{author}{\bibfnamefont{K.~D.} \bibnamefont{{Kokkotas}}},
  \bibinfo{journal}{\prd} \textbf{\bibinfo{volume}{83}}, \bibinfo{eid}{064031}
  (\bibinfo{year}{2011}).

\bibitem[{\citenamefont{{Sekiguchi}
  et~al.}(2011{\natexlab{b}})\citenamefont{{Sekiguchi}, {Kiuchi}, {Kyutoku},
  and {Shibata}}}]{2011PhRvL.107e1102S}
\bibinfo{author}{\bibfnamefont{Y.}~\bibnamefont{{Sekiguchi}}},
  \bibinfo{author}{\bibfnamefont{K.}~\bibnamefont{{Kiuchi}}},
  \bibinfo{author}{\bibfnamefont{K.}~\bibnamefont{{Kyutoku}}},
  \bibnamefont{and}
  \bibinfo{author}{\bibfnamefont{M.}~\bibnamefont{{Shibata}}},
  \bibinfo{journal}{\prl} \textbf{\bibinfo{volume}{107}}, \bibinfo{eid}{051102}
  (\bibinfo{year}{2011}{\natexlab{b}}).

\bibitem[{\citenamefont{{Takami} et~al.}(2011)\citenamefont{{Takami},
  {Rezzolla}, and {Yoshida}}}]{2011MNRAS.416L...1T}
\bibinfo{author}{\bibfnamefont{K.}~\bibnamefont{{Takami}}},
  \bibinfo{author}{\bibfnamefont{L.}~\bibnamefont{{Rezzolla}}},
  \bibnamefont{and}
  \bibinfo{author}{\bibfnamefont{S.}~\bibnamefont{{Yoshida}}},
  \bibinfo{journal}{Mon. Not. R. Astron. Soc.} \textbf{\bibinfo{volume}{416}},
  \bibinfo{pages}{L1} (\bibinfo{year}{2011}).

\end{thebibliography}

\end{document}